\documentclass{article}
\usepackage{amsfonts,amssymb}
\usepackage{stmaryrd}
\usepackage{amsmath}
\usepackage{epsfig}
\usepackage{verbatim}
\usepackage{pstricks}
\usepackage{xy}
\xyoption{all}
\CompileMatrices

\mathcode`\<="4268 
\mathcode`\>="5269 
\mathchardef\gt="313E 
\mathchardef\lt="313C 

\newcommand{\id}{{\rm id}}
\newcommand{\Wee}{\WWW}
\newcommand{\Cee}{\PPP}
\newcommand{\See}{\SSS}
\newcommand{\XC}{\CCc_X}
\newcommand{\CX}{\CCc^X}

\renewcommand{\to}{\xymatrix@C-.5pc{\ar[r]&}}
\newcommand{\ot}{\xymatrix@C-.5pc{& \ar[l]}}
\newcommand{\tto}[1]{\xymatrix@C-.5pc{\ar[r]^{#1}&}}
\newcommand{\oot}[1]{\xymatrix@C-.5pc{&\ar[l]_{#1}}}
\newcommand{\mono}{\xymatrix@C-.5pc{\ar@{>->}[r]&}} 
\newcommand{\epi}{\xymatrix@C-.5pc{\ar@{->>}[r]&}}
\newcommand{\mmono}[1]{\xymatrix@C-.5pc{\ar@{>->}[r]^{#1}&}} 
\newcommand{\eepi}[1]{\xymatrix@C-.5pc{\ar@{->>}[r]^{#1}&}}
\renewcommand{\mapsto}{\xymatrix@C-.5pc{\ar@{|->}[r]&}}
\newcommand{\mmapsto}[1]{\xymatrix@C-.5pc{\ar@{|->}[r]^{#1}&}}
\newcommand{\inclusion}{\xymatrix@C-.5pc{\ar@{^{(}->}[r] &}}
\newcommand{\iinclusion}[1]{\xymatrix@C-.5pc{\ar@{^{(}->}[r]^{#1}&}}

\renewcommand{\Bbb}{\bf}

\newcommand{{\MMM}}{{\cal M}}

\newcommand{{\PPP}}{{\cal P}}

\newcommand{\SSS}{{\cal S}}
\newcommand{\TTT}{{\cal T}}

\newcommand{\WWW}{{\cal W}}

\newcommand{\CCc}{{\Bbb C}}
\newcommand{\DDd}{{\Bbb D}}

\newcommand{\WP}{\mbox{\Large $\wp$}}

 \def\pushright#1{{
    \parfillskip=0pt            
    \widowpenalty=10000         
    \displaywidowpenalty=10000  
    \finalhyphendemerits=0      
    \leavevmode                 
    \unskip                     
    \nobreak                    
    \hfil                       
    \penalty50                  
    \hskip.2em                  
    \null                       
    \hfill                      
    {#1}                        
    \par}}                      

 \def\qed{\pushright{$\square$}\penalty-700 \smallskip}

\newenvironment{prf}[1]{\begin{trivlist} \item[{\bf ~Proof}#1.]}%
{\qed\end{trivlist}} 

\newcommand{\be}[1]{\begin{#1}}

\newcommand{\ee}[1]{\end{#1}}
\newcommand{\ba}[1]{\begin{array}{#1}}
\newcommand{\ea}{\end{array}}
\newcommand{\bea}{\begin{eqnarray}}
\newcommand{\eea}{\end{eqnarray}}
\newcommand{\bear}{\begin{eqnarray*}}
\newcommand{\eear}{\end{eqnarray*}}
\newcommand{\bpr}{\begin{prf}{}}
\newcommand{\epr}{\end{prf}}
\newcommand{\bprf}[1]{\begin{prf}{#1}}
\newcommand{\eprf}{\end{prf}}

\newtheorem{thm}{Theorem}[section]
\newtheorem{defn}[thm]{Definition}
\newtheorem{prop}[thm]{Proposition}
\newtheorem{lemma}[thm]{Lemma}
\newtheorem{corollary}[thm]{Corollary}
\newtheorem{cond}{}[thm]

\newcommand{\adjbis}{\dag}
\newcommand{\adj}{\dag}

\newcommand{\qo}{q}
\newcommand{\cs}{c}
\newcommand{\Cperm}{{\Bbb C}_p}
\newcommand{\Cclas}{{\Bbb C}_\copyright}
\newcommand{\classsub}{\copyright}
\newcommand{\CclasG}{{\Bbb C}_f}
\newcommand{\fsharp}{\WP \varphi}

\newcommand{\bit}{\begin{itemize}}
\newcommand{\eit}{\end{itemize}\par\noindent}
\newcommand{\ben}{\begin{enumerate}}
\newcommand{\een}{\end{enumerate}\par\noindent}
\newcommand{\beq}{\begin{equation}}
\newcommand{\eeq}{\end{equation}\par\noindent}
\newcommand{\beqa}{\begin{eqnarray*}}
\newcommand{\eeqa}{\end{eqnarray*}\par\noindent}
\newcommand{\beqn}{\begin{eqnarray}}
\newcommand{\eeqn}{\end{eqnarray}\par\noindent}

\title{Classical and quantum structuralism\thanks{This work is supported by EPSRD and ONR.}} 
\author{Bob Coecke,  \'Eric Oliver Paquette and Dusko Pavlovic\\
Oxford University and Kestrel Institute}
\date{}
\begin{document}
\maketitle 

\begin{abstract}
In recent work, symmetric dagger-monoidal (SDM) categories have emerged as a convenient categorical formalization of quantum mechanics. The objects represent physical systems, the morphisms physical
operations, whereas the tensors describe
composite systems. Classical data turn out to correspond to
Frobenius algebras with some additional properties. They express the distinguishing capabilities of classical data: in contrast with quantum data, classical data can be copied and deleted. The algebraic approach thus shifts the paradigm of "quantization" of a classical theory to "classicization" of a quantum theory.
Remarkably, the simple SDM framework suffices not only for this conceptual
shift, but even allows us to distinguish the deterministic classical
operations (i.e. functions) from the nondeterministic classical operations
(i.e. relations), and the probabilistic classical operations (stochastic
maps). Moreover, a combination of some basic categorical constructions (due to Kleisli, resp. Grothendieck) with the categorical presentations of quantum states, provides a resource sensitive account of various quantum-classical interactions: of classical control of quantum data, of classical data arising from quantum measurements, as well as of the classical data processing in-between controls and measurements. A salient feature here is
the graphical calculus for categorical quantum mechanics, which allows a purely diagrammatic representation of classical-quantum interaction.
\end{abstract}

\section{Introduction}

Quantum systems are very different from their classical counterparts.
This is captured within the quantum mechanical formalism as follows:
\bit
\item A compound quantum system is not described by a Cartesian structure --- i.e.~a structure in which all the properties of a joint system can be traced back to those of its components --- but by a genuinely linear tensor. Physically, this fact is witnessed by  `entangled states'.
\item A Hilbert space admits many different choices of a basis.  Physically this means that a quantum system admits `incompatible observables' to which no sharp  values can be simultaneously attributed.  Via the eigenvector-eigenvalue connection each of the bases also represents the classical data type corresponding to the observable.  
\eit
These features will  be the key players in a general abstract theory of quantum systems, and in particular, a theory of classical-quantum interaction.  In \cite{Abramsky-Coecke} Abramsky and Coecke axiomatized entangled states in terms of dagger compact categories.  Here, we refer to the objects in such a dagger compact category, each coming with a chosen \em Bell state\em, as \em quantum systems\em.  In \cite{Coecke-Pavlovic} Coecke and Pavlovic axiomatized quantum observables within this context as special dagger Frobenius algebras, to which we refer here as \em classical structures\em, or \em basis structures\em.   

One can think of quantum observables as the `classical interfaces' which provide  (limited) access to a `quantum universe';  we will explore how information gets extracted by measurement at these interfaces, how it gets processed by applying suitable quantum operations, and how these operations are classically controlled.

This paper is structured as follows:
\bit 
\item In section \ref{elements}, we describe the basic structures of quantum categorical semantics, i.e.~the quantum universe, and its classical interfaces. 
\item
In section \ref{Positivity}, we describe the families of morphisms between the classical interfaces, used to control the classical information flows.  They include deterministic operations (functions, permutations), non-deterministic operations (relations) as well as probabilistic operations (stochastic matrices).
\item
In Section \ref{Relativizing}, we describe how  the quantum universe interacts with a fixed classical interface, a classical structure $X$,  how information flows from the quantum universe to $X$, and how $X$ controls the  quantum universe. We consider both closed (pure) and open (mixed) quantum systems.  
We then let $X$ vary over the different classical structures, connected by the morphisms described in section \ref{Positivity}. 
\item Section \ref{Challenge} unifies the above and poses a challenge for future research.
\eit

\paragraph{Related work.}  
Previous structural models for quantum-classical interaction required either additional completeness assumptions, most notably biproducts, to model quantum spectra \cite{Abramsky-Coecke, Sel04,Selinger}, or a second monoidal structure \cite{deLL}.  Our notion of classicality is internal, emergent from symmetric monoidal dagger structure by articulating the local capabilities, in terms of symmetric monoidal dagger language only, and hence formalizable in purely diagrammatic terms. 

\paragraph{More applications.} An earlier version of this paper has been in circulation since 2007 under the title \em Classical and quantum structures \em \cite{OUCL-PRG-RR-08-02}. Besides a complementary formal development, it contains several applications not discussed in this paper.  A more detailed account on these is also in the 2nd author's PhD.~thesis \cite{Paq08}.

\section{Elements of quantum semantics}\label{elements}

\paragraph{Prerequisites.}  
We expect the reader to be familiar with symmetric monoidal categories and also to have some familiarity with compact (closed) categories \cite{KellyLaplaza, Abramsky-Coecke2, Selinger}, as well as with  the corresponding graphical calculus \cite{JoyalStreet, FreydYetter, Selinger}.  
\bigskip

By a symmetric monoidal dagger category we mean a symmetric monoidal category $\CCc$ together with an involutive contravariant identity-on-objects-functor 
\[
(-)^\dagger:\CCc^{op}\to \CCc
\]
which coherently preserves the symmetric monoidal structure.  In particular, this means that the natural isomorphisms $\theta$ of the symmetric monoidal structure are all \em unitary\em, that is, $\theta^{-1}=\theta^\dagger$.

\paragraph{Notation.}
When the confusion seems unlikely, we elide $g \circ f$ to $gf$ and $X\otimes A\otimes B$ to $XAB$.  The tensorial symmetry is denoted by $\sigma : AB \to BA$.  The set of abstract {\em vectors\/}, denoted as $\CCc(A)$, is just the hom-set $\CCc(I, A)$.  An abstract {\em scalar\/} is just a morphism in $\CCc(I,I)$.  
 
\paragraph{Graphical notation.} 
In the graphical calculus, morphisms are drawn upward. The identity on an object $A$ is depicted as 
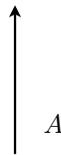
\begin{center}
\ifx\JPicScale\undefined\def\JPicScale{1}\fi
\psset{unit=\JPicScale mm}
\psset{linewidth=0.3,dotsep=1,hatchwidth=0.3,hatchsep=1.5,shadowsize=1,dimen=middle}
\psset{dotsize=0.7 2.5,dotscale=1 1,fillcolor=black}
\psset{arrowsize=1 2,arrowlength=1,arrowinset=0.25,tbarsize=0.7 5,bracketlength=0.15,rbracketlength=0.15}
\begin{pspicture}(0,0)(11,19.8)
\psline{->}(5.9,0)(5.9,19.8)
\rput(11,4){$A$}
\end{pspicture}

\end{center}
A morphism $f:A\rightarrow B$ is depicted as
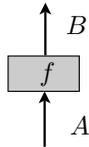
\begin{center}
\ifx\JPicScale\undefined\def\JPicScale{1}\fi
\psset{unit=\JPicScale mm}
\psset{linewidth=0.3,dotsep=1,hatchwidth=0.3,hatchsep=1.5,shadowsize=1,dimen=middle}
\psset{dotsize=0.7 2.5,dotscale=1 1,fillcolor=black}
\psset{arrowsize=1 2,arrowlength=1,arrowinset=0.25,tbarsize=0.7 5,bracketlength=0.15,rbracketlength=0.15}
\begin{pspicture}(0,0)(10.6,19.2)
\psline{->}(5.9,0)(5.9,7.4)
\rput(10.6,2.9){$A$}
\newrgbcolor{userFillColour}{0.8 0.8 0.8}
\pspolygon[linewidth=0.15,fillcolor=userFillColour,fillstyle=solid](1.05,7.4)(10.5,7.4)(10.5,12.1)(1.05,12.1)
\rput(10.2,16.2){$B$}
\rput(6.25,9.7){$f$}
\psline{->}(6,12.2)(6,19.2)
\end{pspicture}

\end{center}
Given two morphisms $f:A\rightarrow B$ and $g:B\rightarrow C$, their composition $g\circ f:A\rightarrow C$ is depicted as
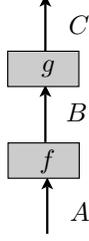
\begin{center}
\ifx\JPicScale\undefined\def\JPicScale{1}\fi
\psset{unit=\JPicScale mm}
\psset{linewidth=0.3,dotsep=1,hatchwidth=0.3,hatchsep=1.5,shadowsize=1,dimen=middle}
\psset{dotsize=0.7 2.5,dotscale=1 1,fillcolor=black}
\psset{arrowsize=1 2,arrowlength=1,arrowinset=0.25,tbarsize=0.7 5,bracketlength=0.15,rbracketlength=0.15}
\begin{pspicture}(0,0)(10.6,31.8)
\psline{->}(6.25,0)(6.25,7.5)
\rput(10.6,2.9){$A$}
\newrgbcolor{userFillColour}{0.8 0.8 0.8}
\pspolygon[linewidth=0.15,fillcolor=userFillColour,fillstyle=solid](1.05,7.4)(10.5,7.4)(10.5,12.1)(1.05,12.1)
\rput(10.2,16.2){$B$}
\rput(6.25,9.7){$f$}
\psline{->}(6,12.2)(6,19.8)
\newrgbcolor{userFillColour}{0.8 0.8 0.8}
\pspolygon[linewidth=0.15,fillcolor=userFillColour,fillstyle=solid](1,19.6)(10.45,19.6)(10.45,24.3)(1,24.3)
\psline{->}(6,24.3)(6,31.8)
\rput(6.4,22){$g$}
\rput(10.5,27.8){$C$}
\end{pspicture}

\end{center}
Given two morphisms $f:A\rightarrow B$ and $g:C\rightarrow D$, their tensor product $f\otimes g: A\otimes C\to B\otimes D$ is depicted as
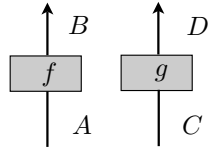
\begin{center}
\ifx\JPicScale\undefined\def\JPicScale{1}\fi
\psset{unit=\JPicScale mm}
\psset{linewidth=0.3,dotsep=1,hatchwidth=0.3,hatchsep=1.5,shadowsize=1,dimen=middle}
\psset{dotsize=0.7 2.5,dotscale=1 1,fillcolor=black}
\psset{arrowsize=1 2,arrowlength=1,arrowinset=0.25,tbarsize=0.7 5,bracketlength=0.15,rbracketlength=0.15}
\begin{pspicture}(0,0)(26,19.3)
\psline(6,0)(6,7.6)
\rput(10.6,2.9){$A$}
\newrgbcolor{userFillColour}{0.8 0.8 0.8}
\pspolygon[linewidth=0.15,fillcolor=userFillColour,fillstyle=solid](1.05,7.4)(10.5,7.4)(10.5,12.1)(1.05,12.1)
\rput(10.2,16.2){$B$}
\rput(6.25,9.7){$f$}
\psline{->}(6,12.2)(6,19.2)
\newrgbcolor{userFillColour}{0.8 0.8 0.8}
\pspolygon[linewidth=0.15,fillcolor=userFillColour,fillstyle=solid](15.8,7.5)(25.25,7.5)(25.25,12.2)(15.8,12.2)
\psline{->}(20.75,12.3)(20.75,19.3)
\rput(21.2,9.9){$g$}
\rput(25.3,3){$C$}
\psline(20.8,0.1)(20.8,7.6)
\rput(26,16.1){$D$}
\end{pspicture}

\end{center}
Finally, given a morphism $f:A\rightarrow B$, the corresponding  morphism $f^\dagger:B\rightarrow A$ is depicted by flipping graphical components upside-down~\cite{Selinger}. 
\bigskip

Below we assume $\CCc$ to be a symmetric monoidal dagger category. We use it to model simple quantum processes,  algorithms and protocols, along the lines of \cite{Abramsky-Coecke,Selinger,Coecke-Pavlovic,Coecke-Duncan, Coecke-Paquette-Perdrix}. 

\subsection{Bell states}

Categorical quantum semantics started with the observation by Abramsky and one of the authors that the \em duality \em (or \em compactness\em) in monoidal dagger categories can be used to model interaction of entangled states and effects (or co-states) \cite{Abramsky-Coecke}. Our Bell state structure thus consists of dualities, modeled in terms of compact structures \cite{KellyLaplaza}.

\be{defn}\em
A {\em compact structure} in any symmetric monoidal category $\CCc$ is a quadruple $(A,A^*,\varepsilon,\eta)$, where the \em pairing \em $A \otimes A^* \tto{\varepsilon} I$ and \em copairing \em $I\tto{\eta} A^*\otimes A$ make the following diagrams commute:
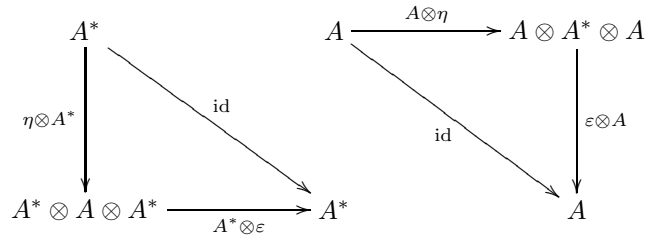
\[
\xymatrix{
A^* \ar[dd]_-{\eta \otimes A^*} \ar[rrdd]^-{\id} &&  A \ar[rr]^-{A\otimes \eta} \ar[rrdd]_-{\id}  
&& A\otimes A^*\otimes A \ar[dd]^-{\varepsilon\otimes A}\\ \\
A^*\otimes A\otimes A^* \ar[rr]_-{A^*\otimes \varepsilon} && A^*    && A}
\]
When $\CCc$ is a symmetric monoidal dagger category, then we define a {\em Bell state} $(A,A^*,\eta)$ on $A$ to be a compact structure $(A,A^*,\eta^\adj\circ\sigma, \eta)$.  

\paragraph{Remark.} 
If $\CCc$ is viewed as a bicategory with one object, then a compact structure makes the 1-cell $A^*$ into a right adjoint of $A$ \cite{KellyLaplaza}. 

\bigskip
A \em dagger compact category \em is a symmetric monoidal  dagger category where each object comes with a chosen Bell state. We call such objects \em quantum systems\em. 
For an arbitrary symmetric monoidal dagger category $\CCc$, we denote by $\CCc_{\qo}$ the category with classical structures $(X,X^*,\eta)$ as objects and with $\CCc_{\qo}((X,X^*,\eta_X), (Y,Y^*,\eta_Y))=\CCc(X,Y)$.
\ee{defn}

\paragraph{Remark.} 
Obviously, rather than introducing  $\CCc_{\qo}$ we could as well have assumed that in the symmetric monoidal dagger category $\CCc$ all objects come with coherently chosen Bell-states.
Constructing  $\CCc_{\qo}$ removes the need to  single out  a specific Bell state for each object.

\paragraph{Notation.}
When the structure is clear from the context, we often leave it implicit. For instance, we write $A$ instead of $(A, A^*,\eta_A)$ in the case of a Bell state; the same also applies to classical structures defined below. We write $(-)^*$ for the contravariant \em transpose endofunctor \em  on a compact category, that is,  for a morphism $f:A\to B$ we have
\bear
f^* & = & B^*\tto{\eta_A B^*}A^* A B^*\tto{A^* f B^*}  A^* B B^*   \tto{A^*\varepsilon_B }A^*.
\eear
We write $(-)_*=(-)^{\adj *}=(-)^{* \adj}$ for the covariant \em conjugate endofunctor\em.
The \em dimension \em of an object $A$ relative to a compact structure is
\[
{\rm dim}(A)=\eta_A\circ\sigma\circ\varepsilon_A\,.
\]

\paragraph{Graphical notation.} 
As usual we represent $\varepsilon$ and $\eta$ as:
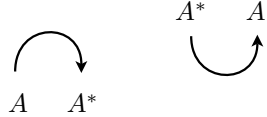
\begin{center}
\ifx\JPicScale\undefined\def\JPicScale{1}\fi
\psset{unit=\JPicScale mm}
\psset{linewidth=0.3,dotsep=1,hatchwidth=0.3,hatchsep=1.5,shadowsize=1,dimen=middle}
\psset{dotsize=0.7 2.5,dotscale=1 1,fillcolor=black}
\psset{arrowsize=1 2,arrowlength=1,arrowinset=0.25,tbarsize=0.7 5,bracketlength=0.15,rbracketlength=0.15}
\begin{pspicture}(0,0)(33.6,12.7)
\psbezier{->}(1.4,4.6)(1.4,11.4)(10.2,11.4)(10.2,4.4)
\psbezier{->}(24.8,9.3)(24.8,2.6)(33.6,2.6)(33.6,9.5)
\rput(10.4,0.4){$A^*$}
\rput(33.4,12.7){$A$}
\rput(1.8,0.4){$A$}
\rput(24.8,12.6){$A^*$}
\end{pspicture}

\end{center}
so that compactness becomes:
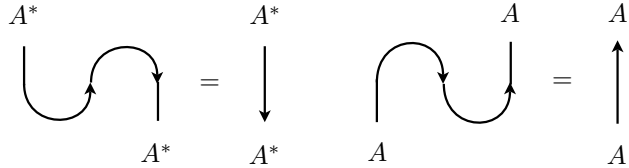
\begin{center}
\ifx\JPicScale\undefined\def\JPicScale{1}\fi
\psset{unit=\JPicScale mm}
\psset{linewidth=0.3,dotsep=1,hatchwidth=0.3,hatchsep=1.5,shadowsize=1,dimen=middle}
\psset{dotsize=0.7 2.5,dotscale=1 1,fillcolor=black}
\psset{arrowsize=1 2,arrowlength=1,arrowinset=0.25,tbarsize=0.7 5,bracketlength=0.15,rbracketlength=0.15}
\begin{pspicture}(0,0)(83.9,19.71)
\psbezier{->}(51.9,10.4)(51.9,17.2)(60.7,17.2)(60.7,10.2)
\psbezier{->}(60.8,10.2)(60.8,3.5)(69.6,3.5)(69.6,10.4)
\psline(51.9,5.11)(51.9,10.81)
\psline(69.7,10.1)(69.7,15.8)
\psline{->}(83.9,4.9)(83.9,16)
\rput(76.6,10.4){$=$}
\psbezier{->}(5,10.2)(5,4.01)(13.8,4.01)(13.8,10.38)
\psbezier{->}(13.9,10.38)(13.9,16.49)(22.7,16.49)(22.7,10.2)
\psline(5,15.2)(5,10.01)
\psline(22.8,10.47)(22.8,5.28)
\psline{->}(37,15.21)(37,5.1)
\rput(29.7,10.2){$=$}
\rput(22.6,0.99){$A^*$}
\rput(37,0.79){$A^*$}
\rput(52,1.11){$A$}
\rput(83.8,0.7){$A$}
\rput(5,19.4){$A^*$}
\rput(37.1,19.3){$A^*$}
\rput(69.6,19.71){$A$}
\rput(83.9,19.61){$A$}
\end{pspicture}

\end{center}
--- the arrows on the identities distinguish the object $A$ from $A^*$.   

\subsection{Classical structure}

Classical structures, first considered for this purpose in \cite{Coecke-Pavlovic}, are described using special commutative Frobenius algebras \cite{LawvereFrobenius, CarboniWalters}.

\be{defn}\em
A\/ {\em Frobenius algebra} in a symmetric monoidal category is an internal monoid 
\[
\xymatrix@=0.44in{
I\ar[r]^{\bot} &X& \ar[l]_{\nabla\ \ } X\otimes X
}
\]  
and an internal comonoid 
\[
\xymatrix@=0.44in{
I & \ar[l]_{\top}X\ar[r]^{\Delta \ \ } &  X\otimes X
}
\]
which together satisfy the {\em Frobenius condition\/} 
\[
\xymatrix{
X\otimes X \ar[dr]^-{\nabla} \ar[dd]_-{\Delta\otimes X} \ar[rr]^{X\otimes \Delta}&& X\otimes X\otimes X \ar[dd]^-{\nabla\otimes X} \\
& X \ar[dr]^-{\Delta} \\
X\otimes X \otimes X \ar[rr]_-{X\otimes \nabla} && X\otimes X
}
\]
A Frobenius algebra is called\/ {\em special}
if
\[
\ \nabla\circ \Delta = \id_X
\]
and it is \em commutative \em if      
\[
\ \sigma \circ \Delta = \Delta\,.
\]
A\/ {\em classical structure} $(X,\nabla,\bot)$ in a dagger symmetric monoidal category $\CCc$ is a commutative special Frobenius algebra for which 
\[
\top = \bot^\adjbis\qquad\mbox{ and }\qquad\Delta = \nabla^\adjbis\,. 
\]
We denote by $\CCc_{\cs}$ the category with classical structures $(X,\nabla,\bot)$ as objects and with $\CCc_{\cs}((X,\nabla_X,\bot_X), (Y,\nabla_Y,\bot_Y))=\CCc(X,Y)$.
\ee{defn}

\paragraph{Example.}  In the category ${\bf FdHilb}$ of finite dimensional Hilbert spaces and linear maps with the tensor product as monoidal structure, classical structures are in bijective correspondence with orthonormal bases. This fact was established by Vicary and two of the authors in \cite{Coecke-Pavlovic-Vicary}. It  supports the name `classical' in terms of standard quantum theory, since given an orthonormal basis $\{|i\rangle\}$ there is a corresponding non-degenerate observable with projector spectrum $\{|i\rangle\langle i|\}$.
Concretely, the bijective correspondence is as follows: each classical structure arises as 
\[
\Delta::|i\rangle\mapsto |i\rangle\otimes |i\rangle\qquad\qquad \top::|i\rangle\mapsto 1
\]
for  some orthonormal basis $\{|i\rangle\}$.  There,  the comultiplication \em copies \em the vectors of the corresponding basis while its unit \em uniformly deletes \em them. Hence classical structure counterfactually addresses  the \em no-cloning \em and \em no-deleting  \em theorems for quantum data \cite{Dieks, WZ, NoDe, AbrClone}.   The papers \cite{Coecke-Pavlovic, Coecke-Pavlovic-Vicary} provide more details on this.

\paragraph{Remark.} 
While the morphisms of $\CCc_{\cs}$   are initially completely oblivious to the classical structures, it will be convenient to have the category $\CCc_{\cs}$ at hand. The gamut of classical categories that we shall analyze in the next section will be extracted from $\CCc_{\cs}$, as the morphisms are constrained to preserve various fragments of classical structure.

\paragraph{Remark.}  If the category $\CCc$ is viewed as a bicategory with a single 0-cell, then the objects of $\CCc$ are 1-cells. In the internal sense of this bicategory, classical structures are just those 1-cells which happen to be both monads and comonads. Such a structure is studied in \cite{EM} --- one of the earliest papers about monads --- in which Eilenberg and Moore introduced the monadic view of universal algebra.

\bigskip
The monoidal structure of the unit $I$ is a canonical classical structure 
$(I,\lambda_I:I\simeq I\otimes I, \id_I)$. Moreover, if $(X,\nabla_X,\bot_X)$ and $(Y,\nabla_X,\bot_X)$ are classical structures, then so is 
$\left(X\otimes Y, \nabla_{X,Y}, \bot_{X,Y}\right)$ where
\[
\nabla_{X,Y}=(\id_X\otimes \sigma\otimes \id_Y)\circ(\nabla_X\otimes \nabla_Y)
\quad\ \ 
\bot_{X,Y}=(\bot_X\otimes \bot_Y)\circ\lambda_I\,.
\]
From this, it follows that $\CCc_{\cs}$ is a symmetric monoidal dagger category. The forgetful functor $\CCc_{\cs}\to\CCc$ is thus full and faithful, monoidal, and preserves the dagger. 

\paragraph{Graphical notation.}  We represent $\Delta, \top, \nabla, \bot$ respectively as:
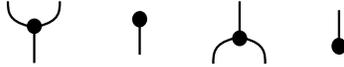
\begin{center}
\ifx\JPicScale\undefined\def\JPicScale{1}\fi
\psset{unit=\JPicScale mm}
\psset{linewidth=0.3,dotsep=1,hatchwidth=0.3,hatchsep=1.5,shadowsize=1,dimen=middle}
\psset{dotsize=0.7 2.5,dotscale=1 1,fillcolor=black}
\psset{arrowsize=1 2,arrowlength=1,arrowinset=0.25,tbarsize=0.7 5,bracketlength=0.15,rbracketlength=0.15}
\begin{pspicture}(0,0)(45.62,8.9)
\rput{90}(4.1,5.59){\psellipse[linestyle=none,fillstyle=solid](0,0)(1.03,1)}
\psbezier(4.74,5.6)(7.5,6.27)(7.5,7.58)(7.5,8.9)
\psbezier(3.36,5.6)(0.6,6.27)(0.6,7.58)(0.6,8.9)
\psline(4.1,0.69)(4.1,5.35)
\psline(18.12,1.8)(18.12,6.39)
\rput{83.46}(18.1,6.5){\psellipse[linestyle=none,fillstyle=solid](0,0)(1.09,1)}
\rput{90}(31.4,3.97){\psellipse[linestyle=none,fillstyle=solid](0,0)(1.05,-1)}
\psbezier(32.04,3.96)(34.8,3.28)(34.8,1.94)(34.8,0.59)
\psbezier(30.66,3.96)(27.9,3.28)(27.9,1.94)(27.9,0.59)
\psline(31.4,8.9)(31.4,4.22)
\psline(44.62,7.72)(44.62,3.01)
\rput{95.2}(44.6,2.9){\psellipse[linestyle=none,fillstyle=solid](0,0)(1.12,-1)}
\end{pspicture}

\end{center}
For example, the Frobenius condition corresponds to:
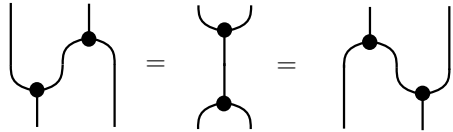
\begin{center}
\ifx\JPicScale\undefined\def\JPicScale{1}\fi
\psset{unit=\JPicScale mm}
\psset{linewidth=0.3,dotsep=1,hatchwidth=0.3,hatchsep=1.5,shadowsize=1,dimen=middle}
\psset{dotsize=0.7 2.5,dotscale=1 1,fillcolor=black}
\psset{arrowsize=1 2,arrowlength=1,arrowinset=0.25,tbarsize=0.7 5,bracketlength=0.15,rbracketlength=0.15}
\begin{pspicture}(0,0)(59.1,17.02)
\rput{90}(4.3,5.43){\psellipse[linestyle=none,fillstyle=solid](0,0)(1.03,1)}
\psbezier(4.94,5.44)(7.7,6.1)(7.6,7.46)(7.7,8.8)
\psbezier(3.56,5.44)(0.8,6.1)(0.8,7.42)(0.8,8.74)
\psline(4.3,0.52)(4.3,5.19)
\psline(14.6,0.52)(14.6,8.9)
\rput{90}(11.2,12.22){\psellipse[linestyle=none,fillstyle=solid](0,0)(1.04,-1)}
\psbezier(11.84,12.2)(14.6,11.53)(14.6,10.18)(14.6,8.83)
\psbezier(10.46,12.2)(7.7,11.53)(7.7,10.18)(7.7,8.83)
\psline(11.2,16.92)(11.2,12.46)
\psline(0.8,8.7)(0.8,17.02)
\rput{90}(55.54,4.98){\psellipse[linestyle=none,fillstyle=solid](0,0)(1.02,-1.02)}
\psbezier(54.9,5)(52.1,5.66)(52.1,7.08)(52.1,8.4)
\psbezier(56.3,5)(59.1,5.66)(59.1,6.98)(59.1,8.3)
\psline(55.55,0.09)(55.55,4.75)
\psline(45.1,0.3)(45.1,8.8)
\rput{90}(48.54,11.78){\psellipse[linestyle=none,fillstyle=solid](0,0)(1.05,1.02)}
\psbezier(47.9,11.76)(45.1,11.09)(45.1,9.74)(45.1,8.39)
\psbezier(49.3,11.76)(52.1,11.09)(52.1,9.74)(52.1,8.39)
\psline(48.55,16.49)(48.55,12.02)
\psline(59.1,8.1)(59.1,16.59)
\rput{90}(29.14,3.63){\psellipse[linestyle=none,fillstyle=solid](0,0)(1.05,1.02)}
\psbezier(28.5,3.61)(25.7,2.94)(25.7,1.59)(25.7,0.24)
\psbezier(29.9,3.61)(32.7,2.94)(32.7,1.59)(32.7,0.24)
\psline(29.2,9)(29.2,4.12)
\rput{90}(29.2,13.38){\psellipse[linestyle=none,fillstyle=solid](0,0)(1.02,-1.01)}
\psbezier(28.55,13.4)(25.75,14.06)(25.75,15.38)(25.75,16.7)
\psbezier(29.95,13.4)(32.75,14.06)(32.75,15.38)(32.75,16.7)
\psline(29.2,8.62)(29.2,13.29)
\rput(20.1,8.7){$=$}
\rput(37.5,8.5){$=$}
\end{pspicture}

\end{center}

The following fact  is discussed in detail in \cite{Lack, Coecke-Paquette}.

\be{prop}\label{prop:spider}
If in graphical representation a morphism generated from classical structure and symmetric monoidal dagger structure is connected, then it is completely characterized by its domain and codomain. So if the domain is $\underbrace{X\otimes\ldots\otimes X}_n$ and the codomain is 
$\underbrace{X\otimes\ldots\otimes X}_m$ then it can be reduced to a ``spider'' with $n$ input and $m$ output wires{\rm:}
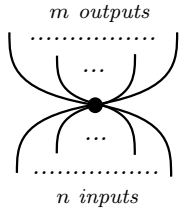
\begin{center}
\medskip
\ifx\JPicScale\undefined\def\JPicScale{1}\fi
\psset{unit=\JPicScale mm}
\psset{linewidth=0.3,dotsep=1,hatchwidth=0.3,hatchsep=1.5,shadowsize=1,dimen=middle}
\psset{dotsize=0.7 2.5,dotscale=1 1,fillcolor=black}
\psset{arrowsize=1 2,arrowlength=1,arrowinset=0.25,tbarsize=0.7 5,bracketlength=0.15,rbracketlength=0.15}
\begin{pspicture}(0,0)(26.6,25.7)
\rput{90}(15.71,13.5){\psellipse[linestyle=none,fillstyle=solid](0,0)(1.05,1.01)}
\psbezier(15.07,13.49)(5.3,11.68)(5.3,8.09)(5.3,4.5)
\psbezier(16.47,13.49)(25.8,11.58)(25.8,7.79)(25.8,4)
\psbezier(14.87,13.67)(4.3,15.59)(4.3,19.39)(4.3,23.2)
\psbezier(16.27,13.67)(26.6,15.59)(26.6,19.39)(26.6,23.2)
\psbezier(16.37,13.57)(21,14.92)(21,17.61)(21,20.3)
\psbezier(15.8,13.7)(10.63,14.98)(10.63,17.54)(10.63,20.1)
\psbezier(16.33,13.6)(20.97,12.26)(20.97,9.58)(20.97,6.9)
\psbezier(15.77,13.47)(10.6,12.19)(10.6,9.65)(10.6,7.1)
\rput(15.5,18.1){...}
\rput(15.8,9.1){...}
\rput(15.5,22.3){................}
\rput(15.9,4.5){................}
\rput(16.3,25.7){{\footnotesize $m$ outputs}}
\rput(16,1.2){{\footnotesize $n$ inputs}}
\end{pspicture}
 
\end{center}
\ee{prop}

All the defining axioms of classical structure follow from this rewriting principle, since all expressions involved have connected graphical representations.

\be{prop}\label{prop:ClassIntoQuant}
Each classical structure $(X,\nabla,\bot)$  induces a `self-dual' Bell state  on $X$ namely 
\[
(X\,,\,X^*=X\,,\, \eta =  \Delta\circ \bot)\,. 
\]
Moreover, by commutativity of the monoid, we have 
\beq\label{no_arrow}
\ \sigma\circ\eta=\eta\,.
\eeq
Hence, the category $\CCc_{\cs}$ of classical structures is dagger compact.  \ee{prop}

Eq.(\ref{no_arrow}) tells us that we can omit the arrows when depicting the Bell state induced by a classical structure.  The equation $\eta =  \Delta\circ \bot$ then  corresponds to:
\begin{center}
\ifx\JPicScale\undefined\def\JPicScale{1}\fi
\psset{unit=\JPicScale mm}
\psset{linewidth=0.3,dotsep=1,hatchwidth=0.3,hatchsep=1.5,shadowsize=1,dimen=middle}
\psset{dotsize=0.7 2.5,dotscale=1 1,fillcolor=black}
\psset{arrowsize=1 2,arrowlength=1,arrowinset=0.25,tbarsize=0.7 5,bracketlength=0.15,rbracketlength=0.15}
\begin{pspicture}(0,0)(33,12.3)
\rput{90}(29.34,6.42){\psellipse[linestyle=none,fillstyle=solid](0,0)(1.03,-1.02)}
\psbezier(28.7,6.44)(25.9,7.1)(25.9,8.43)(25.9,9.76)
\psbezier(30.1,6.44)(32.9,7.1)(32.9,8.43)(32.9,9.76)
\rput{0}(29.41,1.35){\psellipse[linestyle=none,fillstyle=solid](0,0)(1.01,1.01)}
\psline(29.41,6.05)(29.41,1.44)
\rput(18.5,7.1){$=$}
\psbezier(2.11,8.24)(2.11,1.54)(10.91,1.54)(10.91,8.44)
\rput(2.1,11.4){$X$}
\rput(10.9,11.3){$X$}
\rput(25.6,12.2){$X$}
\rput(33,12.3){$X$}
\end{pspicture}

\end{center}

\paragraph{Remark.}  Typically, a quantum system will admit more than one classical structure corresponding to incompatible observables --- we won't explicitly impose this in this paper, and refer the reader to \cite{Coecke-Duncan} for work in this direction.  In the light of proposition \ref{prop:ClassIntoQuant} this means that a Bell state may `factor' in many different ways into a classical structure.   Bell states extracted from different classical structures on the same object may be different, a fact which turns out to be closely related to the self-duality 
of those Bell states  \cite{Coecke-Paquette-Perdrix}.

\paragraph{Remark.}   For some constructions in this paper it is important to rely on a fixed Bell state for each quantum system, hence the dagger compact category $\CCc_{\qo}$. For several other constructions involving a specific classical structure it is important that the Bell state is the one extracted from that classical structure, hence the dagger compact category $\CCc_{\cs}$.

\section{Classical varieties}\label{Positivity}
In this section, we study the categories spanned by classical structures. The largest one is $\CCc_{\cs}$ itself, spanned by classical structures, and all $\CCc$-morphisms between them, ignoring the classical structure.  The smallest nontrivial one is a groupoid, where the morphisms preserve all of the classical structure, i.e.~both monoid and comonoid homomorphisms \cite{Kock}. 
For $\CCc = {\bf FdHilb}$, it then follows from \cite{Coecke-Pavlovic-Vicary} that this groupoid boils down to finite sets and permutations between them.

\subsection{Classical morphisms}\label{sec:clas_morph}

\be{defn}\em
Let $\CCc$ be a dagger category. An endomorphism $e\in\CCc(A,A)$ is {\em positive \/} if there exist a morphism $g:C\to A$ such that 
\beqa
A\tto{e} A & = & A\tto{g^\adjbis}C\tto{g} A\,.
\eeqa
\end{defn}

\be{lemma}
If an endomorphism $e:A\to A$ is positive  then for every `element' $x\in \CCc(A)$ the scalar 
\[
\langle ex\,|\,x\rangle:=(e\circ x)^\adj \circ x
\] 
is also positive. The converse holds if $I$ generates.
\ee{lemma}

\be{defn}\label{pose}\em
Let $\CCc$ be a monoidal dagger category. 
We call a morphism $f:X\to Y$ in $\CCc$ \em classical relative to classical structures $(X,\nabla_X,\bot_X)$ and $(Y,\nabla_Y,\bot_Y)$ \em if the endomorphism
\beq\label{eq:unfold}
XY\tto{\Delta_X Y}XXY\tto{XfY}XYY\tto{X\nabla_Y} XY 
\eeq
is positive, which graphically means
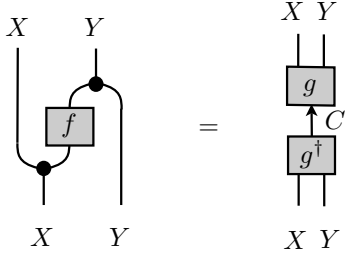
\begin{center}
\ifx\JPicScale\undefined\def\JPicScale{1}\fi
\psset{unit=\JPicScale mm}
\psset{linewidth=0.3,dotsep=1,hatchwidth=0.3,hatchsep=1.5,shadowsize=1,dimen=middle}
\psset{dotsize=0.7 2.5,dotscale=1 1,fillcolor=black}
\psset{arrowsize=1 2,arrowlength=1,arrowinset=0.25,tbarsize=0.7 5,bracketlength=0.15,rbracketlength=0.15}
\begin{pspicture}(0,0)(43.2,31.75)
\psbezier(5.04,10.8)(7.8,11.46)(7.8,12.78)(7.8,14.1)
\psbezier(10.61,21.98)(7.85,21.3)(7.85,19.95)(7.85,18.6)
\newrgbcolor{userFillColour}{0.8 0.8 0.8}
\pspolygon[fillcolor=userFillColour,fillstyle=solid](4.75,18.8)(11,18.8)(11,13.8)(4.75,13.8)
\psline(38.29,5.79)(38.29,10.37)
\psline(41.69,5.71)(41.69,10.3)
\psline(38.23,23.7)(38.24,28.66)
\psline(41.64,23.62)(41.64,28.58)
\psline{->}(40.04,15.02)(40,19.3)
\rput{90}(4.4,10.78){\psellipse[linestyle=none,fillstyle=solid](0,0)(1.02,1)}
\psbezier(3.66,10.8)(0.9,11.46)(0.9,12.78)(0.9,14.1)
\psline(4.4,5.88)(4.4,10.55)
\psline(14.7,5.88)(14.7,18.7)
\rput{90}(11.35,21.98){\psellipse[linestyle=none,fillstyle=solid](0,0)(1.04,-1)}
\psbezier(11.99,21.98)(14.75,21.3)(14.75,19.95)(14.75,18.6)
\psline(11.35,26.7)(11.35,22.24)
\psline(0.9,14.16)(0.95,26.8)
\rput(7.8,16.46){$f$}
\newrgbcolor{userFillColour}{0.8 0.8 0.8}
\pspolygon[fillcolor=userFillColour,fillstyle=solid](36.88,15)(43.12,15)(43.12,10)(36.88,10)
\rput(40.1,12.7){$g^\dagger$}
\newrgbcolor{userFillColour}{0.8 0.8 0.8}
\pspolygon[fillcolor=userFillColour,fillstyle=solid](36.75,24.29)(43.05,24.19)(43.05,19.11)(36.75,19.21)
\rput(39.83,21.52){$g$}
\rput(26.35,15.98){$=$}
\rput(4.2,1.6){$X$}
\rput(14.5,1.6){$Y$}
\rput(0.9,29.6){$X$}
\rput(11.2,29.6){$Y$}
\rput(37.45,31.65){$X$}
\rput(42.05,31.75){$Y$}
\rput(37.9,1.3){$X$}
\rput(42.5,1.4){$Y$}
\rput(43.2,17.2){$C$}
\end{pspicture}

\end{center}

Classical maps are closed under composition:
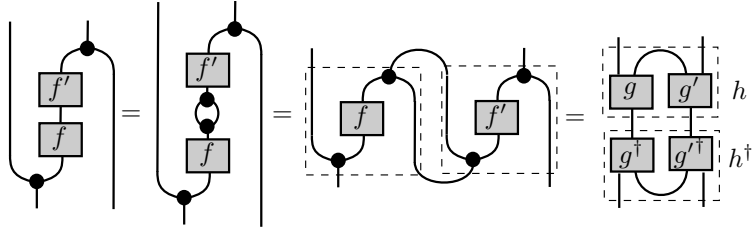
\begin{center}
\ifx\JPicScale\undefined\def\JPicScale{1}\fi
\psset{unit=\JPicScale mm}
\psset{linewidth=0.3,dotsep=1,hatchwidth=0.3,hatchsep=1.5,shadowsize=1,dimen=middle}
\psset{dotsize=0.7 2.5,dotscale=1 1,fillcolor=black}
\psset{arrowsize=1 2,arrowlength=1,arrowinset=0.25,tbarsize=0.7 5,bracketlength=0.15,rbracketlength=0.15}
\begin{pspicture}(0,0)(98.4,30.14)
\psbezier(91.2,7.86)(91.2,3.27)(84.39,3)(84.3,8)
\psbezier(91.1,20.09)(91.1,24.63)(84.29,24.9)(84.2,19.95)
\psline(7.94,16.6)(7.9,13.15)
\rput{90}(27.4,13.47){\psellipse[linestyle=none,fillstyle=solid](0,0)(1.03,1)}
\psline(27.5,9.75)(27.5,13.1)
\rput{90}(27.4,17.04){\psellipse[linestyle=none,fillstyle=solid](0,0)(1.04,-1)}
\psline(27.44,20.45)(27.4,17)
\rput{90}(11.5,23.85){\psellipse[linestyle=none,fillstyle=solid](0,0)(1.04,-1)}
\psbezier(12.14,23.84)(14.9,23.17)(15,21.68)(15,20.35)
\psbezier(10.76,23.84)(8,23.17)(8,21.85)(8,20.52)
\psline(11.54,27.54)(11.5,24.09)
\rput{90}(4.76,6.14){\psellipse[linestyle=none,fillstyle=solid](0,0)(1.03,1)}
\psbezier(5.4,6.15)(8.16,6.82)(8.16,8.13)(8.16,9.45)
\psbezier(4.02,6.15)(1.26,6.82)(1.2,8.33)(1.2,9.65)
\psline(4.76,2.55)(4.76,5.9)
\newrgbcolor{userFillColour}{0.8 0.8 0.8}
\pspolygon[fillcolor=userFillColour,fillstyle=solid](5.11,13.88)(10.73,13.88)(10.73,9.5)(5.11,9.5)
\rput(7.81,11.85){$f$}
\rput(37.4,14.8){$=$}
\psline(14.96,2.45)(15,20.45)
\psline(1.2,9.6)(1.2,27.4)
\pspolygon[linewidth=0.15,linestyle=dashed,dash=1 1](40.5,21.3)(55.8,21.3)(55.8,7)(40.5,7)
\newrgbcolor{userFillColour}{0.8 0.8 0.8}
\pspolygon[fillcolor=userFillColour,fillstyle=solid](5.1,20.53)(10.72,20.53)(10.72,16.15)(5.1,16.15)
\rput(7.8,18.5){$f'$}
\rput{90}(31.1,26.45){\psellipse[linestyle=none,fillstyle=solid](0,0)(1.04,-1)}
\psbezier(31.74,26.44)(34.5,25.77)(34.6,24.28)(34.6,22.95)
\psbezier(30.36,26.44)(27.6,25.77)(27.6,24.45)(27.6,23.12)
\psline(31.14,30.14)(31.1,26.69)
\rput{90}(24.35,3.99){\psellipse[linestyle=none,fillstyle=solid](0,0)(1.03,1)}
\psbezier(24.99,4)(27.75,4.67)(27.75,5.98)(27.75,7.3)
\psbezier(23.61,4)(20.85,4.67)(20.79,6.18)(20.79,7.5)
\psline(24.35,0.4)(24.35,3.75)
\newrgbcolor{userFillColour}{0.8 0.8 0.8}
\pspolygon[fillcolor=userFillColour,fillstyle=solid](24.7,11.73)(30.32,11.73)(30.32,7.35)(24.7,7.35)
\rput(27.4,9.7){$f$}
\psline(34.6,0.1)(34.6,23.05)
\psline(20.79,7.4)(20.9,30)
\newrgbcolor{userFillColour}{0.8 0.8 0.8}
\pspolygon[fillcolor=userFillColour,fillstyle=solid](24.7,23.13)(30.32,23.13)(30.32,18.75)(24.7,18.75)
\rput(27.4,21.1){$f'$}
\rput{90}(27.4,15.2){\psellipse[](0,0)(1.8,-1.5)}
\rput(17.3,15){$=$}
\rput{90}(51.59,20.06){\psellipse[linestyle=none,fillstyle=solid](0,0)(1.04,-1)}
\psbezier(52.23,20.05)(54.99,19.38)(54.99,18.06)(54.99,16.73)
\psbezier(50.85,20.05)(48.09,19.38)(48.09,18.06)(48.09,16.73)
\rput{90}(44.83,8.94){\psellipse[linestyle=none,fillstyle=solid](0,0)(1.03,1)}
\psbezier(45.47,8.95)(48.23,9.62)(48.23,10.93)(48.23,12.25)
\psbezier(44.09,8.95)(41.33,9.62)(41.33,10.93)(41.33,12.25)
\psline(44.83,5.35)(44.83,8.7)
\newrgbcolor{userFillColour}{0.8 0.8 0.8}
\pspolygon[fillcolor=userFillColour,fillstyle=solid](45.18,16.68)(50.8,16.68)(50.8,12.3)(45.18,12.3)
\rput(47.88,14.65){$f$}
\psline(55,8.6)(55.03,16.84)
\psline(41.33,12.26)(41.33,23.85)
\rput{90}(69.51,20.21){\psellipse[linestyle=none,fillstyle=solid](0,0)(1.04,-1)}
\psbezier(70.15,20.2)(72.91,19.53)(72.91,18.21)(72.91,16.88)
\psbezier(68.77,20.2)(66.01,19.53)(66.01,18.21)(66.01,16.88)
\psline(69.55,23.9)(69.51,20.45)
\rput{90}(62.75,9.08){\psellipse[linestyle=none,fillstyle=solid](0,0)(1.03,1)}
\psbezier(63.39,9.1)(66.15,9.77)(66.15,11.08)(66.15,12.4)
\psbezier(62.01,9.1)(59.25,9.77)(59.25,11.08)(59.25,12.4)
\newrgbcolor{userFillColour}{0.8 0.8 0.8}
\pspolygon[fillcolor=userFillColour,fillstyle=solid](63.1,16.83)(68.72,16.83)(68.72,12.45)(63.1,12.45)
\rput(65.8,14.8){$f'$}
\psline(72.95,5.4)(72.95,16.99)
\psline(59.25,12.41)(59.3,20.2)
\psbezier(62.8,8.6)(62.8,5.2)(55.1,5)(55,8.7)
\psbezier(59.3,20.33)(59.3,24.65)(51.8,24.9)(51.7,20.2)
\rput(76.35,14.4){$=$}
\psline(83.91,9.72)(83.91,15.82)
\newrgbcolor{userFillColour}{0.8 0.8 0.8}
\pspolygon[fillcolor=userFillColour,fillstyle=solid](81.09,11.83)(86.73,11.83)(86.73,7.45)(81.09,7.45)
\rput(83.8,9.8){$g^\dagger$}
\newrgbcolor{userFillColour}{0.8 0.8 0.8}
\pspolygon[fillcolor=userFillColour,fillstyle=solid](80.98,20.2)(86.61,20.2)(86.61,15.83)(80.98,15.83)
\rput(83.69,18.18){$g$}
\psline(82.19,2.61)(82.19,7.2)
\psline(82.09,20.36)(82.09,25.32)
\psline(91.71,9.92)(91.71,16.02)
\newrgbcolor{userFillColour}{0.8 0.8 0.8}
\pspolygon[fillcolor=userFillColour,fillstyle=solid](88.89,12.03)(94.53,12.03)(94.53,7.65)(88.89,7.65)
\rput(91.6,10){${g'}^\dagger$}
\newrgbcolor{userFillColour}{0.8 0.8 0.8}
\pspolygon[fillcolor=userFillColour,fillstyle=solid](88.78,20.4)(94.41,20.4)(94.41,16.03)(88.78,16.03)
\rput(91.49,18.38){$g'$}
\psline(93.39,2.74)(93.39,7.32)
\psline(93.28,20.48)(93.28,25.44)
\pspolygon[linewidth=0.15,linestyle=dashed,dash=1 1](58.6,21.5)(73.9,21.5)(73.9,7.2)(58.6,7.2)
\pspolygon[linewidth=0.15,linestyle=dashed,dash=1 1](80,24)(95.3,24)(95.3,14.7)(80,14.7)
\pspolygon[linewidth=0.15,linestyle=dashed,dash=1 1](80.2,12.9)(95.5,12.9)(95.5,3.6)(80.2,3.6)
\rput(98.2,18.3){$h$}
\rput(98.4,9.6){$h^\dagger$}
\end{pspicture}

\end{center}
Since identities are also classical, it follows that classical structures and classical maps form a subcategory $\Cclas$ of $\CCc_{\cs}$.
\ee{defn}

\paragraph{Example.} As explained above, classical structures in {\bf FdHilb} are in one-to-one correspondence with orthonormal bases. If a linear map $f:X\to Y$ is represented as an $n\times m$-matrix in the bases induced by the classical structures $X$ and $Y$, then the operation on $f$  described in (\ref{eq:unfold}) redistributes the entries of this matrix over the diagonal of an $(n\times m)\times(n\times m)$ matrix.  Classical maps in {\bf FdHilb} thus correspond to the matrices with non-negative entries.  They map basis vectors to linear combinations of basis vectors involving only non-negative coefficients.

\be{defn}{\rm\cite{Selinger}}\label{cpos} \em
Let $A$ and $B$ be quantum systems in the dagger compact category $\CCc_{\qo}$.  A morphism $f:A^*A \to B^*B$ in $\CCc_{\qo}$ is called \em completely positive \em if its transpose
\[
\xymatrix{AB^* \ar[rr]^-{\eta_{A^*} AB^*} && AA^*AB^* \ar[rr]^-{AfB^*} && AB^*BB^* \ar[rr]^-{AB^*\varepsilon_B} && A B^*
}
\]
is positive, i.e.
\begin{center}
\ifx\JPicScale\undefined\def\JPicScale{1}\fi
\psset{unit=\JPicScale mm}
\psset{linewidth=0.3,dotsep=1,hatchwidth=0.3,hatchsep=1.5,shadowsize=1,dimen=middle}
\psset{dotsize=0.7 2.5,dotscale=1 1,fillcolor=black}
\psset{arrowsize=1 2,arrowlength=1,arrowinset=0.25,tbarsize=0.7 5,bracketlength=0.15,rbracketlength=0.15}
\begin{pspicture}(0,0)(43.22,30.28)
\psline{<-}(8.49,19.05)(8.49,26.75)
\psline{->}(40.4,12.09)(40.4,18.19)
\psline{<-}(18.19,6.27)(18.19,18.75)
\psline{->}(1.39,14.21)(1.44,26.85)
\newrgbcolor{userFillColour}{0.8 0.8 0.8}
\pspolygon[fillcolor=userFillColour,fillstyle=solid](7.29,18.77)(12.91,18.77)(12.91,14.4)(7.29,14.4)
\rput(9.99,16.75){$f$}
\newrgbcolor{userFillColour}{0.8 0.8 0.8}
\pspolygon[fillcolor=userFillColour,fillstyle=solid](37.58,14.2)(43.22,14.2)(43.22,9.82)(37.58,9.82)
\rput(40.29,12.17){$g^\dagger$}
\newrgbcolor{userFillColour}{0.8 0.8 0.8}
\pspolygon[fillcolor=userFillColour,fillstyle=solid](37.47,22.57)(43.1,22.57)(43.1,18.2)(37.47,18.2)
\rput(40.18,20.55){$g$}
\psline{->}(38.68,4.98)(38.68,9.57)
\psline{<-}(42.08,4.91)(42.08,9.49)
\psline{->}(38.58,22.73)(38.58,27.69)
\psline{<-}(41.97,22.65)(41.97,27.61)
\psbezier{<-}(1.39,14.15)(1.39,8.61)(8.49,8.61)(8.49,14.31)
\psbezier{->}(11.79,19.12)(11.79,24.15)(18.39,24.15)(18.39,18.97)
\psline{->}(11.79,6.55)(11.79,14.25)
\rput(27.89,15.27){$=$}
\rput(11.6,1.48){$A$}
\rput(18,1.38){$B^*$}
\rput(1.3,29.48){$A$}
\rput(8.5,29.58){$B^*$}
\rput(38,0.6){$A$}
\rput(42,0.6){$B^*$}
\rput(38.29,30.28){$A$}
\rput(42.29,30.28){$B^*$}
\end{pspicture}

\end{center}
A completely positive element is called \em mixed state\em. Let $\Cee\CCc$ be the category with the same objects as $\CCc_{\qo}$ and with
\bear
\Cee\CCc(A,B) & = & \{f\in \CCc(A^*A,B^*B)|\ f\mbox{ is completely positive}\}\,.
\eear
\ee{defn}

\paragraph{Example.} In ${\bf FdHilb}$ the abstract notions of complete positivity  
and mixed state coincide with the usual ones \cite{Selinger}, except for the fact that the abstract mixed states  
are not normalised. For example,  in the case that $A=\mathbb{C}$ and $B={\cal H}$ we obtain bipartite states 
\[
\Psi:\mathbb{C}\simeq \mathbb{C}^*\otimes \mathbb{C} \to {\cal H}^*\otimes{\cal H}
\]
for which the transpose (cf.~map-state duality)
\[
\xymatrix{
{\cal H}^*\simeq\mathbb{C}\otimes{\cal H}^*  
\ar[rrr]^-{ 
(\id_{{\cal H}^*} \otimes\varepsilon_{\cal H})\circ(\Psi\otimes  \id_{{\cal H}^*}  )
} &&&
{\cal H}^*\otimes\mathbb{C} \simeq {\cal H}^*
}
\]
is positive, that is, it is a density operator up to a positive real scalar multiple.  So these transposes 
provide the mixed states of the usual quantum mechanical formalism.  General completely positive maps 
\[
f: {\cal H}_1^*\otimes {\cal H}_1\to {\cal H}_2^*\otimes {\cal H}_2
\]
are linear operators which map bipartite states $\Psi:\mathbb{C} \to {\cal H}_1^*\otimes{\cal H}_1$ to bipartite states $f\circ\Psi:\mathbb{C} \to {\cal H}_2^*\otimes{\cal H}_2$. The positivity condition of the transpose of $f$ guarantees that if $\Psi$ represents (via the transpose) a positive operator that  $f\circ\Psi$ also represents a positive operator, and also assures that we indeed have `completely' positive maps.   Below we give an example of a completely positive map, namely decoherence.

\paragraph{Remark.} Complete positivity and the category $\Cee\CCc$ of mixed states and completely positive maps can be defined in any symmetric monoidal category, even in the absence of compactness~\cite{RR}.  
\bigskip

It is easy to see that the constraint on $f$ in Def.~\ref{cpos} is equivalent to:
\smallskip
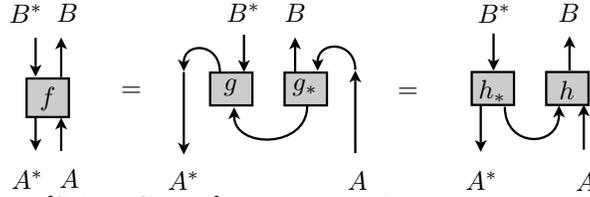
\begin{center}
\ifx\JPicScale\undefined\def\JPicScale{1}\fi
\psset{unit=\JPicScale mm}
\psset{linewidth=0.3,dotsep=1,hatchwidth=0.3,hatchsep=1.5,shadowsize=1,dimen=middle}
\psset{dotsize=0.7 2.5,dotscale=1 1,fillcolor=black}
\psset{arrowsize=1 2,arrowlength=1,arrowinset=0.25,tbarsize=0.7 5,bracketlength=0.15,rbracketlength=0.15}
\begin{pspicture}(0,0)(75.4,19.3)
\psline{<-}(61.08,0.76)(61.08,10.45)
\psline{<-}(1.9,10.77)(1.9,15.93)
\psline{->}(5.3,10.7)(5.3,16)
\psline{<-}(1.9,0.9)(1.9,5.5)
\newrgbcolor{userFillColour}{0.8 0.8 0.8}
\pspolygon[fillcolor=userFillColour,fillstyle=solid](0.7,10.53)(6.32,10.53)(6.32,5.4)(0.7,5.4)
\rput(3.5,7.9){$f$}
\newrgbcolor{userFillColour}{0.8 0.8 0.8}
\pspolygon[fillcolor=userFillColour,fillstyle=solid](35,11.32)(40.63,11.32)(40.63,6.95)(35,6.95)
\rput(37.7,9.3){$g_*$}
\newrgbcolor{userFillColour}{0.8 0.8 0.8}
\pspolygon[fillcolor=userFillColour,fillstyle=solid](25.1,11.32)(30.72,11.32)(30.72,6.95)(25.1,6.95)
\rput(27.8,9.3){$g$}
\psline{->}(36.4,11.41)(36.4,16)
\psline{<-}(21.6,0.4)(21.6,11.39)
\psline{<-}(29.6,11.4)(29.6,16.36)
\psbezier{<-}(28.3,6.7)(28.3,1.16)(38,1.16)(38,6.86)
\psline{->}(5.3,0.9)(5.3,5.33)
\rput(14.6,8.8){$=$}
\psbezier{<-}(21.6,11.48)(21.6,15.4)(26.4,15.4)(26.4,11.36)
\psline{->}(44.4,0.6)(44.4,11.59)
\psbezier{->}(44.4,11.68)(44.4,15.6)(39.4,15.6)(39.4,11.56)
\newrgbcolor{userFillColour}{0.8 0.8 0.8}
\pspolygon[fillcolor=userFillColour,fillstyle=solid](69.78,11.39)(75.4,11.39)(75.4,7.01)(69.78,7.01)
\rput(72.5,9){$h$}
\newrgbcolor{userFillColour}{0.8 0.8 0.8}
\pspolygon[fillcolor=userFillColour,fillstyle=solid](59.88,11.39)(65.5,11.39)(65.5,7.01)(59.88,7.01)
\rput(62.6,8.8){$h_*$}
\psline{->}(72.88,11.58)(72.88,16.16)
\psline{<-}(62.78,11.56)(62.78,16.52)
\psbezier{->}(64.28,6.76)(64.28,1.23)(71.48,1.23)(71.48,6.93)
\rput(51.4,8.6){$=$}
\psline{->}(74.78,1.06)(74.78,7.06)
\rput(0.8,-2.9){$A^*$}
\rput(6.4,-2.9){$A$}
\rput(0.6,19.2){$B^*$}
\rput(6.1,19.2){$B$}
\rput(29.5,19.3){$B^*$}
\rput(36.4,19.3){$B$}
\rput(21.6,-3.1){$A^*$}
\rput(44.7,-3.1){$A$}
\rput(61.2,-3.1){$A^*$}
\rput(75,-3.2){$A$}
\rput(62.9,19.2){$B^*$}
\rput(72.8,19.3){$B$}
\end{pspicture}

\end{center}
\medskip\noindent
As explained in \cite{Selinger, Coecke-Paquette-Perdrix}, in order to keep the graphs of the compact structure planar, we swap the wires of the conjugate morphisms:
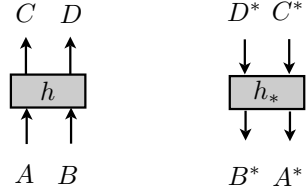
\begin{center}
\ifx\JPicScale\undefined\def\JPicScale{1}\fi
\psset{unit=\JPicScale mm}
\psset{linewidth=0.3,dotsep=1,hatchwidth=0.3,hatchsep=1.5,shadowsize=1,dimen=middle}
\psset{dotsize=0.7 2.5,dotscale=1 1,fillcolor=black}
\psset{arrowsize=1 2,arrowlength=1,arrowinset=0.25,tbarsize=0.7 5,bracketlength=0.15,rbracketlength=0.15}
\begin{pspicture}(0,0)(39.95,22.36)
\newrgbcolor{userFillColour}{0.8 0.8 0.8}
\pspolygon[fillcolor=userFillColour,fillstyle=solid](0.95,13.89)(10.85,13.89)(10.85,9.51)(0.95,9.51)
\rput(5.85,11.76){$h$}
\psline{->}(2.95,13.92)(2.95,18.91)
\psline{->}(8.85,13.96)(8.85,18.95)
\psline{->}(3.05,4.65)(3.05,9.46)
\psline{->}(8.95,4.7)(8.95,9.51)
\rput(2.75,0.61){$A$}
\rput(8.55,0.54){$B$}
\rput(2.95,22.01){$C$}
\rput(8.95,21.91){$D$}
\newrgbcolor{userFillColour}{0.8 0.8 0.8}
\pspolygon[fillcolor=userFillColour,fillstyle=solid](30.05,13.69)(39.95,13.69)(39.95,9.31)(30.05,9.31)
\rput(34.95,11.56){$h_*$}
\psline{<-}(32.05,13.91)(32.05,18.58)
\psline{<-}(37.95,13.94)(37.95,18.61)
\psline{<-}(32.15,5.11)(32.15,9.08)
\psline{<-}(38.05,5.01)(38.05,8.98)
\rput(32.05,0.41){$B^*$}
\rput(32.05,22.23){$D^*$}
\rput(37.85,0.31){$A^*$}
\rput(37.85,22.36){$C^*$}
\end{pspicture}

\end{center}

Since each classical structure $(X,\nabla,\bot)$ induces a self-dual compact structure $(X,X,\eta^\dagger,\eta)$ where $\eta = \Delta\circ \bot$ and hence $\sigma\circ\eta=\eta$, the notion of complete positivity applies to the morphisms in the form 
\[
f:XX\to YY
\]
between the classical structures $X$ and $Y$.  The following proposition is then easily proven in graphical language.

\be{prop}\label{posIfCpos}
For a morphism $f:X\to Y$ between classical structures $X$ and $Y$ the following statements are equivalent\,{\rm:}
\bit
\item $f:X\to Y$ is classical\,{\rm;}
\item 
$f_\Xi=XX\tto{\nabla}X\tto{f}Y\tto{\Delta}YY$
is completely positive\,{\rm;}
\item  
$f=g_\classsub=X\tto{\Delta}XX\tto{g}YY\tto{\nabla}Y$
for a  completely positive map $g:XX\to YY$ which  satisfies
\beq\label{decoherent}
\Xi_Y\circ g=g\circ \Xi_X= g
\eeq
for $\Xi_X= \Delta_X\circ\nabla_X$ and $\Xi_Y= \Delta_Y\circ\nabla_Y$.
\eit
\ee{prop}

\begin{defn}\em
A morphism 
\[
\Xi_X= \Delta_X\circ\nabla_X:XX\to XX
\]
induced by a classical structure on $X$ is called a \em decoherence\em. \em Decoherent morphisms \em are completely positive maps that preserve decoherences,  like in {\rm(\ref{decoherent})}.  The subcategory of $\Cee(\CCc_{\cs})$ consisting of all classical structures  with decoherent completely positive maps is denoted by $\CCc_\Xi$.
\end{defn}

\paragraph{Example.} In ${\bf FdHilb}$ a mixed states is decoherent for a classical structure --- that is, an orthonormal basis --- if its matrix representation in that basis is diagonal.  Indeed, decoherences are completely positive maps which, when applied to a density matrix, erase the non-diagonal elements \cite{Coecke-Pavlovic}.  This justifies their name: they maximally destroy coherence.   Physically, this means that these states correspond to a probability distribution on the basis vectors, up to a positive real scalar multiple.  Since we interpret these basis vectors as classical data, decoherent mixed states correspond to probability distributions on classical data, up to a positive real scalar multiple.  A completely positive map is decoherent if it maps mixed states with diagonal matrices to mixed states with diagonal matrices. Hence they map probability distributions on classical data to probability distributions on classical data, all again up to positive real scalar multiples.  In Section \ref{sec:stochmaps} we define normalised probability distributions on classical data and corresponding mappings, i.e.~stochastic maps.
\bigskip

Graphically, complete positivity of $f_\Xi$ means
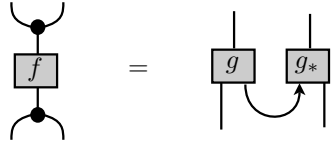
\begin{center}
\ifx\JPicScale\undefined\def\JPicScale{1}\fi
\psset{unit=\JPicScale mm}
\psset{linewidth=0.3,dotsep=1,hatchwidth=0.3,hatchsep=1.5,shadowsize=1,dimen=middle}
\psset{dotsize=0.7 2.5,dotscale=1 1,fillcolor=black}
\psset{arrowsize=1 2,arrowlength=1,arrowinset=0.25,tbarsize=0.7 5,bracketlength=0.15,rbracketlength=0.15}
\begin{pspicture}(0,0)(42.82,18.97)
\rput{90}(4.1,4){\psellipse[linestyle=none,fillstyle=solid](0,0)(1.04,-1)}
\psbezier(4.74,3.99)(7.5,3.32)(7.5,2)(7.5,0.67)
\psbezier(3.36,3.99)(0.6,3.32)(0.6,2)(0.6,0.67)
\psline(4.1,10.67)(4.1,4.24)
\rput{90}(4.1,15.66){\psellipse[linestyle=none,fillstyle=solid](0,0)(1.04,1)}
\psbezier(4.74,15.67)(7.5,16.34)(7.5,17.65)(7.5,18.97)
\psbezier(3.36,15.67)(0.6,16.34)(0.6,17.65)(0.6,18.97)
\psline(4.1,9.03)(4.1,15.42)
\psline(28.5,2.03)(28.5,11.72)
\newrgbcolor{userFillColour}{0.8 0.8 0.8}
\pspolygon[fillcolor=userFillColour,fillstyle=solid](1.05,12.03)(6.67,12.03)(6.67,7.65)(1.05,7.65)
\rput(3.75,10){$f$}
\rput(17.5,10){$=$}
\newrgbcolor{userFillColour}{0.8 0.8 0.8}
\pspolygon[fillcolor=userFillColour,fillstyle=solid](37.2,12.66)(42.82,12.66)(42.82,8.28)(37.2,8.28)
\rput(39.9,10.62){$g_*$}
\newrgbcolor{userFillColour}{0.8 0.8 0.8}
\pspolygon[fillcolor=userFillColour,fillstyle=solid](27.3,12.66)(32.92,12.66)(32.92,8.28)(27.3,8.28)
\rput(30,10.62){$g$}
\psline(40.3,12.84)(40.3,17.42)
\psline(30.2,12.83)(30.2,17.78)
\psbezier{->}(31.7,8.02)(31.7,2.5)(38.9,2.5)(38.9,8.2)
\psline(42.2,2.33)(42.2,8.32)
\end{pspicture}

\end{center}
As in Proposition \ref{prop:spider}, decoherences are graphically depicted as 
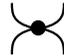
\begin{center}
\ifx\JPicScale\undefined\def\JPicScale{1}\fi
\psset{unit=\JPicScale mm}
\psset{linewidth=0.3,dotsep=1,hatchwidth=0.3,hatchsep=1.5,shadowsize=1,dimen=middle}
\psset{dotsize=0.7 2.5,dotscale=1 1,fillcolor=black}
\psset{arrowsize=1 2,arrowlength=1,arrowinset=0.25,tbarsize=0.7 5,bracketlength=0.15,rbracketlength=0.15}
\begin{pspicture}(0,0)(7.9,7.25)
\psbezier(5.14,3.95)(7.9,4.61)(7.9,5.93)(7.9,7.25)
\psbezier(3.76,3.78)(1,3.1)(1,1.75)(1,0.4)
\rput{90}(4.5,3.93){\psellipse[linestyle=none,fillstyle=solid](0,0)(1.02,1)}
\psbezier(3.76,3.95)(1,4.61)(1,5.93)(1,7.25)
\psbezier(5.14,3.78)(7.9,3.1)(7.9,1.75)(7.9,0.4)
\end{pspicture}

\end{center}

Given a classical morphism $f:X\to Y$, a completely positive map of the form
\[
f_\Xi=\Delta_Y\circ f\circ\nabla_X
\]
is decoherent. Indeed, since classical structures are special, we have 
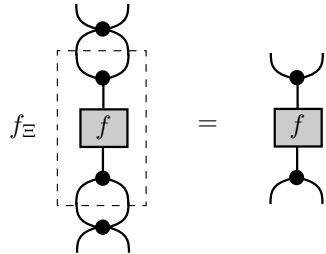
\begin{center}
\ifx\JPicScale\undefined\def\JPicScale{1}\fi
\psset{unit=\JPicScale mm}
\psset{linewidth=0.3,dotsep=1,hatchwidth=0.3,hatchsep=1.5,shadowsize=1,dimen=middle}
\psset{dotsize=0.7 2.5,dotscale=1 1,fillcolor=black}
\psset{arrowsize=1 2,arrowlength=1,arrowinset=0.25,tbarsize=0.7 5,bracketlength=0.15,rbracketlength=0.15}
\begin{pspicture}(0,0)(43.27,33.76)
\psline(14.04,14.84)(14.04,10.94)
\psbezier(14.64,30.46)(17.4,31.12)(17.4,32.44)(17.4,33.76)
\psbezier(13.26,30.29)(10.5,29.61)(10.5,28.26)(10.5,26.9)
\rput{90}(14,30.44){\psellipse[linestyle=none,fillstyle=solid](0,0)(1.02,1)}
\psbezier(13.26,30.46)(10.5,31.12)(10.5,32.44)(10.5,33.76)
\psbezier(14.64,30.29)(17.4,29.61)(17.4,28.26)(17.4,26.9)
\newrgbcolor{userFillColour}{0.8 0.8 0.8}
\pspolygon[fillcolor=userFillColour,fillstyle=solid](11.05,19.74)(17.3,19.74)(17.3,14.74)(11.05,14.74)
\rput(14.1,17.4){$f$}
\rput{90}(14.04,23.91){\psellipse[linestyle=none,fillstyle=solid](0,0)(1.02,1)}
\psbezier(14.68,23.88)(17.44,24.54)(17.44,25.86)(17.44,27.18)
\psbezier(13.3,23.88)(10.54,24.54)(10.54,25.86)(10.54,27.18)
\psline(14.1,23.74)(14.1,19.84)
\psbezier(13.26,10.46)(10.5,9.78)(10.5,8.43)(10.5,7.08)
\rput{90}(14,10.61){\psellipse[linestyle=none,fillstyle=solid](0,0)(1.02,1)}
\psbezier(14.64,10.46)(17.4,9.78)(17.4,8.43)(17.4,7.08)
\rput{90}(14.04,4.09){\psellipse[linestyle=none,fillstyle=solid](0,0)(1.02,1)}
\psbezier(14.68,4.06)(17.44,4.72)(17.44,6.04)(17.44,7.36)
\psbezier(13.3,4.06)(10.54,4.72)(10.54,6.04)(10.54,7.36)
\psbezier(13.36,4.04)(10.6,3.36)(10.6,2.01)(10.6,0.66)
\psbezier(14.74,4.04)(17.5,3.36)(17.5,2.01)(17.5,0.66)
\pspolygon[linewidth=0.15,linestyle=dashed,dash=1 1](7.98,27.54)(19.78,27.54)(19.78,6.94)(7.98,6.94)
\rput(3.47,17.54){$f_\Xi$}
\psline(39.86,14.9)(39.86,11)
\newrgbcolor{userFillColour}{0.8 0.8 0.8}
\pspolygon[fillcolor=userFillColour,fillstyle=solid](36.88,19.8)(43.12,19.8)(43.12,14.8)(36.88,14.8)
\rput(39.92,17.46){$f$}
\rput{90}(39.86,23.97){\psellipse[linestyle=none,fillstyle=solid](0,0)(1.02,1)}
\psbezier(40.5,23.94)(43.27,24.6)(43.27,25.92)(43.27,27.24)
\psbezier(39.12,23.94)(36.36,24.6)(36.36,25.92)(36.36,27.24)
\psline(39.92,23.8)(39.92,19.9)
\psbezier(39.08,10.52)(36.32,9.84)(36.32,8.49)(36.32,7.14)
\rput{90}(39.82,10.68){\psellipse[linestyle=none,fillstyle=solid](0,0)(1.02,1)}
\psbezier(40.46,10.52)(43.22,9.84)(43.22,8.49)(43.22,7.14)
\rput(28,17.7){=}
\end{pspicture}

\end{center}
Conversely, if a completely positive map $g$ is decoherent then
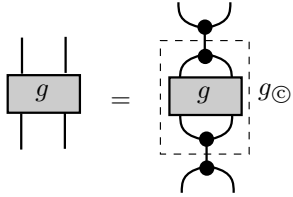
\begin{center}
\ifx\JPicScale\undefined\def\JPicScale{1}\fi
\psset{unit=\JPicScale mm}
\psset{linewidth=0.3,dotsep=1,hatchwidth=0.3,hatchsep=1.5,shadowsize=1,dimen=middle}
\psset{dotsize=0.7 2.5,dotscale=1 1,fillcolor=black}
\psset{arrowsize=1 2,arrowlength=1,arrowinset=0.25,tbarsize=0.7 5,bracketlength=0.15,rbracketlength=0.15}
\begin{pspicture}(0,0)(36.3,26)
\psline(27.14,7.9)(27.14,4)
\psbezier(27.78,7.8)(30.54,8.46)(30.54,9.78)(30.54,11.1)
\psbezier(26.6,4.03)(23.84,3.35)(23.84,2)(23.84,0.65)
\rput{90}(27.14,7.78){\psellipse[linestyle=none,fillstyle=solid](0,0)(1.02,1)}
\psbezier(26.4,7.8)(23.64,8.46)(23.64,9.78)(23.64,11.1)
\psbezier(27.98,4.03)(30.74,3.35)(30.74,2)(30.74,0.65)
\rput{90}(27.18,3.91){\psellipse[linestyle=none,fillstyle=solid](0,0)(1.02,1)}
\psline(26.9,22.8)(26.9,18.9)
\psbezier(27.54,22.7)(30.3,23.36)(30.3,24.68)(30.3,26)
\psbezier(26.36,18.93)(23.6,18.25)(23.6,16.9)(23.6,15.55)
\rput{90}(26.9,22.68){\psellipse[linestyle=none,fillstyle=solid](0,0)(1.02,1)}
\psbezier(26.16,22.7)(23.4,23.36)(23.4,24.68)(23.4,26)
\psbezier(27.74,18.93)(30.5,18.25)(30.5,16.9)(30.5,15.55)
\rput{90}(26.94,18.81){\psellipse[linestyle=none,fillstyle=solid](0,0)(1.02,1)}
\pspolygon[linewidth=0.15,linestyle=dashed,dash=1 1](21,20.9)(32.8,20.9)(32.8,5.8)(21,5.8)
\rput(15.7,12.7){=}
\newrgbcolor{userFillColour}{0.8 0.8 0.8}
\pspolygon[fillcolor=userFillColour,fillstyle=solid](22.2,16)(31.8,16)(31.8,11)(22.2,11)
\rput(26.8,13.7){$g$}
\rput(36.3,13.9){$g_\copyright$}
\newrgbcolor{userFillColour}{0.8 0.8 0.8}
\pspolygon[fillcolor=userFillColour,fillstyle=solid](0.7,16.3)(10.3,16.3)(10.3,11.3)(0.7,11.3)
\rput(5.3,14){$g$}
\psline(8.4,16.6)(8.4,21.4)
\psline(2.6,16.5)(2.6,21.3)
\psline(8.3,6.5)(8.3,11.3)
\psline(2.5,6.4)(2.5,11.2)
\end{pspicture}

\end{center}
so it indeed has the desired form, and that $g_\classsub$ is a classical morphism follows from $g$ being  completely positive.

\begin{corollary}\label{cor:embed}
The category $\Cclas$ of classical structures and classical morphisms is isomorphic to the category $\CCc_\Xi$ of classical structures and decoherent morphisms{\rm:}
\[
\xymatrix@=0.54in{
\Cclas \ar@/^0.8em/[r]^{(-)_\Xi} & \CCc_\Xi \ar@/^0.8em/[l]^{(-)_\classsub}
}
\]
\end{corollary}

\paragraph{Example.} In the case of ${\bf FdHilb}$ the isomorphism $(-)_\Xi$ of corollary \ref{cor:embed} takes a column vector with positive real entries and maps it on on a diagonal matrix with these entries on the diagonal, i.e.~a mixed state. Similarly, it maps matrices with positive real entries on a map which takes these diagonal mixed states to other diagonal mixed states in the same way as the initial matrix transforms the underlying vectors.

\be{defn} \em
A morphism in a dagger compact category is \em real \em if 
\[
f_*=f\,.
\]
\ee{defn}

\be{prop}
The inclusion functor
\[
\Cclas \inclusion \CCc_{\cs}
\]
preserves dagger symmetric monoidal structure. Hence $\Cclas$ is a symmetric monoidal dagger category.  Since for classical structure $(X,\nabla, \bot)$ the two morphisms $\nabla$ and $\bot$ are both classical morphisms relative to $(X,\nabla, \bot)$, $\Cclas$ inherits classical structures from $\CCc_{\cs}$ along this inclusion. Hence $\Cclas$ is also dagger compact. Moreover, all morphisms in $\Cclas$ are real. 
\ee{prop}

For instance, that the comultiplication of a classical structure is itself classical relative to that classical structure holds by Proposition \ref{prop:spider}: 
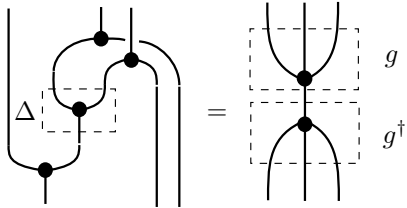
\begin{center}
\ifx\JPicScale\undefined\def\JPicScale{1}\fi
\psset{unit=\JPicScale mm}
\psset{linewidth=0.3,dotsep=1,hatchwidth=0.3,hatchsep=1.5,shadowsize=1,dimen=middle}
\psset{dotsize=0.7 2.5,dotscale=1 1,fillcolor=black}
\psset{arrowsize=1 2,arrowlength=1,arrowinset=0.25,tbarsize=0.7 5,bracketlength=0.15,rbracketlength=0.15}
\begin{pspicture}(0,0)(53.5,28.98)
\rput{90}(18.4,21.28){\psellipse[linestyle=none,fillstyle=solid](0,0)(1.04,-1)}
\psbezier(19.04,21.27)(21.8,20.6)(21.8,19.28)(21.8,17.95)
\psbezier(17.66,21.27)(14.9,20.6)(14.9,19.28)(14.9,17.95)
\psline(18.4,28.2)(18.4,21.52)
\rput{90}(7,6.61){\psellipse[linestyle=none,fillstyle=solid](0,0)(1.04,1)}
\psbezier(7.64,6.63)(11.5,7.3)(11.5,8.61)(11.5,9.93)
\psbezier(6.26,6.63)(2,7.3)(2.1,8.58)(2.1,9.9)
\psline(7,2.1)(7,6.38)
\rput{90}(11.5,14.7){\psellipse[linestyle=none,fillstyle=solid](0,0)(1.04,1)}
\psbezier(12.14,14.71)(14.9,15.38)(14.9,16.69)(14.9,18.01)
\psbezier(10.76,14.71)(8,15.38)(8,16.8)(8.1,19.7)
\psline(11.5,10.18)(11.5,14.46)
\rput{90}(14.4,24.1){\psellipse[linestyle=none,fillstyle=solid](0,0)(1.04,-1)}
\psbezier(15.04,24.09)(16.4,24.1)(16.8,24.1)(17.6,23.9)
\psbezier(13.66,24.09)(10.1,23.9)(8.1,21.7)(8.1,19.8)
\psline(14.4,28.3)(14.4,24.34)
\psbezier(19.44,23.72)(22.1,23.6)(24.8,20.9)(24.8,17.9)
\psline(2.1,9.9)(2.1,28.1)
\psline(21.8,17.7)(21.8,1.7)
\psline(24.8,17.7)(24.8,1.7)
\pspolygon[linewidth=0.15,linestyle=dashed,dash=1 1](6.6,17.4)(16.3,17.4)(16.3,11.7)(6.6,11.7)
\rput(4.3,14.4){$\Delta$}
\rput(29.8,14){=}
\rput{90}(41.4,18.86){\psellipse[linestyle=none,fillstyle=solid](0,0)(1.04,1)}
\psbezier(42.04,18.88)(45.9,20.93)(45.9,24.94)(45.9,28.98)
\psbezier(40.66,18.88)(36.4,20.93)(36.5,24.85)(36.5,28.89)
\psline(41.5,2.5)(41.5,12.5)
\psbezier(42.14,12.84)(46,10.74)(46,6.64)(46,2.5)
\psbezier(40.76,12.84)(36.5,10.74)(36.6,6.73)(36.6,2.59)
\psline(41.44,18.88)(41.44,28.88)
\rput{90}(41.5,18.76){\psellipse[linestyle=none,fillstyle=solid](0,0)(1.04,1)}
\rput{90}(41.5,12.73){\psellipse[linestyle=none,fillstyle=solid](0,0)(1.04,1)}
\psline(41.5,12.8)(41.5,18.4)
\pspolygon[linewidth=0.15,linestyle=dashed,dash=1 1](34.2,25.4)(48.6,25.4)(48.6,17.3)(34.2,17.3)
\pspolygon[linewidth=0.15,linestyle=dashed,dash=1 1](34.3,15.6)(48.7,15.6)(48.7,7.5)(34.3,7.5)
\rput(53,21.4){$g$}
\rput(53.5,11.4){$g^\dagger$}
\end{pspicture}

\end{center}
Similarly, we prove that classical maps are real:
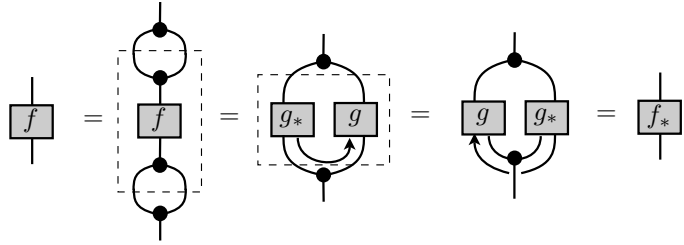
\begin{center}
\ifx\JPicScale\undefined\def\JPicScale{1}\fi
\psset{unit=\JPicScale mm}
\psset{linewidth=0.3,dotsep=1,hatchwidth=0.3,hatchsep=1.5,shadowsize=1,dimen=middle}
\psset{dotsize=0.7 2.5,dotscale=1 1,fillcolor=black}
\psset{arrowsize=1 2,arrowlength=1,arrowinset=0.25,tbarsize=0.7 5,bracketlength=0.15,rbracketlength=0.15}
\begin{pspicture}(0,0)(90.12,32.57)
\psbezier(68.9,9.8)(73.4,10.8)(73.4,12.98)(73.4,15.11)
\psbezier{->}(67.3,9.8)(63.1,10.8)(62.7,12.98)(62.7,15.11)
\psline(3.89,11)(3.89,22.59)
\newrgbcolor{userFillColour}{0.8 0.8 0.8}
\pspolygon[fillcolor=userFillColour,fillstyle=solid](1,18.83)(6.62,18.83)(6.62,14.45)(1,14.45)
\rput(3.7,16.8){$f$}
\rput(12,16.8){$=$}
\rput{90}(20.87,28.88){\psellipse[linestyle=none,fillstyle=solid](0,0)(1.04,-1)}
\psbezier(21.51,28.87)(24.27,28.2)(24.27,26.88)(24.27,25.55)
\psbezier(20.13,28.87)(17.37,28.2)(17.37,26.88)(17.37,25.55)
\psline(20.91,32.57)(20.87,29.12)
\rput{90}(20.91,22.46){\psellipse[linestyle=none,fillstyle=solid](0,0)(1.03,1)}
\psbezier(21.55,22.47)(24.31,23.14)(24.31,24.45)(24.31,25.77)
\psbezier(20.17,22.47)(17.41,23.14)(17.41,24.45)(17.41,25.77)
\psline(20.91,18.87)(20.91,22.22)
\newrgbcolor{userFillColour}{0.8 0.8 0.8}
\pspolygon[fillcolor=userFillColour,fillstyle=solid](18,18.93)(23.62,18.93)(23.62,14.55)(18,14.55)
\rput(20.7,16.9){$f$}
\rput{90}(20.87,10.85){\psellipse[linestyle=none,fillstyle=solid](0,0)(1.04,-1)}
\psbezier(21.51,10.84)(24.27,10.17)(24.27,8.85)(24.27,7.52)
\psbezier(20.13,10.84)(17.37,10.17)(17.37,8.85)(17.37,7.52)
\psline(20.91,14.54)(20.87,11.09)
\rput{90}(20.91,4.43){\psellipse[linestyle=none,fillstyle=solid](0,0)(1.04,1)}
\psbezier(21.55,4.44)(24.31,5.11)(24.31,6.42)(24.31,7.74)
\psbezier(20.17,4.44)(17.41,5.11)(17.41,6.42)(17.41,7.74)
\psline(20.91,0.84)(20.91,4.19)
\newrgbcolor{userFillColour}{0.8 0.8 0.8}
\pspolygon[fillcolor=userFillColour,fillstyle=solid](35.7,19.1)(41.32,19.1)(41.32,14.72)(35.7,14.72)
\rput(38.4,17.07){$g_*$}
\newrgbcolor{userFillColour}{0.8 0.8 0.8}
\pspolygon[fillcolor=userFillColour,fillstyle=solid](44.1,19.13)(49.72,19.13)(49.72,14.75)(44.1,14.75)
\rput(46.8,17.1){$g$}
\rput{90}(42.6,24.63){\psellipse[linestyle=none,fillstyle=solid](0,0)(1.04,-1)}
\psbezier(43,24.62)(48,23.51)(48,21.31)(48,19.1)
\psbezier(42.3,24.62)(37.3,23.51)(37.3,21.31)(37.3,19.1)
\psline(42.64,28.32)(42.6,24.87)
\rput{90}(42.6,9.56){\psellipse[linestyle=none,fillstyle=solid](0,0)(1,1)}
\psbezier(43,9.56)(48,10.64)(48,12.75)(48,14.88)
\psbezier(42.3,9.56)(37.3,10.64)(37.3,12.75)(37.3,14.88)
\psline(42.64,6)(42.6,9.32)
\psbezier{->}(39.2,14.5)(39.2,10.53)(46.1,10.27)(46.1,14.5)
\rput(30.1,17.1){$=$}
\newrgbcolor{userFillColour}{0.8 0.8 0.8}
\pspolygon[fillcolor=userFillColour,fillstyle=solid](61.1,19.33)(66.72,19.33)(66.72,14.95)(61.1,14.95)
\rput(63.8,17.3){$g$}
\newrgbcolor{userFillColour}{0.8 0.8 0.8}
\pspolygon[fillcolor=userFillColour,fillstyle=solid](69.5,19.36)(75.12,19.36)(75.12,14.98)(69.5,14.98)
\rput(72.2,17.33){$g_*$}
\rput{90}(68,24.86){\psellipse[linestyle=none,fillstyle=solid](0,0)(1.04,-1)}
\psbezier(68.4,24.85)(73.4,23.74)(73.4,21.54)(73.4,19.33)
\psbezier(67.7,24.85)(62.7,23.74)(62.7,21.54)(62.7,19.33)
\psline(68.04,28.55)(68,25.1)
\rput{90}(68,11.8){\psellipse[linestyle=none,fillstyle=solid](0,0)(1,1)}
\psline(68,6.4)(68,11.57)
\psbezier(64.6,14.73)(64.6,10.76)(71.5,10.5)(71.5,14.73)
\rput(55.5,17.33){$=$}
\psline(87.39,11.6)(87.39,23.19)
\newrgbcolor{userFillColour}{0.8 0.8 0.8}
\pspolygon[fillcolor=userFillColour,fillstyle=solid](84.5,19.43)(90.12,19.43)(90.12,15.05)(84.5,15.05)
\rput(87.2,17.4){$f_*$}
\rput(80.2,17.5){$=$}
\pspolygon[linewidth=0.15,linestyle=dashed,dash=1 1](15.3,26.1)(26.4,26.1)(26.4,7.3)(15.3,7.3)
\pspolygon[linewidth=0.15,linestyle=dashed,dash=1 1](33.9,22.9)(51.4,22.9)(51.4,10.9)(33.9,10.9)
\end{pspicture}

\end{center}
where we used (co)commutativity of the (co)multiplication.

\subsection{Relations}

Given classical structures $X$ and $Y$, we define the \em convolution monoid \em  
\[
\left(\CCc(X,Y),\star,\iota_{XY}\right)
\]
by
\[
f\star g = \nabla_Y\circ (f_*\otimes g)\circ \Delta_X\\
\qquad\qquad
\iota_{XY} = \top_Y\circ \bot_X\,.
\]

\paragraph{Remark.}
Given the convolution monoid on $\CCc(X)$ induced by classical structure $X$ we can recover the inner-product of $x,y\in\CCc(X)$ as 
\beqa
\langle x\,|\,y\rangle&=& I\tto{x\star y}X\tto{\top} I\,.  
\eeqa
Moreover, for morphisms $f,g\in\CCc(X, Y)$ we have by compactness that 
\[
f\star g=(\epsilon X)\circ\Bigl(X \bigl(((X f)\circ\eta)\star ( (X g)\circ\eta)\bigr) \Bigr)  
\]
where $(X f)\circ\eta, (X g)\circ\eta  \in\CCc(X\otimes Y)$, that is 
\begin{center}
\ifx\JPicScale\undefined\def\JPicScale{1}\fi
\psset{unit=\JPicScale mm}
\psset{linewidth=0.3,dotsep=1,hatchwidth=0.3,hatchsep=1.5,shadowsize=1,dimen=middle}
\psset{dotsize=0.7 2.5,dotscale=1 1,fillcolor=black}
\psset{arrowsize=1 2,arrowlength=1,arrowinset=0.25,tbarsize=0.7 5,bracketlength=0.15,rbracketlength=0.15}
\begin{pspicture}(0,0)(65.3,26.2)
\psline(50.4,6.74)(50.4,15.6)
\psbezier(30.83,6.34)(31.1,0.7)(38.26,0.85)(38.33,6.4)
\newrgbcolor{userFillColour}{0.8 0.8 0.8}
\pspolygon[fillcolor=userFillColour,fillstyle=solid](1.7,15.2)(7.32,15.2)(7.32,10.82)(1.7,10.82)
\rput(4.4,13.17){$f$}
\newrgbcolor{userFillColour}{0.8 0.8 0.8}
\pspolygon[fillcolor=userFillColour,fillstyle=solid](10.1,15.23)(15.72,15.23)(15.72,10.85)(10.1,10.85)
\rput(12.8,13.2){$g$}
\rput{90}(8.6,20.73){\psellipse[linestyle=none,fillstyle=solid](0,0)(1.04,-1)}
\psbezier(9,20.72)(14,19.61)(14,17.41)(14,15.2)
\psbezier(8.3,20.72)(3.3,19.61)(3.3,17.41)(3.3,15.2)
\psline(8.6,24.2)(8.6,20.97)
\rput{90}(8.6,5.66){\psellipse[linestyle=none,fillstyle=solid](0,0)(1,1)}
\psbezier(9,5.66)(14,6.7)(14,8.74)(14,10.8)
\psbezier(8.3,5.66)(3.3,6.7)(3.3,8.74)(3.3,10.8)
\psline(8.6,1.9)(8.6,5.42)
\rput{90}(34.7,19.74){\psellipse[linestyle=none,fillstyle=solid](0,0)(1.04,-1)}
\rput(21.2,13.1){$=$}
\psbezier(38.2,11)(38.2,13.36)(42.83,13.42)(42.9,15.54)
\psbezier(42.9,10.94)(42.9,13.3)(38.27,13.36)(38.2,15.48)
\newrgbcolor{userFillColour}{0.8 0.8 0.8}
\pspolygon[fillcolor=userFillColour,fillstyle=solid](35.18,10.88)(40.8,10.88)(40.8,6.5)(35.18,6.5)
\rput(37.88,8.85){$f$}
\newrgbcolor{userFillColour}{0.8 0.8 0.8}
\pspolygon[fillcolor=userFillColour,fillstyle=solid](47.58,10.87)(53.2,10.87)(53.2,6.49)(47.58,6.49)
\rput(50.28,8.84){$g$}
\psbezier(42.93,6.41)(43.2,0.77)(50.36,0.92)(50.43,6.47)
\psline(30.9,6.17)(30.9,15.7)
\psline(42.9,6.27)(42.9,10.8)
\psbezier(30.83,15.4)(31.1,21.1)(38.13,21.1)(38.2,15.4)
\psbezier(42.93,15.4)(43.2,21.1)(50.36,21.1)(50.43,15.4)
\psline(26.68,0.6)(26.59,19.6)
\psbezier(34.96,19.7)(34.7,25.1)(26.62,25.2)(26.56,19.8)
\psline(46.6,20.3)(46.6,26.2)
\rput{90}(46.6,19.64){\psellipse[linestyle=none,fillstyle=solid](0,0)(1.04,-1)}
\pspolygon[linewidth=0.15,linestyle=dashed,dash=0.8 1.2](0.8,22.2)(16.3,22.2)(16.3,4)(0.8,4)
\rput(17.4,24.2){$f\star g$}
\pspolygon[linewidth=0.15,linestyle=dashed,dash=0.8 1.2](29.7,21.6)(54.7,21.6)(54.7,0.8)(29.7,0.8)
\rput(65.3,24){$((Xf)\circ\eta)\star ((Xg)\circ\eta)$}
\end{pspicture}

\end{center}
Hence, knowing how $-\star-$ acts on states implies knowing how it acts on morphisms.  Speciality of classical structure just means that $\id_X$ is an idempotent element of the convolution monoid $\CCc(X, X)$. In general it is, of course, not the only idempotent.

\begin{defn}\em
A morphism $r\in \CCc_{\cs}(X,Y)$ is a {\em relation} if it is an idempotent of the convolution monoid, that is, 
\bear
r & = & r\star r\ = \ \nabla_Y \circ (r_*\otimes r) \circ \Delta_X\,.
\eear
\end{defn}

\bigskip
Every commutative monoid is a semilattice iff it is idempotent.  Therefore,   convolution of relations is written as the intersection operation 
\[
r\wedge s = r\star s\,. 
\]
As usual, the induced partial order is  
\[
r\leq s\ \iff\ r = r\wedge s\,.
\]

\be{prop}
All relations are classical morphisms, and hence also real. 
If $(X,\nabla, \bot)$ is a classical structure then $\nabla$ and $\bot$ are relations relative to this classical structure.
\ee{prop}

That relations are classical is established as follows:
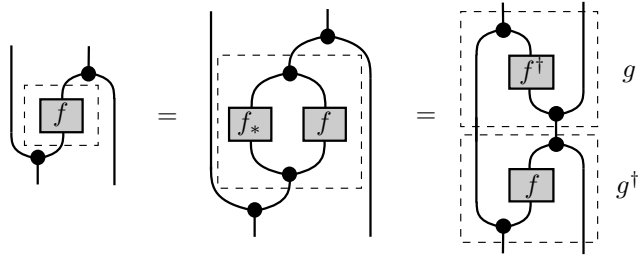
\begin{center}
\ifx\JPicScale\undefined\def\JPicScale{1}\fi
\psset{unit=\JPicScale mm}
\psset{linewidth=0.3,dotsep=1,hatchwidth=0.3,hatchsep=1.5,shadowsize=1,dimen=middle}
\psset{dotsize=0.7 2.5,dotscale=1 1,fillcolor=black}
\psset{arrowsize=1 2,arrowlength=1,arrowinset=0.25,tbarsize=0.7 5,bracketlength=0.15,rbracketlength=0.15}
\begin{pspicture}(0,0)(88.1,33.4)
\rput{90}(16.22,23.81){\psellipse[linestyle=none,fillstyle=solid](0,0)(1.04,-1)}
\psbezier(16.86,23.8)(19.62,23.13)(19.62,21.81)(19.62,20.48)
\psbezier(15.48,23.8)(12.72,23.13)(12.72,21.81)(12.72,20.48)
\psline(16.26,27.5)(16.22,24.05)
\rput{90}(9.46,12.68){\psellipse[linestyle=none,fillstyle=solid](0,0)(1.03,1)}
\psbezier(10.1,12.7)(12.86,13.37)(12.86,14.68)(12.86,16)
\psbezier(8.72,12.7)(5.96,13.37)(5.96,14.68)(5.96,16)
\psline(9.46,9.1)(9.46,12.45)
\newrgbcolor{userFillColour}{0.8 0.8 0.8}
\pspolygon[fillcolor=userFillColour,fillstyle=solid](9.81,20.43)(15.43,20.43)(15.43,16.05)(9.81,16.05)
\rput(12.51,18.4){$f$}
\rput(26.71,17.9){$=$}
\newrgbcolor{userFillColour}{0.8 0.8 0.8}
\pspolygon[fillcolor=userFillColour,fillstyle=solid](44.8,19.3)(50.42,19.3)(50.42,14.92)(44.8,14.92)
\rput(47.5,17.26){$f$}
\newrgbcolor{userFillColour}{0.8 0.8 0.8}
\pspolygon[fillcolor=userFillColour,fillstyle=solid](34.9,19.3)(40.52,19.3)(40.52,14.92)(34.9,14.92)
\rput(37.6,17.26){$f_*$}
\psline(19.66,9)(19.66,20.59)
\psline(5.96,16.01)(5.96,27.6)
\rput{90}(42.9,10.46){\psellipse[linestyle=none,fillstyle=solid](0,0)(1.03,1)}
\psbezier(43.6,10.47)(47.6,11.38)(47.6,13.15)(47.6,14.94)
\psbezier(42.3,10.47)(37.8,11.38)(37.8,13.15)(37.8,14.94)
\rput{90}(43,23.86){\psellipse[linestyle=none,fillstyle=solid](0,0)(1.07,-1)}
\psbezier(43.7,23.84)(47.7,22.9)(47.7,21.08)(47.7,19.24)
\psbezier(42.4,23.84)(37.9,22.9)(37.9,21.08)(37.9,19.24)
\rput{90}(48,28.76){\psellipse[linestyle=none,fillstyle=solid](0,0)(1.06,-1)}
\psbezier(48.7,28.74)(53.7,27.62)(53.7,25.44)(53.7,23.24)
\psbezier(47.4,28.74)(42.9,27.8)(42.9,25.98)(42.9,24.14)
\psline(48,32.44)(48,28.99)
\rput{90}(38.3,5.72){\psellipse[linestyle=none,fillstyle=solid](0,0)(1.03,1)}
\psbezier(39,5.73)(43,6.64)(43,8.41)(43,10.2)
\psbezier(37.7,5.73)(32.5,6.89)(32.5,9.16)(32.5,11.44)
\psline(38.3,2.13)(38.3,5.48)
\psline(53.7,23.24)(53.7,2.14)
\psline(32.5,32.14)(32.5,11.04)
\rput{0}(71.34,29.61){\psellipse[linestyle=none,fillstyle=solid](0,0)(1.04,1.04)}
\psbezier(70.68,29.6)(67.81,28.93)(67.81,27.61)(67.81,26.28)
\psbezier(72.11,29.6)(74.98,28.93)(74.98,27.61)(74.98,26.28)
\psline(71.3,33.3)(71.34,29.85)
\rput{0}(78.37,18.48){\psellipse[linestyle=none,fillstyle=solid](0,0)(1.04,-1.04)}
\psbezier(77.7,18.5)(74.84,19.17)(74.84,20.48)(74.84,21.8)
\psbezier(79.14,18.5)(82.01,19.17)(82.01,20.48)(82.01,21.8)
\psline(78.37,14.9)(78.37,18.25)
\newrgbcolor{userFillColour}{0.8 0.8 0.8}
\pspolygon[fillcolor=userFillColour,fillstyle=solid](78.01,26.23)(72.16,26.23)(72.16,21.85)(78.01,21.85)
\rput(75.2,24.2){$f^\dagger$}
\psline(67.77,14.8)(67.77,26.39)
\psline(82.01,21.81)(82.01,33.4)
\rput{0}(71.35,3.54){\psellipse[linestyle=none,fillstyle=solid](0,0)(1.04,-1.02)}
\psbezier(70.69,3.55)(67.82,4.21)(67.82,5.5)(67.82,6.8)
\psbezier(72.12,3.55)(74.99,4.21)(74.99,5.5)(74.99,6.8)
\psline(71.31,-0.07)(71.35,3.31)
\rput{-0}(78.38,14.43){\psellipse[linestyle=none,fillstyle=solid](0,0)(1.04,1.01)}
\psbezier(77.72,14.41)(74.85,13.76)(74.85,12.48)(74.85,11.19)
\psbezier(79.15,14.41)(82.02,13.76)(82.02,12.48)(82.02,11.19)
\newrgbcolor{userFillColour}{0.8 0.8 0.8}
\pspolygon[fillcolor=userFillColour,fillstyle=solid](78.02,6.85)(72.18,6.85)(72.18,11.14)(78.02,11.14)
\rput(75.21,8.84){$f$}
\psline(67.78,18.03)(67.78,6.69)
\psline(82.02,11.18)(82.02,-0.17)
\rput(61.1,18.1){$=$}
\pspolygon[linewidth=0.15,linestyle=dashed,dash=1 1](7.7,22.3)(17.5,22.3)(17.5,14.3)(7.7,14.3)
\pspolygon[linewidth=0.15,linestyle=dashed,dash=1 1](33.4,26.3)(52.5,26.3)(52.5,8.6)(33.4,8.6)
\pspolygon[linewidth=0.15,linestyle=dashed,dash=1 1](65.7,32.1)(83.9,32.1)(83.9,16.8)(65.7,16.8)
\pspolygon[linewidth=0.15,linestyle=dashed,dash=1 1](65.7,15.7)(83.9,15.7)(83.9,1.4)(65.7,1.4)
\rput(88,9.1){$g^\dagger$}
\rput(88.1,23.8){$g$}
\end{pspicture}

\end{center}
where we relied again on prop.~\ref{prop:spider}.

\paragraph{Remark.} 
In general, relations are not closed under composition in $\CCc$ however, they are closed under the tensor.  We show below how composition in $\CCc$ induces a new composition on relations.

\paragraph{Example.}
In ${\bf FdHilb}$, the abstract relations as defined above correspond to the matrices $f$ such that $f_{ij} = f_{ij}^2$. Since the only idempotent complex numbers are $0$ and $1$, these matrices correspond to the usual matrix representation of a binary relation, in this case relating basis vectors in Hilbert spaces.  Relative to these bases they are the classical maps involving only $0$'s and $1$'s as coefficients.  However, this family of matrices is not closed under composition in ${\bf FdHilb}$, for example, 
\[
\left(\begin{array}{cc}
1 & 1
\end{array}\right)
\circ
\left(\begin{array}{c}
1 \\ 1
\end{array}\right)
= 2 \ \,
\]
is not a relation.  
However, when considering relational composition instead, we obtain:   
\[
\left(\begin{array}{cc} 
1 & 1
\end{array}\right) \circ_{_r}
\left(\begin{array}{c}
1 \\ 1
\end{array}\right)
= 1\,.
\]
we do obtain a category ${\bf FdHilb}_r$ which is isomorphic to  ${\bf FRel}$, the category of finite sets and relations with the Cartesian product as monoidal structure.  The key difference of composition in ${\bf FdHilb}$ and in ${\bf FRel}$ is that in ${\bf FdHilb}$ we rely on the field structure of $\mathbb{C}$ while for composition in ${\bf FRel}$ we rely on the semiring (or rig) structure of the Booleans.  

Rather than a subcategory, ${\bf FdHilb}_r$ can be viewed as a quotient category of ${\bf FdHilb}_\classsub$.   The quotient ${\bf FdHilb}_\classsub \to {\bf FdHilb}_r$ is identity on the objects, and it maps each matrix of non-negative reals into a matrix of $0$s and $1$s, keeping the $0$s, and mapping all positive reals to $1$.

\bigskip
In general, setting\footnote{For clarity of the argument we assume all meets involved to be finite, as it is for example the case in ${\bf FdHilb}$.} 
\[
r_f = \bigwedge \left\{ t=t\star t \in\CCc_\classsub(X, Y)
\bigm| f = f\star t\right\}
\]
for  classical maps $f:X\to Y$, we can define an equivalence relation
\[
\ f \sim g \ \iff\ r_f=r_g\,.
\]
When restricting to the subcategory of $\CCc_\classsub$ generated by all relations, which we denote by $\DDd$,  the equivalence turns out to be a congruence with respect to composition, and hence we obtain a quotient 
\[
\ R:\DDd\to \CCc_r\,.
\]
Explicitly, the objects of $\CCc_r$ are the same as those of $\CCc_{\cs}$, the
morphisms are the equivalence classes for $\sim$, and  composition is inherited. 

This means that  for the equivalence classes respectively containing relations $s\in \CCc_\classsub(X, Y)$ and $r\in \CCc_\classsub(Y, Z)$, composition in $\mathbb{C}_r$ yields the equivalence class which contains the relation
\[
r\hspace{1mm} \tilde{\circ}\hspace{1mm} s:=\bigwedge \left\{ t=t\star t \in\CCc_\classsub(X, Z)
\bigm| r\circ s = (r\circ s) \star t\right\}\,.
\]

\paragraph{Notation.}  When confusion is unlikely it will be convenient to refer by $\mathbb{C}_r$ not only to the codomain of the above constructed quotient but also to the category which has relations as morphisms and $\tilde{\circ}$ as composition.  Obviously both categories are isomorphic.

\begin{prop}\label{prop:cartbicat1}
Classical structures and relations between them constitute a symmetric monoidal dagger category   $\CCc_r$.  It is moreover a locally thin bicategory for the partial ordering $\leq$.
\end{prop}

\begin{prop}\label{prop:cartbicat2}
In $\CCc_r$ all morphisms are lax comonoid homomorphisms with respect to the partial ordering $\leq$.
\end{prop}

For $r\in\CCc_r(X,Y)$ to be a  lax comonoid homomorphism means 
\[
\Delta_Y\circ r \leq (r\otimes r)\circ \Delta_X\qquad\mbox{and}\qquad \top_Y\circ r \leq \top_X\,.
\]
Unfolding the definition of $\leq$ the first inequality becomes an equality
\[
\Delta_Y\circ r =\Delta^\dagger_{Y\otimes Y}\circ \Bigl( \bigl(\Delta_Y\circ r\bigr) \otimes \bigl((r\otimes r)\circ \Delta_X\bigr)  \Bigr) \circ \Delta_X
\]
which indeed holds since graphically we have
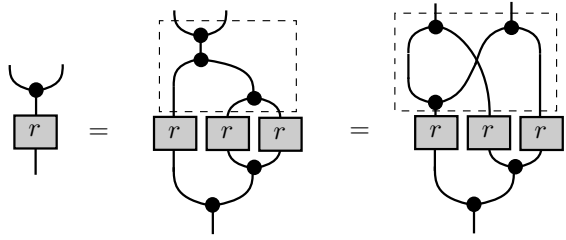
\begin{center}
\ifx\JPicScale\undefined\def\JPicScale{1}\fi
\psset{unit=\JPicScale mm}
\psset{linewidth=0.3,dotsep=1,hatchwidth=0.3,hatchsep=1.5,shadowsize=1,dimen=middle}
\psset{dotsize=0.7 2.5,dotscale=1 1,fillcolor=black}
\psset{arrowsize=1 2,arrowlength=1,arrowinset=0.25,tbarsize=0.7 5,bracketlength=0.15,rbracketlength=0.15}
\begin{pspicture}(0,0)(75.22,31.2)
\psline(72.2,15)(72.2,24.25)
\psline(58.3,14.51)(58.3,17.86)
\psline(23.5,15.3)(23.5,19.9)
\rput{90}(27.08,26.76){\psellipse[linestyle=none,fillstyle=solid](0,0)(1.03,1)}
\psbezier(27.72,26.77)(30.48,27.44)(30.48,28.75)(30.48,30.07)
\psbezier(26.34,26.77)(23.58,27.44)(23.58,28.75)(23.58,30.07)
\psline(27.08,23.17)(27.08,26.52)
\rput{90}(5.2,19.48){\psellipse[linestyle=none,fillstyle=solid](0,0)(1.03,1)}
\psbezier(5.84,19.5)(8.74,20.19)(8.74,21.54)(8.74,22.9)
\psbezier(4.46,19.5)(1.7,20.17)(1.7,21.48)(1.7,22.8)
\psline(5.2,15.9)(5.2,19.25)
\psline(5.14,8.2)(5.14,15.05)
\rput(13.5,13.7){=}
\pspolygon[linewidth=0.15,linestyle=dashed,dash=1 1](21.6,28.7)(39.8,28.7)(39.8,16.6)(21.6,16.6)
\newrgbcolor{userFillColour}{0.8 0.8 0.8}
\pspolygon[fillcolor=userFillColour,fillstyle=solid](2.32,16.08)(7.94,16.08)(7.94,11.7)(2.32,11.7)
\rput(5.02,14.05){$r$}
\rput{0}(27.1,23.49){\psellipse[linestyle=none,fillstyle=solid](0,0)(1,-0.98)}
\psbezier(27.74,23.47)(34.1,22.54)(34.1,20.73)(34.1,18.9)
\psbezier(26.36,23.47)(23.5,22.75)(23.5,21.33)(23.5,19.9)
\rput{90}(34.18,9.18){\psellipse[linestyle=none,fillstyle=solid](0,0)(1.03,1)}
\psbezier(34.82,9.2)(37.72,9.89)(37.72,11.24)(37.72,12.6)
\psbezier(33.44,9.2)(30.68,9.87)(30.68,11.18)(30.68,12.5)
\psline(23.5,8.6)(23.5,11.95)
\rput{0}(34.2,18.36){\psellipse[linestyle=none,fillstyle=solid](0,0)(1,-0.98)}
\psbezier(34.84,18.35)(37.7,17.69)(37.7,16.4)(37.7,15.1)
\psbezier(33.46,18.35)(30.6,17.65)(30.6,16.28)(30.6,14.9)
\newrgbcolor{userFillColour}{0.8 0.8 0.8}
\pspolygon[fillcolor=userFillColour,fillstyle=solid](27.88,15.88)(33.5,15.88)(33.5,11.5)(27.88,11.5)
\rput(30.58,13.85){$r$}
\newrgbcolor{userFillColour}{0.8 0.8 0.8}
\pspolygon[fillcolor=userFillColour,fillstyle=solid](20.88,15.9)(26.5,15.9)(26.5,11.52)(20.88,11.52)
\rput(23.58,13.87){$r$}
\newrgbcolor{userFillColour}{0.8 0.8 0.8}
\pspolygon[fillcolor=userFillColour,fillstyle=solid](34.88,15.78)(40.5,15.78)(40.5,11.4)(34.88,11.4)
\rput(37.58,13.75){$r$}
\rput{90}(28.66,4.18){\psellipse[linestyle=none,fillstyle=solid](0,0)(1.04,1)}
\psbezier(29.3,4.2)(34,5.11)(34,6.9)(34,8.7)
\psbezier(27.92,4.2)(23.5,5.22)(23.5,7.2)(23.5,9.2)
\psline(28.7,0.3)(28.7,3.65)
\rput(48.22,13.8){=}
\pspolygon[linewidth=0.15,linestyle=dashed,dash=1 1](52.9,29.7)(74.4,29.7)(74.4,16.7)(52.9,16.7)
\rput{90}(68.9,9.28){\psellipse[linestyle=none,fillstyle=solid](0,0)(1.03,1)}
\psbezier(69.54,9.3)(72.44,9.99)(72.44,11.34)(72.44,12.7)
\psbezier(68.16,9.3)(65.4,9.97)(65.4,11.28)(65.4,12.6)
\psline(58.22,8.7)(58.22,12.05)
\newrgbcolor{userFillColour}{0.8 0.8 0.8}
\pspolygon[fillcolor=userFillColour,fillstyle=solid](62.6,15.98)(68.22,15.98)(68.22,11.6)(62.6,11.6)
\rput(65.3,13.95){$r$}
\newrgbcolor{userFillColour}{0.8 0.8 0.8}
\pspolygon[fillcolor=userFillColour,fillstyle=solid](55.6,16)(61.22,16)(61.22,11.62)(55.6,11.62)
\rput(58.3,13.97){$r$}
\newrgbcolor{userFillColour}{0.8 0.8 0.8}
\pspolygon[fillcolor=userFillColour,fillstyle=solid](69.6,15.88)(75.22,15.88)(75.22,11.5)(69.6,11.5)
\rput(72.3,13.85){$r$}
\rput{90}(63.38,4.28){\psellipse[linestyle=none,fillstyle=solid](0,0)(1.03,1)}
\psbezier(64.02,4.3)(68.72,5.21)(68.72,7)(68.72,8.8)
\psbezier(62.64,4.3)(58.22,5.32)(58.22,7.3)(58.22,9.3)
\psline(63.42,0.4)(63.42,3.75)
\rput{90}(58.26,17.84){\psellipse[linestyle=none,fillstyle=solid](0,0)(1.03,1)}
\psbezier(58.9,17.86)(61.8,18.55)(62.5,20.7)(63.7,23.1)
\psbezier(57.52,17.86)(54.7,18.53)(54.7,19.84)(54.7,21.16)
\rput{0}(58.3,27.86){\psellipse[linestyle=none,fillstyle=solid](0,0)(1,-0.98)}
\psbezier(58.94,27.85)(61.8,27.19)(65.6,22.8)(65.5,16)
\psbezier(57.56,27.85)(54.7,27.15)(54.7,25.78)(54.7,24.4)
\rput{0}(68.4,27.76){\psellipse[linestyle=none,fillstyle=solid](0,0)(1,-0.98)}
\psbezier(69.34,27.75)(72.2,27.09)(72.2,25.6)(72.2,24.3)
\psbezier(67.66,27.75)(64.8,27.11)(64.3,24.9)(63.6,22.9)
\psline(58.3,27.65)(58.3,31)
\psline(68.4,27.85)(68.4,31.2)
\psline(54.7,20.95)(54.7,24.3)
\end{pspicture}

\end{center}
where the first step uses $r=r\star r=r\star (r\star r)$ and the second one uses Proposition \ref{prop:spider}. Similarly, the second inequality becomes the equality
\[
\top_Y\circ r=\Delta^\dagger_I\circ\bigl((\top_Y\circ r)\otimes\top_X\bigr) \circ\Delta_X
\]
which also holds since 
\begin{center}
\ifx\JPicScale\undefined\def\JPicScale{1}\fi
\psset{unit=\JPicScale mm}
\psset{linewidth=0.3,dotsep=1,hatchwidth=0.3,hatchsep=1.5,shadowsize=1,dimen=middle}
\psset{dotsize=0.7 2.5,dotscale=1 1,fillcolor=black}
\psset{arrowsize=1 2,arrowlength=1,arrowinset=0.25,tbarsize=0.7 5,bracketlength=0.15,rbracketlength=0.15}
\begin{pspicture}(0,0)(29.3,16.22)
\rput{90}(21.1,15.18){\psellipse[linestyle=none,fillstyle=solid](0,0)(1.03,1)}
\psline(21.1,11.6)(21.1,14.95)
\rput{90}(4.96,13.28){\psellipse[linestyle=none,fillstyle=solid](0,0)(1.03,1)}
\psline(4.96,9.7)(4.96,13.05)
\psline(4.98,2.78)(4.98,9.28)
\rput(13.8,8.31){=}
\newrgbcolor{userFillColour}{0.8 0.8 0.8}
\pspolygon[fillcolor=userFillColour,fillstyle=solid](1.98,10.3)(7.6,10.3)(7.6,5.92)(1.98,5.92)
\rput(4.68,8.27){$r$}
\rput{90}(24.76,6.6){\psellipse[linestyle=none,fillstyle=solid](0,0)(1.03,1)}
\psbezier(25.4,6.62)(28.3,7.31)(28.3,8.66)(28.3,10.02)
\psbezier(24.02,6.62)(21.26,7.29)(21.26,8.6)(21.26,9.92)
\psline(24.8,0.57)(24.8,6.07)
\psline(28.3,10.12)(28.3,15)
\newrgbcolor{userFillColour}{0.8 0.8 0.8}
\pspolygon[fillcolor=userFillColour,fillstyle=solid](18.48,13.1)(24.1,13.1)(24.1,8.72)(18.48,8.72)
\rput(21.18,11.07){$r$}
\rput{90}(28.3,15.1){\psellipse[linestyle=none,fillstyle=solid](0,0)(1.03,1)}
\end{pspicture}

\end{center}
where we again used Proposition \ref{prop:spider}.

\begin{prop}\label{prop:cartbicat3}
If $(X,\Delta, \top)$ is a classical structure then $\Delta$ and $\top$ have right adjoints in $\CCc_r$ with respect to the partial ordering $\leq$, namely $\nabla=\Delta^\dagger$ and $\bot=\top^\dagger$.
\end{prop}

For $r\in\CCc_r(X,Y)$ and $s\in\CCc_r(Y,X)$ we have $r\dashv s$ iff 
\[
\id_X\leq s\circ r
\qquad\mbox{and}\qquad
r\circ s\leq\id_Y\,
\]
that is, 
\[
\id_X=\Delta^\dagger_X\circ \bigl( \id_X \otimes (s\circ r)  \bigr) \circ\Delta_X
\ \ \, \mbox{and}\ \ \,
r\circ s=\Delta^\dagger_Y\circ \bigl((r\circ s)  \otimes \id_Y \bigr) \circ\Delta_Y\,.
\]
In the graphical language this means 
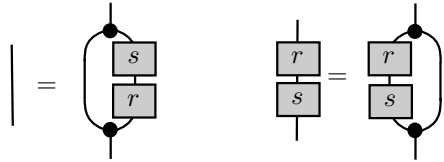
\begin{center}
\ifx\JPicScale\undefined\def\JPicScale{1}\fi
\psset{unit=\JPicScale mm}
\psset{linewidth=0.3,dotsep=1,hatchwidth=0.3,hatchsep=1.5,shadowsize=1,dimen=middle}
\psset{dotsize=0.7 2.5,dotscale=1 1,fillcolor=black}
\psset{arrowsize=1 2,arrowlength=1,arrowinset=0.25,tbarsize=0.7 5,bracketlength=0.15,rbracketlength=0.15}
\begin{pspicture}(0,0)(58.62,21.17)
\psline(18,8.1)(18,13.3)
\psline(11.3,8.5)(11.3,13.7)
\psline(1.74,4.8)(1.6,15.6)
\rput(6.3,10){=}
\rput{90}(14.72,4.27){\psellipse[linestyle=none,fillstyle=solid](0,0)(1.04,1)}
\psbezier(15.1,4.09)(18.1,5)(18.1,6.79)(18.1,8.59)
\psline(14.76,0.39)(14.76,3.74)
\rput(44.9,11.18){=}
\psbezier(14.9,3.99)(11.36,4.9)(11.3,6.8)(11.3,8.6)
\rput{0}(14.71,17.43){\psellipse[linestyle=none,fillstyle=solid](0,0)(1,-1)}
\psbezier(15.1,17.6)(18.1,16.73)(18.1,15)(18.1,13.27)
\psline(14.75,21.17)(14.75,17.94)
\psbezier(14.9,17.7)(11.3,16.82)(11.3,15.1)(11.3,13.36)
\psline(39.5,2.8)(39.6,18.9)
\newrgbcolor{userFillColour}{0.8 0.8 0.8}
\pspolygon[fillcolor=userFillColour,fillstyle=solid](15.08,10.28)(20.7,10.28)(20.7,5.9)(15.08,5.9)
\rput(17.78,8.25){$r$}
\newrgbcolor{userFillColour}{0.8 0.8 0.8}
\pspolygon[fillcolor=userFillColour,fillstyle=solid](15.1,15.88)(20.72,15.88)(20.72,11.5)(15.1,11.5)
\rput(17.8,13.85){$s$}
\newrgbcolor{userFillColour}{0.8 0.8 0.8}
\pspolygon[fillcolor=userFillColour,fillstyle=solid](36.9,15.9)(42.52,15.9)(42.52,11.52)(36.9,11.52)
\rput(39.6,13.87){$r$}
\newrgbcolor{userFillColour}{0.8 0.8 0.8}
\pspolygon[fillcolor=userFillColour,fillstyle=solid](36.9,10.48)(42.52,10.48)(42.52,6.1)(36.9,6.1)
\rput(39.6,8.45){$s$}
\psline(58.6,8.1)(58.6,13.3)
\psline(51.82,8.4)(51.82,13.6)
\rput{90}(55.24,4.17){\psellipse[linestyle=none,fillstyle=solid](0,0)(1.04,1)}
\psbezier(55.62,3.99)(58.62,4.9)(58.62,6.69)(58.62,8.49)
\psline(55.28,0.29)(55.28,3.64)
\psbezier(55.42,3.89)(51.88,4.8)(51.82,6.7)(51.82,8.5)
\rput{0}(55.23,17.33){\psellipse[linestyle=none,fillstyle=solid](0,0)(1,-1)}
\psbezier(55.62,17.5)(58.62,16.63)(58.62,14.9)(58.62,13.17)
\psline(55.27,21.07)(55.27,17.84)
\psbezier(55.42,17.6)(51.82,16.72)(51.82,15)(51.82,13.26)
\newrgbcolor{userFillColour}{0.8 0.8 0.8}
\pspolygon[fillcolor=userFillColour,fillstyle=solid](49.06,15.9)(54.68,15.9)(54.68,11.52)(49.06,11.52)
\rput(51.76,13.87){$r$}
\newrgbcolor{userFillColour}{0.8 0.8 0.8}
\pspolygon[fillcolor=userFillColour,fillstyle=solid](49.18,10.2)(54.8,10.2)(54.8,5.82)(49.18,5.82)
\rput(51.88,8.17){$s$}
\end{pspicture}

\end{center}
and for the specific case of $\delta\dashv \delta^\dagger$ these become
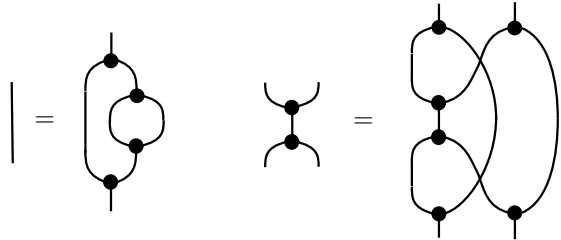
\begin{center}
\ifx\JPicScale\undefined\def\JPicScale{1}\fi
\psset{unit=\JPicScale mm}
\psset{linewidth=0.3,dotsep=1,hatchwidth=0.3,hatchsep=1.5,shadowsize=1,dimen=middle}
\psset{dotsize=0.7 2.5,dotscale=1 1,fillcolor=black}
\psset{arrowsize=1 2,arrowlength=1,arrowinset=0.25,tbarsize=0.7 5,bracketlength=0.15,rbracketlength=0.15}
\begin{pspicture}(0,0)(74.1,31.2)
\psline(11.3,11.2)(11.3,19.3)
\psline(1.64,9.77)(1.5,20.57)
\rput(5.9,15.37){=}
\rput{90}(18.04,12.08){\psellipse[linestyle=none,fillstyle=solid](0,0)(1.03,1)}
\psbezier(18.68,12.1)(21.7,12.79)(21.7,14.14)(21.7,15.5)
\psbezier(17.3,12.1)(14.54,12.77)(14.54,14.08)(14.54,15.4)
\rput{0}(18.16,18.76){\psellipse[linestyle=none,fillstyle=solid](0,0)(1,-0.98)}
\psbezier(18.8,18.75)(21.66,18.09)(21.66,16.8)(21.66,15.5)
\psbezier(17.42,18.75)(14.56,18.05)(14.56,16.68)(14.56,15.3)
\rput{90}(14.66,7.28){\psellipse[linestyle=none,fillstyle=solid](0,0)(1.03,1)}
\psbezier(15.05,7.1)(18.04,8.01)(18.04,9.8)(18.04,11.6)
\psline(14.7,3.4)(14.7,6.75)
\rput(48.3,15.2){=}
\rput{90}(58.26,17.84){\psellipse[linestyle=none,fillstyle=solid](0,0)(1.03,1)}
\psbezier(58.9,17.86)(61.8,18.55)(62.5,20.7)(63.7,23.1)
\psbezier(57.52,17.86)(54.7,18.53)(54.7,19.84)(54.7,21.16)
\rput{0}(58.3,27.86){\psellipse[linestyle=none,fillstyle=solid](0,0)(1,-0.98)}
\psbezier(58.94,27.85)(61.8,27.26)(66,21.93)(65.9,15.6)
\psbezier(57.56,27.85)(54.7,27.15)(54.7,25.78)(54.7,24.4)
\rput{0}(68.4,27.76){\psellipse[linestyle=none,fillstyle=solid](0,0)(1,-0.98)}
\psbezier(69.34,27.75)(73.1,26.16)(74.1,20.43)(74.1,15.7)
\psbezier(67.66,27.75)(64.8,27.11)(64.3,24.9)(63.6,22.9)
\psline(58.3,27.65)(58.3,31)
\psline(68.4,27.85)(68.4,31.2)
\psline(54.7,20.95)(54.7,24.3)
\psline(58.3,14.2)(58.3,17.55)
\rput{90}(58.26,13.28){\psellipse[linestyle=none,fillstyle=solid](0,0)(1.05,-1)}
\psbezier(58.9,13.27)(61.8,12.57)(62.5,10.37)(63.7,7.93)
\psbezier(57.52,13.27)(54.7,12.59)(54.7,11.25)(54.7,9.9)
\rput{0}(58.3,3.07){\psellipse[linestyle=none,fillstyle=solid](0,0)(1,1)}
\psbezier(58.94,3.08)(61.8,3.71)(66,9.38)(65.9,16.1)
\psbezier(57.56,3.08)(54.7,3.8)(54.7,5.19)(54.7,6.6)
\rput{0}(68.4,3.18){\psellipse[linestyle=none,fillstyle=solid](0,0)(1,1)}
\psbezier(69.34,3.19)(73.1,4.84)(74.1,10.79)(74.1,15.7)
\psbezier(67.66,3.19)(64.8,3.84)(64.3,6.09)(63.6,8.13)
\psline(58.3,3.29)(58.3,-0.13)
\psline(68.4,3.08)(68.4,-0.33)
\psline(54.7,10.12)(54.7,6.7)
\psbezier(14.85,7)(11.3,7.91)(11.3,9.7)(11.3,11.5)
\rput{0}(14.71,23.42){\psellipse[linestyle=none,fillstyle=solid](0,0)(1,-1)}
\psbezier(15.1,23.6)(18.1,22.73)(18.1,21)(18.1,19.27)
\psline(14.75,27.17)(14.75,23.94)
\psbezier(14.9,23.7)(11.3,22.82)(11.3,21.1)(11.3,19.36)
\rput{90}(38.76,17.18){\psellipse[linestyle=none,fillstyle=solid](0,0)(1.03,1)}
\psbezier(38.02,17.2)(35.2,17.87)(35.2,19.18)(35.2,20.5)
\psline(38.8,13.54)(38.8,16.89)
\rput{90}(38.76,12.62){\psellipse[linestyle=none,fillstyle=solid](0,0)(1.05,-1)}
\psbezier(38.02,12.61)(35.2,11.93)(35.2,10.59)(35.2,9.25)
\psbezier(39.12,17.27)(42.3,17.94)(42.3,19.25)(42.3,20.57)
\psbezier(39.12,12.68)(42.3,12)(42.3,10.66)(42.3,9.32)
\end{pspicture}

\end{center}
which hold by Proposition \ref{prop:spider}. Similarly, for $\varepsilon\dashv\varepsilon^\dagger$ we have
\begin{center}
\ifx\JPicScale\undefined\def\JPicScale{1}\fi
\psset{unit=\JPicScale mm}
\psset{linewidth=0.3,dotsep=1,hatchwidth=0.3,hatchsep=1.5,shadowsize=1,dimen=middle}
\psset{dotsize=0.7 2.5,dotscale=1 1,fillcolor=black}
\psset{arrowsize=1 2,arrowlength=1,arrowinset=0.25,tbarsize=0.7 5,bracketlength=0.15,rbracketlength=0.15}
\begin{pspicture}(0,0)(42.1,21.17)
\psline(11.3,8.5)(11.3,13.7)
\psline(1.74,4.8)(1.6,15.6)
\rput(6.3,10){=}
\rput{90}(18.1,9.07){\psellipse[linestyle=none,fillstyle=solid](0,0)(1.03,1)}
\rput{0}(18.16,12.76){\psellipse[linestyle=none,fillstyle=solid](0,0)(1,-0.98)}
\rput{90}(14.72,4.27){\psellipse[linestyle=none,fillstyle=solid](0,0)(1.04,1)}
\psbezier(15.1,4.09)(18.1,5)(18.1,6.79)(18.1,8.59)
\psline(14.76,0.39)(14.76,3.74)
\rput(35.6,9.8){=}
\psbezier(14.9,3.99)(11.36,4.9)(11.3,6.8)(11.3,8.6)
\rput{0}(14.71,17.43){\psellipse[linestyle=none,fillstyle=solid](0,0)(1,-1)}
\psbezier(15.1,17.6)(18.1,16.73)(18.1,15)(18.1,13.27)
\psline(14.75,21.17)(14.75,17.94)
\psbezier(14.9,17.7)(11.3,16.82)(11.3,15.1)(11.3,13.36)
\rput{90}(30.46,12){\psellipse[linestyle=none,fillstyle=solid](0,0)(1.03,1)}
\psline(30.5,8.35)(30.5,11.7)
\rput{90}(30.46,7.43){\psellipse[linestyle=none,fillstyle=solid](0,0)(1.06,-1)}
\rput{90}(41.1,11.97){\psellipse[linestyle=none,fillstyle=solid](0,0)(1.03,1)}
\psline(41.14,8.32)(41.14,11.67)
\rput{90}(41.1,7.4){\psellipse[linestyle=none,fillstyle=solid](0,0)(1.06,-1)}
\end{pspicture}

\end{center}
which again hold by Proposition \ref{prop:spider}.

By Propositions \ref{prop:cartbicat1}, and~\ref{prop:cartbicat3}, the category $\CCc_r$ of classical structures and relations is a  \em cartesian bicategory of relations \em in the sense of Carboni and Walters \cite{CarboniWalters}. More specifically, it is a \em dagger cartesian bicategory of relations\em.

\begin{corollary}\label{col:cartbicat}
For a symmetric monoidal category $\CCc$ the corresponding category $\CCc_r$ 
is a  dagger  cartesian bicategory of relations, that is, 
\bit
\item a symmetric monoidal locally posetal dagger bicategory, 
\item in which every object comes with a classical structure, 
\item in which each morphism is a lax comonoid homomorphism,
\item and in which the classical structures $(\Delta,\bot)$ are 
left adjoint to $(\Delta^\dagger,\bot^\dagger)$, with respect to the partial ordering $\leq$.
\eit
\end{corollary}

\paragraph{Example.} The canonical example in \cite{CarboniWalters}  of a cartesian bicategory of relations is ${\bf FRel}$. As shown above, we recover this example as $\mathbf{FdHilb}_r$, which has  sets of basis vectors of orthonormal bases in $\mathbf{FdHilb}$ as objects, and ordinary relations between these as morphisms.     

\paragraph{Remark.} Surprisingly, $\mathbf{FRel}_r\not=\mathbf{FRel}$!
Indeed, recently  it was shown by Edwards and one of the authors that on the two elements set in $\mathbf{FRel}$ there is not only one, but there are two very different kinds of classical structures  \cite{Spek}. Setting $I:=\{*\}$ and $I\!\! I:=\{0,1\}$, the `expected' classical structure on $I\!\! I$ is
\[
\Delta_Z:I\!\! I\rightarrow I\!\! I\times I\!\! I::\left\{\begin{array}{l} 0\sim (0,0)\\ 1\sim (1,1)\end{array}\right.\ \ \ \mbox{and}\ \ \  \top_{Z}:I\!\! I\rightarrow I::\left\{\begin{array}{l} 0\sim * \\ 1\sim *\end{array}\right..
\]
This is not the only one, also
\[
\Delta_X:I\!\! I\rightarrow I\!\! I\times I\!\! I::\left\{\begin{array}{l} 0\sim (0,1),(1,0)\\ 1\sim (0,0),(1,1)\end{array}\right.\ \ \ \mbox{and}\ \ \  \top_{X}:I\!\! I\rightarrow I:: 0\sim *.
\]
is a classical structure, which has very different properties than the above one, for example, it only has one `classical point' --- see \cite{Spek} for the definition of this.  All the classical structures in ${\bf FRel}$, of which there are plenty,  have been classified by one of the authors in~\cite{Pavlovic09}. These `non-standard' classical structures result in some fascinating facts which we intend to present elsewhere.  For example, consider the relation $r:I\rightarrow I\!\! I::*\sim 0$.  Then we have
$\Delta_X\circ r:I\rightarrow I\!\! I\times I\!\! I:: *\sim (0,1),(1,0)$ while
$(r\otimes r)\circ\lambda_I:I\rightarrow I\!\! I\times I\!\! I:: *\sim (0,0)$.  Hence $r$ does not seem to be a lax comonoid homomorphism, contradicting Corollary \ref{col:cartbicat}. What resolves this is the fact that  $r\star_{_X} r=*\sim 1\not=r$. That is, as strange as it may sound, $r$ \em is not a relation relative to \em $(I\!\! I, \Delta_X, \top_X)$!  But on the other hand,  $r':I\rightarrow I\!\! I::*\sim 1$ does satisfy $r'\star_{_X} r'=*\sim 1=r$ so it is a relation.  Now we have 
$\Delta_X\circ r':I\rightarrow I\!\! I\times I\!\! I:: *\sim (0,0),(1,1)$ while
$(r'\otimes r')\circ\lambda_I:I\rightarrow I\!\! I\times I\!\! I:: *\sim (1,1)$.  Since $r'$ being a lax comonoid homomorphism requires $\Delta_X\circ r'=*\sim (0,0),(1,1)$ to be below $(r'\otimes r')\circ\lambda_I=*\sim (1,1)$ in the partial order on relations, this again seems to be in contradiction with Corollary \ref{col:cartbicat}. What resolves this is the fact that  the partial order on relations depends on the classical structure relative to which we define it, and indeed, 
$\{(0,0),(1,1)\}\leq_X \{(1,1)\}$!

\paragraph{Remark.}  Our definition differs from the one in \cite{CarboniWalters} in that we do not assume that on each object there is no other classical structure which is also a comonoid homomorphism.  While we do not know of a counterexample, we were not able to prove that relative to a fixed local order induced by chosen classical structures, these classical structures are the only ones that have right adjoints relative to it.   In particular, the classical structures discussed in \cite{Spek} which differ from the ones that provide ${\bf FRel}$ with the structure of a dagger  cartesian bicategory of relations do not admit right adjoint with respect to the local ordering in ${\bf FRel}$, since the structure maps are proper relations, not functions (cf.~Definition 1.5 and Lemma 2.5 in \cite{CarboniWalters}).

\paragraph{Remark.}
From the above it follows that being a  lax comonoid homomorphism means that $r=r^{\star 3}=r\star (r\star r)$, which is a strictly weaker condition than $r=r^{\star 2}=r\star r$, the defining equation for relations.  For example, while there are only two complex numbers $c\in\mathbb{C}$ satisfying $c=c^2$, $0$ and $1$, there are three satisfying $c=c^3$, $0$, $1$ and $-1$. As a consequence, not all lax comonoid homomorphisms are relations.

\subsection{Functions}

Following Carboni and Walters in \cite{CarboniWalters}, the preservation of the parts of the comonoid structure corresponds to the familiar properties of relations. A relation $r\in \CCc_r(X,Y)$ is called
\bit
\item {\it single-valued} if $\Delta_Y \circ r = (r\otimes r)\circ \Delta_X$\,;
\item {\it total}  if $\top_Y \circ r = \top_X$\,;
\item {\it function}  if it is both total and single-valued.
\eit

\paragraph{Example.}
In {\bf FdHilb}, these notions correspond to the standard ones:
\begin{itemize}
\item the matrix of a single-valued relation has at most one 1 in each column, while the remaining entries must be 0\,;
\item the matrix of  a total relation has at least one 1 in each column\,;
\item the matrix of a function has exactly one 1 in each column.
\end{itemize}
\begin{prop}
The following are equivalent for $f\in\CCc_c(X,Y)${\rm :}
\bit
\item $f$ is a function i.e.~a total single-valued relation\,{\rm;}
\item $f$ is a real comonoid homomorphism, i.e.~we have that $f=f_*$, and that the following two diagrams commute\,{\rm:}
\beq\label{eq:Cong_Funct_pre}
\xymatrix@=0.54in{
X\ar[r]^{\Delta_X}\ar[d]_{f} & XX\ar[d]^{f\otimes f}\\
Y\ar[r]_{\Delta_Y} & YY
}
\qquad\quad
\xymatrix@=0.54in{
X\ar[r]^{\top_X}\ar[d]_{f} & I\ar[d]^{\id_I}\\
Y\ar[r]_{\top_Y} & I
}
\eeq
\item 
the following two diagrams commute\,{\rm:}
\beq\label{eq:Cong_Funct}
\xymatrix@=0.54in{
X\ar[r]^{\Delta_X}\ar[d]_{f} & XX\ar[d]^{f_*\otimes f}\\
Y\ar[r]_{\Delta_Y} & YY
}
\qquad\quad
\xymatrix@=0.54in{
X\ar[r]^{\top_X}\ar[d]_{f} & I\ar[d]^{\id_I}\\
Y\ar[r]_{\top_Y} & I
}
\eeq
\eit
\end{prop}

Restricting to comonoid homomorphisms makes the comonoid components of the classical structures into natural transformations 
\[
\Delta: X\to X\otimes X\qquad \mbox{ and } \qquad\top :X\to I\,. 
\]
It is easy to see that a tensor with such natural transformation is just a cartesian product from which~\cite{Fox}:

\be{prop}
The category $\CCc_f$ of classical structures and functions is cartesian. The inclusion functor
\[
\CCc_f \inclusion \CCc_{\cs}
\]
maps cartesian products of $\CCc_f$ to symmetric monoidal structure of $\CCc_\cs$. 
\end{prop}

\subsection{Permutations}

\begin{lemma}
If a relation $r$ has an inverse $r'$ then $r$ is a function.
\end{lemma}

Indeed, one easily verifies that if a relation $r$ has an inverse $r'$, then $r\dashv r'$, so  by Definition 1.5 and Theorem 1.6 in \cite{CarboniWalters} it is a function.

\begin{lemma}
If both $r$ and $r^\dagger$ are functions, then they are invertible.
\end{lemma}

Indeed, if both $r$ and $r^\dagger$ are functions then $r$ is both a monoid and a comonoid homomorphism, and in  \cite{Kock} it was shown that a morphism between Frobenius algebras which is both a monoid and a comonoid homomorphism must be invertible. 

\begin{lemma}
If $r$ and $r^\dagger$ are functions, then $r^\dagger\circ r = \id$.
\end{lemma}
Indeed, we have
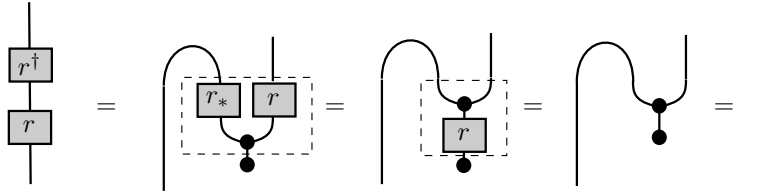
\begin{center}
\ifx\JPicScale\undefined\def\JPicScale{1}\fi
\psset{unit=\JPicScale mm}
\psset{linewidth=0.3,dotsep=1,hatchwidth=0.3,hatchsep=1.5,shadowsize=1,dimen=middle}
\psset{dotsize=0.7 2.5,dotscale=1 1,fillcolor=black}
\psset{arrowsize=1 2,arrowlength=1,arrowinset=0.25,tbarsize=0.7 5,bracketlength=0.15,rbracketlength=0.15}
\begin{pspicture}(0,0)(102.7,25)
\psline(4.9,0.8)(4.7,25)
\psbezier(22.6,13.9)(22.6,20.68)(30.1,21.14)(30.1,13.9)
\newrgbcolor{userFillColour}{0.8 0.8 0.8}
\pspolygon[fillcolor=userFillColour,fillstyle=solid](2,10.53)(7.62,10.53)(7.62,6.15)(2,6.15)
\rput(4.7,8.5){$r$}
\newrgbcolor{userFillColour}{0.8 0.8 0.8}
\pspolygon[fillcolor=userFillColour,fillstyle=solid](2.1,18.83)(7.72,18.83)(7.72,14.45)(2.1,14.45)
\rput(4.8,16.8){$r^\dagger$}
\rput{90}(33.7,6.42){\psellipse[linestyle=none,fillstyle=solid](0,0)(1.03,1)}
\psbezier(34.34,6.43)(37.1,7.1)(37.1,8.41)(37.1,9.73)
\psbezier(32.96,6.43)(30.2,7.1)(30.2,8.41)(30.2,9.73)
\psline(33.7,2.83)(33.7,6.18)
\newrgbcolor{userFillColour}{0.8 0.8 0.8}
\pspolygon[fillcolor=userFillColour,fillstyle=solid](27.1,14.13)(32.72,14.13)(32.72,9.75)(27.1,9.75)
\rput(29.8,12.1){$r_*$}
\newrgbcolor{userFillColour}{0.8 0.8 0.8}
\pspolygon[fillcolor=userFillColour,fillstyle=solid](34.49,14.18)(40.11,14.18)(40.11,9.8)(34.49,9.8)
\rput(37.19,12.15){$r$}
\rput{90}(33.7,3.4){\psellipse[linestyle=none,fillstyle=solid](0,0)(1.03,1)}
\psline(37.1,14.5)(37.1,20.4)
\psline(22.6,-0.1)(22.6,13.8)
\psbezier(51.59,14.84)(51.59,21.4)(59.09,21.84)(59.09,14.84)
\rput{90}(62.56,11.48){\psellipse[linestyle=none,fillstyle=solid](0,0)(1.03,1)}
\psbezier(63.2,11.5)(66.1,12.19)(66.1,13.54)(66.1,14.9)
\psbezier(61.82,11.5)(59.06,12.17)(59.06,13.48)(59.06,14.8)
\psline(62.56,7.9)(62.56,11.25)
\rput{90}(62.5,3.3){\psellipse[linestyle=none,fillstyle=solid](0,0)(1.03,1)}
\psline(66.1,15)(66.1,20.9)
\psline(51.6,0.8)(51.6,14.7)
\psline(62.5,3.7)(62.5,7.05)
\newrgbcolor{userFillColour}{0.8 0.8 0.8}
\pspolygon[fillcolor=userFillColour,fillstyle=solid](59.78,9.48)(65.4,9.48)(65.4,5.1)(59.78,5.1)
\rput(62.48,7.45){$r$}
\psbezier(77.51,14.4)(77.51,21.14)(85.01,21.59)(85.01,14.4)
\rput{90}(88.48,11.24){\psellipse[linestyle=none,fillstyle=solid](0,0)(1.03,1)}
\psbezier(89.12,11.25)(92.02,11.95)(92.02,13.32)(92.02,14.7)
\psbezier(87.74,11.25)(84.98,11.92)(84.98,13.23)(84.98,14.55)
\psline(88.48,7.65)(88.48,11)
\rput{90}(88.5,7.04){\psellipse[linestyle=none,fillstyle=solid](0,0)(1.03,1)}
\psline(92.02,14.75)(92.02,20.65)
\psline(77.52,0.55)(77.52,14.45)
\psline(102.7,1.3)(102.7,20.4)
\rput(97.1,10.9){=}
\rput(71.7,11){=}
\rput(45.4,11){=}
\rput(15,10.9){=}
\pspolygon[linewidth=0.15,linestyle=dashed,dash=1 1](25,15)(42.3,15)(42.3,4.8)(25,4.8)
\pspolygon[linewidth=0.15,linestyle=dashed,dash=1 1](56.8,14.6)(68.2,14.6)(68.2,4.6)(56.8,4.6)
\end{pspicture}

\end{center}
where we relied on eq.(\ref{eq:Cong_Funct}) and on the assumption that $r^\dagger$ is total. 
So if both $r$ and $r^\dagger$ are functions then they must be unitary.  

\begin{prop}\label{prop:def_perm}
The following are equivalent for $r\in\CCc_r(X,Y)$\,{\rm:}
\bit
\item $r$ and $r^\dagger$ are functions\,{\rm;}
\item $r$ has an inverse $r'$\,{\rm;}
\item $r$ is unitary.
\eit
\end{prop}

\begin{defn}\em
A \em permutation \em is a relation which satisfies the equivalent conditions of Proposition \ref{prop:def_perm}.
\end{defn}

We denote by $\Cperm$ the groupoid of classical structures  and permutations.
The cartesian category $\CCc_f$ is in general not self-dual, so it is not a dagger category. The category $\Cperm$, on the other hand, has a degenerate dagger, mapping each  $f:X\to Y$ to its inverse $f^\dagger:Y\to X$.  

\subsection{Stochastic maps}\label{sec:stochmaps}

In Section \ref{sec:clas_morph} we defined classical morphisms, and in ${\bf FdHilb}$ these classical morphisms maps probability distributions to probability distributions, up to a positive real scalar.  We now define the normalised counterpart.

\be{defn}\em
A total classical morphism $s\in \CCc_{\cs}(X,Y)$ is called {\em stochastic}. It is {\em doubly stochastic} if both $s$ and $s^\adjbis$ are stochastic.  Denote by $\CCc_s$ the category of classical structures and stochastic morphisms.  
\ee{defn}

The inclusion $\CCc_s\inclusion \Cclas$ is both functorial and monoidal, but obviously,  $\CCc_s$ does not inherit dagger structure nor compact structure in general.
If $h:A\to B$ is doubly stochastic then ${\rm dim}(A)={\rm dim}(B)$.

For two objects of the same dimension we denote by $\CCc_{ds}(A,B)$ the set of all doubly stochastic maps of type $A\to B$. 

Since all relations are positive, permutations are doubly stochastic. 

\begin{defn}
A morphism $f:X_1\to X_2$ is \em majorized \em by a morphism $g:Y_1\to Y_2$ if there exist doubly stochastic maps $h_1:X_1\to Y_1$ and $h_2:X_2\to Y_2$ such that $g=h_2\circ f\circ h_1^\dagger$.
\end{defn}

\begin{prop}
Majorization is a preordering on $\bigcup_{XY}\CCc_s(X,Y)$.
\end{prop}

\paragraph{Example.} In ${\bf FdHilb}$ all the concepts defined in this section coincide with the usual ones.  That is, states in ${\bf FdHilb}_s({\cal H})$ are probability distributions and maps in ${\bf FdHilb}_s({\cal H},{\cal H}')$ send probability distributions to probability distributions.  Majorization on $\bigcup_{{\cal H}{\cal H}'}{\bf FdHilb}_s({\cal H},{\cal H}')$ extends the usual notion of majorisation  which is typically only defined for probability distributions \cite{AlbertiUhlmann, Nielsen}.

\subsection{Hierarchy of classical varieties}
We order all classical varieties extracted from $\CCc_{\cs}$ that is: permutations $\CCc_p$, functions $\CCc_f$, stochastic maps $\CCc_s$, relations $\CCc_r$, and classical morphisms $\Cclas$.  

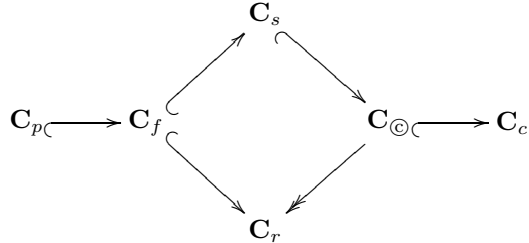
\[
\xymatrix@=0.36in{  
&& \CCc_s \ar@{_{(}->}[dr] \\
\CCc_p \ar@{_{(}->}[r] & \CCc_f \ar@{_{(}->}[ur] \ar@{^{(}->}[dr] && {\Cclas} \ar@{_{(}->}[r]  \ar@{->>}[dl] & \CCc_{\cs}\\ 
&& \CCc_r
}
\]

\section{Relativizing over classical interfaces}\label{Relativizing}
Every object $X$ in a symmetric monoidal dagger category $\CCc$ induces an endofunctor $X\otimes(-):\CCc\to\CCc$. The endofunctors $F:\CCc\to \CCc$ that are in this form can be recognized by their \em strength\em, {\em viz\/} the natural isomorphism $F(A\otimes B) \cong FA\otimes B$. Indeed, every monoidal category is equivalent with the category of strong endofunctors on it. Extending this correspondence, a monoid structure on $X$ induces a monad structure on the corresponding endofunctor $X\otimes (-)$; a comonoid structure on $X$ corresponds to a comonad structure on it. Since a classical structure $X$ carries both a monoid and a comonoid structure, the induced endofunctor $X\otimes (-)$ is both a monad and a comonad. The structure of such correspondences, and the particular logical meaning of the comonads $X\otimes (-)$ was analyzed in \cite{Pavlovic}. 

The case when $\CCc$ is a cartesian category, i.e.~when the tensor $\otimes$ is the cartesian product $\times$, goes back to the early days of categorical logic: the comonads in the form $X\times (-)$ were analysed by Lambek and Scott  already in \cite{LambekScott}. In general, the Kleisli category $\XC$ induced by a comonad $X\otimes (-) :\CCc\to \CCc$, induced by a comonoid object $X$ in a monoidal category $\CCc$, captures the data flows relative to the data type $X$. 

When $\CCc$ is a symmetric monoidal \em dagger \em category, then these data flows can be construed as quantum flows relative to the classical data of type $X$. 
Commutative Frobenius algebra structure of $X$  assures  that $\XC$ is also a symmetric monoidal dagger category \cite{Pavlovic08}.

\subsection{Indexing over a classical structure}

In the sequel, we often abbreviate $X\otimes (-)$ to $X(-)$.   Recall that the Kleisli category $\XC$, induced by the comonad 
\[
X(-):\CCc\to \CCc
\]
corresponding to the comonoid 
\[
\xymatrix{I&  \ar[l]_-{\top}X  \ar[r]^-{\Delta}& X\otimes X}\,, 
\]
consists of
\begin{itemize}
\item the same objects as $\CCc$
\item $\XC(A,B) = \CCc(XA,B)$
\item $\id_A = XA\tto{\top A}A$,
\item given $f:XA\to B$ and $g:XB\to C$, the composite is
\bear
g\circ_X f & = & XA\tto{\Delta A} XXA\tto{Xf}XB\tto{g} C\,.
\eear
\end{itemize}
The monoidal structure of $\XC$ is
\begin{itemize}
\item for $A,B\in|\CCc|$ the tensor $A\otimes_X B = A\otimes B$ is the same as in $\CCc$
\item for $f: XA\to B$ and $h:XC\to D$, the tensor is
\bear
f\otimes_X h & = & XAC\tto{\Delta AC} XXAC\tto{X\sigma C}XAXC\tto{f\otimes h}BD
\eear
\item the monoid unit is $I_X = I$, but the scalars in $\XC$ are $\CCc(X,I)$.
\end{itemize}
The dagger structure of $\XC$ makes the full use of the classical  structure of $X$:
\begin{itemize}
\item given $f\in \XC(A,B)$, i.e.~an arrow $f\in \CCc(XA,B)$, $f^{\adjbis_X}\in \XC(B,A)$ is defined to be the transpose of its adjoint $f^\adjbis\in \CCc(B,XA)$, i.e.
\bear
f^{\adjbis_X} & = & XB \tto{Xf^\adjbis} XXA\tto{\varepsilon A} A\,.
\eear
\end{itemize}
Composition, tensor and dagger in $\CCc_X$ respectively depict as:
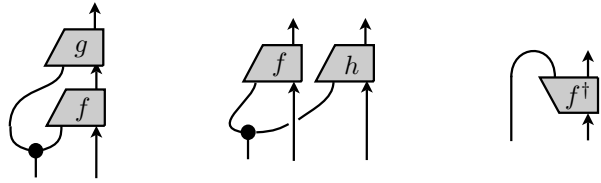
\begin{center}
\ifx\JPicScale\undefined\def\JPicScale{1}\fi
\psset{unit=\JPicScale mm}
\psset{linewidth=0.3,dotsep=1,hatchwidth=0.3,hatchsep=1.5,shadowsize=1,dimen=middle}
\psset{dotsize=0.7 2.5,dotscale=1 1,fillcolor=black}
\psset{arrowsize=1 2,arrowlength=1,arrowinset=0.25,tbarsize=0.7 5,bracketlength=0.15,rbracketlength=0.15}
\begin{pspicture}(0,0)(81.1,24.72)
\rput{90}(6.36,4.98){\psellipse[linestyle=none,fillstyle=solid](0,0)(1.03,1)}
\psbezier(7,5)(9.76,5.67)(9.76,6.98)(9.76,8.3)
\psbezier(5.62,5)(2.86,5.67)(3,6.88)(3,8.2)
\psline(6.36,1.4)(6.36,4.75)
\psline{<-}(14.41,8.27)(14.4,1.3)
\psline{<-}(14.41,16.72)(14.41,13.1)
\newrgbcolor{userFillColour}{0.8 0.8 0.8}
\psline[fillcolor=userFillColour,fillstyle=solid](15.53,8.27)
(15.53,13.1)
(10.37,13.1)
(7.8,8.27)(15.53,8.27)
\rput(12.6,10.5){$f$}
\psline{<-}(14.21,24.72)(14.21,21.1)
\newrgbcolor{userFillColour}{0.8 0.8 0.8}
\psline[fillcolor=userFillColour,fillstyle=solid](15.33,16.27)
(15.33,21.1)
(10.17,21.1)
(7.6,16.27)(15.33,16.27)
\rput(12.4,18.5){$g$}
\psbezier(2.9,8.2)(3,14)(9.8,12.4)(10,16.2)
\psline{<-}(50.18,22.62)(50.18,19)
\newrgbcolor{userFillColour}{0.8 0.8 0.8}
\psline[fillcolor=userFillColour,fillstyle=solid](51.3,14.17)
(51.3,19)
(46.14,19)
(43.57,14.17)(51.3,14.17)
\rput(48.37,16.4){$h$}
\psline{<-}(40.58,22.62)(40.58,19)
\newrgbcolor{userFillColour}{0.8 0.8 0.8}
\psline[fillcolor=userFillColour,fillstyle=solid](41.7,14.17)
(41.7,19)
(36.54,19)
(33.97,14.17)(41.7,14.17)
\rput(38.77,16.4){$f$}
\psline{<-}(40.67,14.16)(40.66,3.6)
\psline{<-}(50.37,14.27)(50.36,3.71)
\rput{90}(34.57,7.4){\psellipse[linestyle=none,fillstyle=solid](0,0)(1.03,1)}
\psbezier(41.27,9.1)(43.37,10.6)(45.87,11.8)(45.93,14.08)
\psbezier(33.83,7.42)(29.27,8.2)(35.37,11.7)(35.57,14)
\psline(34.57,3.82)(34.57,7.17)
\psbezier(35.67,7.4)(38.17,7.5)(38.97,8)(39.97,8.5)
\newrgbcolor{userFillColour}{0.8 0.8 0.8}
\psline[fillcolor=userFillColour,fillstyle=solid](81.1,14.7)
(81.1,10)
(75.94,10)
(73.37,14.7)(81.1,14.7)
\rput(78.5,12.3){$f^\dagger$}
\psline{<-}(79.77,18.35)(79.77,14.73)
\psline{<-}(79.77,9.95)(79.77,6.33)
\psbezier(69.57,14.73)(69.57,19.23)(75.37,19.53)(75.37,14.73)
\psline(69.57,14.93)(69.57,6.23)
\end{pspicture}

\end{center}

The Kleisli category $\CX$ for the monad $X(-):\CCc\to \CCc$ has the hom-sets $\CX(A,B) = \CCc(A,XB)$, and a structure dual to the above. The duality is in fact formal, by transposing the $X$-type:

\be{prop}\label{Kleisliso}
The 
identity-on-objects-functors
\[
\xymatrix@=0.54in{
\XC \ar@/^0.8em/[r]^{(-)^X} & \CX \ar@/^0.8em/[l]^{(-)_X}
}
\]
with
\[
\Bigl(XA\tto{f}B\Bigr)\  \stackrel{\ \ (-)^X}{\mapsto}\  \Bigl(A\tto{\eta A}XXA\tto{Xf} XB\Bigr)
\]
and 
\[
\Bigl(A\tto{f}XB\Bigr)\  \stackrel{\ \ (-)_X}{\mapsto}\  \Bigl(XA\tto{Xf}XXB\tto{\varepsilon B} B\Bigr)
\]
make the Kleisli categories $\XC$ and $\CX$ isomorphic.
\ee{prop}

The usual Kleisli adjunction 
\[
\xymatrix@=0.54in{
\CCc \ar@/^0.8em/[r]^{F}="a" & \XC \ar@/^0.8em/[l]^{U}="b"\ar@{}|\bot "a";"b"
}
\]
where
\beqa
A&\stackrel{F}{\mapsto}&A\\
\Bigl(A\tto{f}B\Bigr)&\stackrel{}{\mapsto}& \Bigl(XA\tto{Xf}XB \tto{\top B}B\Bigr)
\eeqa
and
\beqa
A&\stackrel{U}{\mapsto} & XA\\
\Bigl(XA\tto{f}B\Bigr)& \stackrel{}{\mapsto} & \Bigl(XA\tto{\Delta A}XXA \tto{Xf }XB\Bigr)
\eeqa
can now be graphically presented as follows:
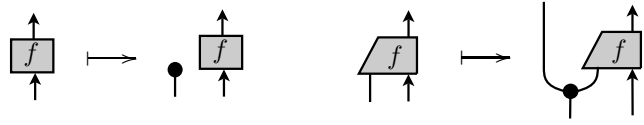
\begin{center}
\ifx\JPicScale\undefined\def\JPicScale{1}\fi
\psset{unit=\JPicScale mm}
\psset{linewidth=0.3,dotsep=1,hatchwidth=0.3,hatchsep=1.5,shadowsize=1,dimen=middle}
\psset{dotsize=0.7 2.5,dotscale=1 1,fillcolor=black}
\psset{arrowsize=1 2,arrowlength=1,arrowinset=0.25,tbarsize=0.7 5,bracketlength=0.15,rbracketlength=0.15}
\begin{pspicture}(0,0)(85.63,15.63)
\psline{<-}(54.74,14.52)(54.74,10.9)
\newrgbcolor{userFillColour}{0.8 0.8 0.8}
\psline[fillcolor=userFillColour,fillstyle=solid](55.86,6.07)
(55.86,10.9)
(50.7,10.9)
(48.13,6.07)(55.86,6.07)
\rput(52.93,8.3){$f$}
\psline(72.73,15.63)(72.73,6.93)
\newrgbcolor{userFillColour}{0.8 0.8 0.8}
\pspolygon[fillcolor=userFillColour,fillstyle=solid](1.9,10.63)(7.52,10.63)(7.52,6.25)(1.9,6.25)
\rput(4.6,8.6){$f$}
\psline{<-}(5.1,6.17)(5.1,2.55)
\psline{<-}(4.9,14.17)(4.9,10.55)
\rput(15.3,8.1){$\mapsto$}
\rput{90}(23.7,6.58){\psellipse[linestyle=none,fillstyle=solid](0,0)(1.03,1)}
\psline(23.7,3)(23.7,6.35)
\newrgbcolor{userFillColour}{0.8 0.8 0.8}
\pspolygon[fillcolor=userFillColour,fillstyle=solid](27,11.1)(32.62,11.1)(32.62,6.72)(27,6.72)
\rput(29.7,9.07){$f$}
\psline{<-}(30.2,6.64)(30.2,3.02)
\psline{<-}(30,14.64)(30,11.02)
\psline{<-}(54.73,6)(54.73,2.38)
\psline(49.63,2.3)(49.63,6)
\rput(65,8.2){$\mapsto$}
\psline{<-}(84.51,15.25)(84.51,11.63)
\newrgbcolor{userFillColour}{0.8 0.8 0.8}
\psline[fillcolor=userFillColour,fillstyle=solid](85.63,6.8)
(85.63,11.63)
(80.47,11.63)
(77.9,6.8)(85.63,6.8)
\rput(82.7,9.03){$f$}
\psline{<-}(84.5,6.73)(84.53,0.53)
\rput{0}(76.26,3.72){\psellipse[linestyle=none,fillstyle=solid](0,0)(1.04,-1.02)}
\psbezier(75.6,3.73)(72.73,4.39)(72.73,5.68)(72.73,6.98)
\psbezier(77.03,3.73)(79.9,4.39)(79.9,5.68)(79.9,6.98)
\psline(76.22,0.11)(76.26,3.49)
\end{pspicture}

\end{center}

\noindent It is easy to see that $F$ is a monoidal functor, but $U$ is not. The monoidal functor $F$ enables to interpret of quantum flows which do not involve the classical data type $X$.

\begin{prop}{\rm\cite{Pavlovic08}}
For $X$ a classical structure, the Kleisli category $\XC$ inherits all classical structures from $\CCc$ along $F$, 
and therefore
\bit
\item if $\CCc$ is dagger compact then so is $\CCc_X$\,{\rm ;} 
\item we can consider $(\CCc_X)_Y$  and $(\CCc_X)^Y$, for which we have
\[
(\CCc_X)_Y\simeq  (\CCc_X)^Y\simeq  \CCc_{X\otimes Y}. 
\]
\eit
\end{prop}

All constructions available in dagger compact categories thus lift to $\CCc_X$, but now relative to the classical data type $X$. For example, the inner-product of $x,y\in\CCc_X(A)$ is
\bear
<x\, |\,y>_X & \!\!=\!\! & x^{\adjbis_X}\circ_Xy\\
& = & X\tto{\Delta}XX\tto{X y} XA\tto{X x^\adjbis} XX\tto\varepsilon I
\eear
whereas the transpose and conjugate of $f\in \CCc_X(A,B)$ are
\bear
\!\!\!\!f^{\ast_X} & \!\!=\!\! & (A^*\otimes_X \varepsilon_B)\circ_X(A^*\otimes_X f \otimes_X B^*)\circ_X(\eta_A \otimes_X B^*)
\\
& \!\!=\!\! &
XB^*\!\tto{X\eta_A B^*\!\!}XA^*\!AB^*\!\tto{\sigma A B^*\!\!}A^*\!XAB^*\!\tto{A^*\!fB^*\!}\!\!A^*\!BB^*\!\tto{A^*\varepsilon_B} \!A^*\!
\\ \\
\!\!\!\!f_{\ast_X} 
& \!\!=\!\! & 
\left(f^{\adjbis_X}\right)^{\ast_X}\\
& \!\!=\!\! &
XA^*\!\tto{X\eta_B A^*\!\!}XB^*\!BA^*\!\tto{\sigma B A^*\!\!}B^*\!XBA^*\!\tto{B^*\!Xf^\dagger\! A^*\!}\!\!B^*\!XXAA^*\!\tto{B^*\varepsilon_X\varepsilon_A} \!B^*\!\\
& \!\!=\!\! & f^*
\eear
or pictorially
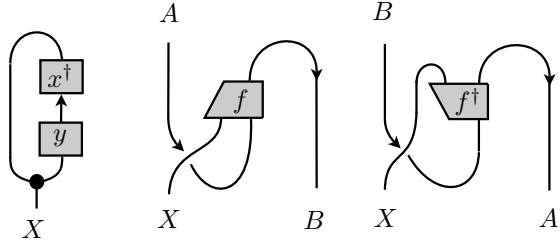
\begin{center}
\ifx\JPicScale\undefined\def\JPicScale{1}\fi
\psset{unit=\JPicScale mm}
\psset{linewidth=0.3,dotsep=1,hatchwidth=0.3,hatchsep=1.5,shadowsize=1,dimen=middle}
\psset{dotsize=0.7 2.5,dotscale=1 1,fillcolor=black}
\psset{arrowsize=1 2,arrowlength=1,arrowinset=0.25,tbarsize=0.7 5,bracketlength=0.15,rbracketlength=0.15}
\begin{pspicture}(0,0)(75.16,30.4)
\psbezier(3.46,23.4)(3.46,28.74)(10.36,29.1)(10.36,23.4)
\psbezier{->}(35.29,21.1)(35.29,27.66)(44.19,28.1)(44.19,21.1)
\psbezier{->}(24.46,16.3)(24.46,14.4)(25.06,13.2)(26.56,11.7)
\newrgbcolor{userFillColour}{0.8 0.8 0.8}
\psline[fillcolor=userFillColour,fillstyle=solid](37.02,16.17)
(37.02,21)
(31.86,21)
(29.29,16.17)(37.02,16.17)
\rput(34.09,18.4){$f$}
\psbezier(31.49,16.2)(31.49,11.2)(24.86,12.6)(24.56,6)
\psbezier(27.46,10)(29.83,4.9)(35.86,5)(35.46,16)
\psline(44.16,21.4)(44.19,6.8)
\psline(24.49,26.2)(24.46,16.2)
\psbezier{->}(65.8,20.8)(65.8,27.36)(75.16,27.4)(75.16,20.4)
\psbezier{->}(53.26,16)(53.26,14.4)(54.16,13.3)(55.36,12.1)
\newrgbcolor{userFillColour}{0.8 0.8 0.8}
\psline[fillcolor=userFillColour,fillstyle=solid](66.96,20.7)
(66.96,16)
(61.8,16)
(59.23,20.7)(66.96,20.7)
\rput(64.3,18.3){$f^\dagger$}
\psbezier(57.46,17)(57.32,9.8)(53.36,12.2)(53.26,6.6)
\psbezier(65.76,11.6)(65.66,6)(59.96,5)(56.36,11.2)
\psline(75.13,20.7)(75.13,6.5)
\psline(53.26,26)(53.26,16)
\psbezier(57.46,20.8)(57.46,23.99)(61.3,24.2)(61.3,20.8)
\psline(57.46,16.8)(57.46,21.1)
\psline{->}(10.16,11.5)(10.25,19.29)
\newrgbcolor{userFillColour}{0.8 0.8 0.8}
\pspolygon[fillcolor=userFillColour,fillstyle=solid](7.36,15.53)(12.98,15.53)(12.98,11.15)(7.36,11.15)
\rput(10.06,13.5){$y$}
\newrgbcolor{userFillColour}{0.8 0.8 0.8}
\pspolygon[fillcolor=userFillColour,fillstyle=solid](7.46,23.83)(13.08,23.83)(13.08,19.45)(7.46,19.45)
\rput(10.16,21.8){$x^\dagger$}
\rput{90}(6.96,7.68){\psellipse[linestyle=none,fillstyle=solid](0,0)(1.03,1)}
\psbezier(7.6,7.7)(10.36,8.37)(10.36,9.68)(10.36,11)
\psbezier(6.22,7.7)(3.46,8.37)(3.46,9.68)(3.46,11)
\psline(6.96,4.1)(6.96,7.45)
\psline(3.46,10.8)(3.46,23.3)
\psline(65.76,11.7)(65.76,16)
\rput(6.56,1.4){$X$}
\rput(24.56,2.7){$X$}
\rput(53.36,2.8){$X$}
\rput(75.06,2.9){$A$}
\rput(24.56,30.1){$A$}
\rput(52.96,30.4){$B$}
\rput(43.86,2.6){$B$}
\end{pspicture}

\end{center}

\subsubsection{Mixing and convex closure}
Let $p$ be a \em stochastic state\em, that is, $p\in\CCc_s(I,X)$ where $X$ is a classical structure.  Then define mappings
\[
\omega_p^{A,B}: \CCc_X(A,B)\to \CCc(A,B)
\]
with
\[
\Bigl(XA\tto{f}B\Bigr)\  \stackrel{\ \ \omega_p^{A,B}}{\mapsto}\  \Bigl(A\tto{p A}XA\tto{f} B\Bigr)
\]
and we call these operations \em mixing \em or \em convex combining\em. We call morphisms in the range of  $\omega_p^{A,B}$ \em mixtures \em or \em convex combinations\em.  

We will sometimes omit the superscripts and just write $\omega_p$ and $\omega_p(f)$.

\begin{prop}
Both stochastic maps and doubly stochastic maps are `convex closed', that is, respectively, 
for $p\in \CCc_{\cs}(X)$,
\[
\begin{array}{ccc}
f\in(\CCc_X)_s(A,B) & \Longrightarrow & \omega_p^{A,B}(f)\in\CCc_s(A,B)\vspace{1.5mm}\\
f\in(\CCc_X)_{ds}(A,B) & \Longrightarrow & \omega_p^{A,B}(f)\in\CCc_{ds}(A,B)
\end{array}
\]
Hence, convex combinations of permutations are doubly stochastic.
\end{prop}

This family of conceptually meaningful mappings is neither monoidal nor functorial. 
It connects quantum categorical semantics with the framework of convex theories \cite[and references therein]{BBLW}.

\subsubsection{Sum}

As a slight variation of the above we define mappings
\[
\sum_X: \CCc_X(A,B)\to \CCc(A,B)
\]
with
\[
\Bigl(XA\tto{f}B\Bigr)\  \stackrel{\  \sum_X}{\mapsto}\  \Bigl(A\tto{\bot A}XA\tto{f} B\Bigr)
\]
and we call these operations \em sum\em.  This notion of sum for $\CCc_X$ induces a notion of sum for $\CCc^X$ which we denote by $\sum^X$.

\paragraph{Example.}  We can now write 
\[
\langle x\, |\, y\rangle = \sum x \star y\,.
\]
In ${\bf FdHilb}$, if $x=(x_1,\ldots, x_n)$ and $y=(y_1,\ldots, y_n)$ are coordinate vectors relative to a chosen basis, then we have 
\[
x \star y=(\bar{x}_1y_1,\ldots, \bar{x}_ny_n)
\]
and 
\[
\sum x \star y=\bar{x}_1y_1+\ldots+ \bar{x}_ny_n\,.
\]

\subsection{Indexed pure quantum states and operations}

The $\Wee$-construction\footnote{Was originally called ${\cal WP}roj$-construction.} for dagger compact categories is introduced in \cite{deLL} to factor out redundant global phases, and involves a passage from vector representation to density matrix representation for pure states.   The category $\Wee \CCc$ consists of
the same objects as $\CCc_{\qo}$ and  morphism $f\in \Wee \CCc (A,B)$ are of the form $f = \varphi_*\otimes\varphi$ for some $\varphi\in\CCc_{\qo}(A,B)$.

\begin{defn}{\rm\cite{deLL}}\label{def:PureStatesOpp} \em
A \em pure quantum state \em is an element $\psi\in \Wee \CCc(A)$. A \em pure quantum evolution \em is a unitary operation $U\in \Wee \CCc(A,B)$.
\end{defn}

We will now see that applying the $\Wee$-construction to $\XC$ brings in important complementary features.  The category $\Wee \XC$ consists of
\begin{itemize}
\item the same objects as $\CCc_{\qo}$\,;
\item morphism $f\in \Wee\XC(A,B)$ are of the form 
\bear
f &= & \varphi_{\ast_X}\otimes_X \varphi\\
& = & A^*XA\tto{A^*\Delta A} A^*XXA \tto{\varphi_*\otimes\varphi}B^*B
\eear
for some $\varphi \in \CCc\left(XA, B\right)$, that is graphically:
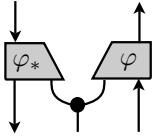
\begin{center}
\ifx\JPicScale\undefined\def\JPicScale{1}\fi
\psset{unit=\JPicScale mm}
\psset{linewidth=0.3,dotsep=1,hatchwidth=0.3,hatchsep=1.5,shadowsize=1,dimen=middle}
\psset{dotsize=0.7 2.5,dotscale=1 1,fillcolor=black}
\psset{arrowsize=1 2,arrowlength=1,arrowinset=0.25,tbarsize=0.7 5,bracketlength=0.15,rbracketlength=0.15}
\begin{pspicture}(0,0)(19.3,18.5)
\rput{90}(9.6,4.68){\psellipse[linestyle=none,fillstyle=solid](0,0)(1.04,1)}
\psbezier(10.24,4.7)(13.14,5.37)(13.14,6.68)(13.14,8)
\psbezier(8.86,4.7)(6.1,5.37)(6.1,6.68)(6.1,8)
\psline(9.6,1.1)(9.6,4.45)
\newrgbcolor{userFillColour}{0.8 0.8 0.8}
\psline[fillcolor=userFillColour,fillstyle=solid](19.3,8.2)
(19.3,13.03)
(14.14,13.03)
(11.57,8.2)(19.3,8.2)
\rput(16.37,10.43){$\varphi$}
\newrgbcolor{userFillColour}{0.8 0.8 0.8}
\psline[fillcolor=userFillColour,fillstyle=solid](-0.02,8.07)
(-0.02,12.9)
(5.12,12.9)
(7.68,8.07)(-0.02,8.07)
\rput(2.9,10.3){$\varphi_*$}
\psline{<-}(1.3,13)(1.3,18.5)
\psline{<-}(1.2,0.9)(1.2,8)
\psline{->}(17.7,1.1)(17.7,8.2)
\psline{->}(17.8,12.9)(17.8,18.4)
\end{pspicture}

\end{center}
\end{itemize}
The isomorphism $\XC\cong \CX$ induces the isomorphism $\Wee\XC\cong \Wee\CX$ and the morphism $f^X \in \Wee\CX(A,B)$ corresponding to the above morphism $f \in \Wee\XC(A,B)$ is in the form
\bear
f  &= & \pi_{\ast_X}\otimes^X \pi\\
& = & A^*A\tto{\pi_*\otimes\pi} B^*XXB \tto{B^*\nabla B}B^*B
\eear
where $\pi = \varphi^X$, that is graphically:
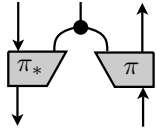
\begin{center}
\ifx\JPicScale\undefined\def\JPicScale{1}\fi
\psset{unit=\JPicScale mm}
\psset{linewidth=0.3,dotsep=1,hatchwidth=0.3,hatchsep=1.5,shadowsize=1,dimen=middle}
\psset{dotsize=0.7 2.5,dotscale=1 1,fillcolor=black}
\psset{arrowsize=1 2,arrowlength=1,arrowinset=0.25,tbarsize=0.7 5,bracketlength=0.15,rbracketlength=0.15}
\begin{pspicture}(0,0)(19.3,19.3)
\psline{<-}(1.2,2.8)(1.2,8.05)
\psline{->}(17.9,2.7)(17.9,7.95)
\rput{0}(9.6,15.9){\psellipse[linestyle=none,fillstyle=solid](0,0)(1,-0.98)}
\psbezier(10.24,15.88)(13.14,15.25)(13.14,14.01)(13.14,12.76)
\psbezier(8.86,15.88)(6.1,15.25)(6.1,14.01)(6.1,12.76)
\psline(9.6,19.29)(9.6,16.12)
\newrgbcolor{userFillColour}{0.8 0.8 0.8}
\psline[fillcolor=userFillColour,fillstyle=solid](19.3,12.57)
(19.3,8)
(14.14,8)
(11.57,12.57)(19.3,12.57)
\rput(16.37,10.43){$\pi$}
\newrgbcolor{userFillColour}{0.8 0.8 0.8}
\psline[fillcolor=userFillColour,fillstyle=solid](-0.02,12.69)
(-0.02,8.12)
(5.12,8.12)
(7.68,12.69)(-0.02,12.69)
\rput(2.9,10.58){$\pi_*$}
\psline{<-}(1.3,12.72)(1.3,19.3)
\psline{->}(17.8,12.6)(17.8,19.18)
\end{pspicture}

\end{center}

\paragraph{Remark.} The above pictures display the important fact that the tensors $\otimes_X$ of $\CCc_X$ and $\otimes^X$ effectively {\em correlate\/} the components over the classical interface $X$, in contrast with the tensor $\otimes$ of $\CCc$, which leaves them separate,
as in
\begin{center}
\ifx\JPicScale\undefined\def\JPicScale{1}\fi
\psset{unit=\JPicScale mm}
\psset{linewidth=0.3,dotsep=1,hatchwidth=0.3,hatchsep=1.5,shadowsize=1,dimen=middle}
\psset{dotsize=0.7 2.5,dotscale=1 1,fillcolor=black}
\psset{arrowsize=1 2,arrowlength=1,arrowinset=0.25,tbarsize=0.7 5,bracketlength=0.15,rbracketlength=0.15}
\begin{pspicture}(0,0)(16.82,12.09)
\psline{<-}(3.1,8.2)(3.09,11.99)
\newrgbcolor{userFillColour}{0.8 0.8 0.8}
\pspolygon[fillcolor=userFillColour,fillstyle=solid](0.2,8.23)(5.82,8.23)(5.82,3.85)(0.2,3.85)
\rput(2.9,6.2){$\nu_*$}
\psline{<-}(3,-0.1)(2.99,3.69)
\psline{->}(14.1,8.3)(14.09,12.09)
\newrgbcolor{userFillColour}{0.8 0.8 0.8}
\pspolygon[fillcolor=userFillColour,fillstyle=solid](11.2,8.33)(16.82,8.33)(16.82,3.95)(11.2,3.95)
\rput(13.9,6.3){$\nu$}
\psline{->}(14,0)(13.99,3.79)
\end{pspicture}

\end{center}

\begin{defn}{\rm\cite{Coecke-Pavlovic}} \label{def:PureMeasurement} \em
A \em pure quantum measurement on $A$ of quantity $X$ \em is a self-adjoint  morphism $m= \pi_{\ast^X}\otimes^X\pi\in\Wee\CX(A)$ with 
\bit
\item $\pi$ an Eilenberg-Moore coalgebra for comonad $X\otimes-: \CCc\to \CCc$.
\eit
We refer to this last condition as $m$ being \em spectral\em.
Explicitly we have 
\bear
XA\tto{\pi^\dagger} A &=& XA\tto{X\pi} XXA \tto{\varepsilon A}A
\eear
\bear
A\tto{\pi} XA \tto{\Delta A}XXA&=&A\tto{\pi} XA \tto{X\pi}XXA
\eear
\bear
A\tto{\pi} XA \tto{\top A}A&=&\id_A
\eear
that is,  graphically:
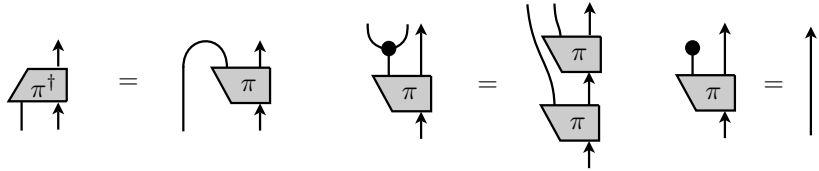
\begin{center}
\ifx\JPicScale\undefined\def\JPicScale{1}\fi
\psset{unit=\JPicScale mm}
\psset{linewidth=0.3,dotsep=1,hatchwidth=0.3,hatchsep=1.5,shadowsize=1,dimen=middle}
\psset{dotsize=0.7 2.5,dotscale=1 1,fillcolor=black}
\psset{arrowsize=1 2,arrowlength=1,arrowinset=0.25,tbarsize=0.7 5,bracketlength=0.15,rbracketlength=0.15}
\begin{pspicture}(0,0)(107.93,22.1)
\psline(3,9.7)(3,5)
\newrgbcolor{userFillColour}{0.8 0.8 0.8}
\psline[fillcolor=userFillColour,fillstyle=solid](36.03,13.37)
(36.03,8.67)
(30.87,8.67)
(28.3,13.37)(36.03,13.37)
\rput(33.2,11){$\pi$}
\psline{<-}(34.7,17.02)(34.7,13.4)
\psline{<-}(34.7,8.62)(34.7,5)
\psbezier(24.5,13.4)(24.5,17.9)(30.3,18.2)(30.3,13.4)
\psline(24.5,13.6)(24.5,4.9)
\newrgbcolor{userFillColour}{0.8 0.8 0.8}
\psline[fillcolor=userFillColour,fillstyle=solid](9,8.9)
(9,13.2)
(3.84,13.2)
(1.27,8.9)(9,8.9)
\rput(6,11){$\pi^\dagger$}
\psline{<-}(7.87,17.15)(7.87,13.53)
\psline{<-}(7.87,8.75)(7.87,5.13)
\rput(17.2,11.3){=}
\newrgbcolor{userFillColour}{0.8 0.8 0.8}
\psline[fillcolor=userFillColour,fillstyle=solid](57.44,12.27)
(57.44,7.57)
(52.28,7.57)
(49.71,12.27)(57.44,12.27)
\rput(54.61,9.9){$\pi$}
\psline{<-}(56.11,19.27)(56.11,12.3)
\psline{<-}(56.11,7.52)(56.11,3.9)
\rput{90}(51.81,15.92){\psellipse[linestyle=none,fillstyle=solid](0,0)(1.03,1)}
\psbezier(51.07,15.94)(49.01,16.61)(49.01,17.92)(49.01,19.24)
\psline(51.81,12.34)(51.81,15.69)
\psbezier(51.17,15.54)(54.31,16.27)(54.31,17.7)(54.31,19.14)
\newrgbcolor{userFillColour}{0.8 0.8 0.8}
\psline[fillcolor=userFillColour,fillstyle=solid](79.77,8.37)
(79.77,3.67)
(74.61,3.67)
(72.04,8.37)(79.77,8.37)
\rput(76.94,6){$\pi$}
\psline{<-}(78.44,12.8)(78.44,8.4)
\psline{<-}(78.44,3.62)(78.44,0)
\newrgbcolor{userFillColour}{0.8 0.8 0.8}
\psline[fillcolor=userFillColour,fillstyle=solid](80,17.5)
(80,12.8)
(74.84,12.8)
(72.27,17.5)(80,17.5)
\rput(77.17,15.13){$\pi$}
\psbezier(73.84,8.4)(73.74,13.72)(70.84,14.84)(70.24,21.8)
\psline{<-}(78.54,22.1)(78.54,17.7)
\rput(65,11){=}
\newrgbcolor{userFillColour}{0.8 0.8 0.8}
\psline[fillcolor=userFillColour,fillstyle=solid](97.83,12.37)
(97.83,7.67)
(92.67,7.67)
(90.1,12.37)(97.83,12.37)
\rput(95,10){$\pi$}
\psline{<-}(96.5,19.37)(96.5,12.4)
\psline{<-}(96.5,7.62)(96.5,4)
\rput{90}(92.2,16.03){\psellipse[linestyle=none,fillstyle=solid](0,0)(1.03,1)}
\psline(92.2,12.44)(92.2,15.79)
\psline{<-}(107.93,18.84)(107.93,4.37)
\rput(103,11){=}
\psbezier(74.64,17.6)(74.54,19.1)(73.94,19)(73.84,21.9)
\end{pspicture}

\end{center}
\end{defn}
A \em pure quantum measurement on $A$ of quantity $X$ \em is a morphism $e\in\Wee\CX(A,I)$ which is such that 
$e^\dagger\circ e$ is a self-adjoint and spectral.

\paragraph{Remark.} 
The 3rd equation can also be rewritten as 
\beq\label{eq:spectra1}
\sum_X \pi_X\,=\,\id\,.
\eeq
When syntactically distinguishing between identical index types as $X$ and $X'$ and setting $\delta_{X,X'}$ for $\nabla\otimes A$ the 2nd one can be rewritten as 
\beq\label{eq:spectra2}
\pi_X \pi_{X'}\,=\, \pi_X \delta_{X,X'}
\eeq
and the 1st can be rewritten as 
\beq\label{eq:spectra3}
\pi_X^{\adj_X}\,=\,\pi_X\,,
\eeq
so we recover an analogue of the properties of a projector spectrum. 
\paragraph{Example.} It was shown in \cite{Coecke-Pavlovic} that definition \ref{def:PureMeasurement}, instantiated to the category  ${\bf FdHilb}$, captures the usual pure measurements, i.e.~spectra of orthogonal projectors indexed by basis vectors of the classical structure $X$, which now represent the outcomes of the measurement. The argument showing this is based on eqs.(\ref{eq:spectra1}, \ref{eq:spectra2}, \ref{eq:spectra3}). 

\paragraph{Example.}  Each classical structure induces a canonical measurement, namely, the one obtained by taking $\pi$ to be $\Delta_X$.  This is a measurement against the `basis' specified by the classical structure $X$. 

\bigskip
The following definition captures the idea that certain operations may depend on values of some classical data type $X$, for example, the outcomes of previously performed measurements.  This is for example essential in teleportation-like protocols and measurement-based quantum computational schemes.  

\begin{defn}\em
An \em operation controlled by $X$ \em is a morphism in $\Wee\CCc_X$.  
\end{defn}

In particular do we have:
\bit
\item  \em pure quantum states controlled by $X$ \em are $\psi\in \Wee \CCc_X(A)$\,;
\item \em pure quantum evolution controlled by $X$ \em are unitary $U\in \Wee \CCc_X(A,B)$\,;
\item \em pure quantum measurements on $A$ of quantity $Y$ controlled by quantity $X$ \em are self-adjoint  spectral  $m\in \Wee(\CCc^Y)_X(A,A)$.
\eit
The pure quantum states and operations of defn.~\ref{def:PureStatesOpp} can be represented in $\Wee \CCc_X$ via the canonical functor 
\[
(-)_X: \Wee\CCc\to\Wee\CCc_X
\]
with
\beqa
A&\stackrel{(-)_X}{\mapsto}&A\\
\Bigl(A^*A\tto{f}B^*B\Bigr)&\stackrel{}{\mapsto}& \Bigl(A^*XA\tto{A^*\top A}A^*A \tto{f}B^*B\Bigr)
\eeqa
and the pure quantum measurement of quantity $X$ of defn.~\ref{def:PureMeasurement} can be represented in  $\Wee \CCc_X$ along the isomorphism $(-)_X:\Wee\CX\cong \Wee\XC$.  

With the above definitions, we can now model quantum protocols in $\Wee \CCc_X$, assuming that $X$ is the space of classical data, and interpreting the classical flows in terms of the Kleisli structure.  For example, the teleportation protocol is simply the composite
\bear
\TTT & = & A\tto{A(\eta_A)_X} AA^*A \tto{m_X A}A\tto{U}A
\eear

where $I \tto{(\eta_A)_X}A^*A$ is the pairing in $\CCc_X$,  $U\in \Wee \CCc^X(A,A)$ is the appropriate unitary, and $m\in\Wee \CCc^X(AA^*,I)$ is the transpose (coname) of the conjugate of $U$, obtained by composing in $\CCc^X$,
\bear
m &=& AA^*\tto{AU_*}AA^*\tto{\epsilon}I\,.
\eear
Showing that $\TTT =   \id_A$ boils down to an easy exercise using the compositionality lemmas for compact categories \cite{Abramsky-Coecke}.  In the graphical language of the dagger compact category $\Wee \CCc^X$ this proof is just
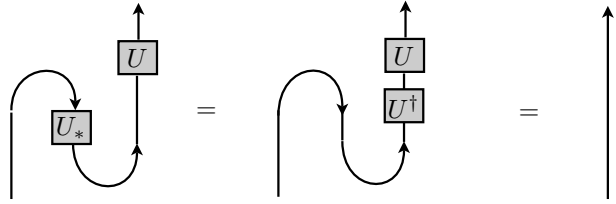
\begin{center}
\ifx\JPicScale\undefined\def\JPicScale{1}\fi
\psset{unit=\JPicScale mm}
\psset{linewidth=0.3,dotsep=1,hatchwidth=0.3,hatchsep=1.5,shadowsize=1,dimen=middle}
\psset{dotsize=0.7 2.5,dotscale=1 1,fillcolor=black}
\psset{arrowsize=1 2,arrowlength=1,arrowinset=0.25,tbarsize=0.7 5,bracketlength=0.15,rbracketlength=0.15}
\begin{pspicture}(0,0)(80.3,27.4)
\psbezier{->}(1,12.8)(1,19.36)(9.5,19.8)(9.5,12.8)
\psline{<-}(80.3,26.7)(80.3,1)
\rput(69.8,12.33){=}
\newrgbcolor{userFillColour}{0.8 0.8 0.8}
\pspolygon[fillcolor=userFillColour,fillstyle=solid](6.44,12.7)(11.56,12.7)(11.56,8.32)(6.44,8.32)
\rput(8.9,10.4){$U_*$}
\psbezier{->}(9.2,8.4)(9.2,1.27)(17.7,0.8)(17.7,8.4)
\newrgbcolor{userFillColour}{0.8 0.8 0.8}
\pspolygon[fillcolor=userFillColour,fillstyle=solid](15.24,22)(20.36,22)(20.36,17.62)(15.24,17.62)
\rput(17.7,19.7){$U$}
\psline(17.7,8.3)(17.7,17.4)
\psline{->}(17.9,22.2)(17.9,27.1)
\psline(1,1)(1,12.5)
\rput(26.9,12.6){=}
\psbezier{->}(36.5,12.1)(36.5,19.6)(45,20.1)(45,12.1)
\psbezier{->}(45,8.7)(45,1.57)(53.2,1.1)(53.2,8.7)
\newrgbcolor{userFillColour}{0.8 0.8 0.8}
\pspolygon[fillcolor=userFillColour,fillstyle=solid](50.74,22.3)(55.86,22.3)(55.86,17.92)(50.74,17.92)
\rput(53.2,20){$U$}
\psline(53.2,8.6)(53.2,17.7)
\psline{->}(53.4,22.5)(53.4,27.4)
\psline(36.5,1.3)(36.5,12.8)
\newrgbcolor{userFillColour}{0.8 0.8 0.8}
\pspolygon[fillcolor=userFillColour,fillstyle=solid](50.64,15.6)(55.76,15.6)(55.76,11.22)(50.64,11.22)
\rput(53.1,13.3){$U^\dagger$}
\psline(45,13.3)(45,8.8)
\end{pspicture}

\end{center}
So the diagrammatic proof of teleportation with classical data flow in $\Wee \CCc^X$  is just the diagrammatic proof of post-selected (or conditional) teleportation in $\Wee \CCc$. In the graphical language of the dagger compact category $\CCc_{\qo}$ this unfolds to the following diagrammatic equation:
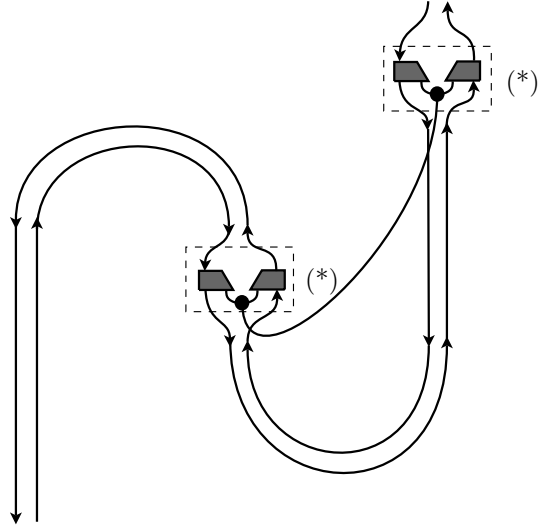
\begin{center}
\ifx\JPicScale\undefined\def\JPicScale{1}\fi
\psset{unit=\JPicScale mm}
\psset{linewidth=0.3,dotsep=1,hatchwidth=0.3,hatchsep=1.5,shadowsize=1,dimen=middle}
\psset{dotsize=0.7 2.5,dotscale=1 1,fillcolor=black}
\psset{arrowsize=1 2,arrowlength=1,arrowinset=0.25,tbarsize=0.7 5,bracketlength=0.15,rbracketlength=0.15}
\begin{pspicture}(0,0)(77.1,71.4)
\psbezier{<-}(9.7,41.1)(9.7,58.18)(40.5,59.83)(40.5,41.6)
\psbezier{->}(12.5,41.1)(12.7,55)(38.02,56.4)(38.02,40.2)
\psbezier{->}(38.2,26.21)(38.2,4)(67,1.8)(67,26.81)
\psbezier{<-}(40.5,26.3)(40.5,7.65)(64.59,5.7)(64.59,25.59)
\rput{90}(65.68,59.08){\psellipse[linestyle=none,fillstyle=solid](0,0)(1.03,0.98)}
\psbezier(66.05,59)(67.77,59.35)(67.77,60.04)(67.77,60.74)
\psbezier(65.24,59)(63.61,59.35)(63.61,60.04)(63.61,60.74)
\newrgbcolor{userFillColour}{0.4 0.4 0.4}
\psline[fillcolor=userFillColour,fillstyle=solid](71.4,60.85)
(71.4,63.39)
(68.36,63.39)
(66.84,60.85)(71.4,60.85)
\newrgbcolor{userFillColour}{0.4 0.4 0.4}
\psline[fillcolor=userFillColour,fillstyle=solid](60,60.78)
(60,63.32)
(63.03,63.32)
(64.54,60.78)(60,60.78)
\psline{<-}(64.6,25.4)(64.5,54.4)
\psline{->}(67,26.5)(67,55.3)
\psbezier(39.8,31.3)(39.8,17.18)(65.7,39.83)(65.7,58.8)
\psline{<-}(9.7,1.8)(9.7,41.47)
\psline{->}(12.5,2.15)(12.5,42.7)
\rput{90}(39.72,31.29){\psellipse[linestyle=none,fillstyle=solid](0,0)(1.03,0.98)}
\psbezier(40.1,31.21)(41.81,31.56)(41.81,32.25)(41.81,32.95)
\psbezier(39.28,31.21)(37.66,31.56)(37.66,32.25)(37.66,32.95)
\newrgbcolor{userFillColour}{0.4 0.4 0.4}
\psline[fillcolor=userFillColour,fillstyle=solid](45.44,33.05)
(45.44,35.6)
(42.4,35.6)
(40.88,33.05)(45.44,33.05)
\newrgbcolor{userFillColour}{0.4 0.4 0.4}
\psline[fillcolor=userFillColour,fillstyle=solid](34.04,32.99)
(34.04,35.53)
(37.08,35.53)
(38.59,32.99)(34.04,32.99)
\pspolygon[linewidth=0.15,linestyle=dashed,dash=1 1](58.6,65.4)(72.9,65.4)(72.9,56.8)(58.6,56.8)
\pspolygon[linewidth=0.15,linestyle=dashed,dash=1 1](32.3,38.7)(46.6,38.7)(46.6,30.1)(32.3,30.1)
\rput(77.1,60.5){(*)}
\rput(50.4,34.1){(*)}
\psbezier{->}(38.07,41.3)(38,38.49)(35,39.1)(34.8,35.5)
\psbezier{<-}(40.5,41.8)(40.5,37.3)(44.5,39.6)(44.3,35.5)
\psbezier{<-}(38.17,25.4)(38.1,29.13)(35.1,28.32)(34.9,33.1)
\psbezier{->}(40.5,26.1)(40.5,30)(44.5,28.8)(44.4,33.1)
\psbezier{<-}(64.42,54.2)(64.35,57.35)(60.93,56.66)(60.7,60.7)
\psbezier{->}(66.95,54.55)(66.95,59.09)(70.7,56.77)(70.51,60.9)
\psbezier{->}(64.51,71.4)(64.43,67.68)(60.93,68.16)(60.7,63.4)
\psbezier{<-}(67.1,71.4)(67.1,66.05)(70.79,68.37)(70.6,63.5)
\end{pspicture}

\end{center}
\begin{center}
\ifx\JPicScale\undefined\def\JPicScale{1}\fi
\psset{unit=\JPicScale mm}
\psset{linewidth=0.3,dotsep=1,hatchwidth=0.3,hatchsep=1.5,shadowsize=1,dimen=middle}
\psset{dotsize=0.7 2.5,dotscale=1 1,fillcolor=black}
\psset{arrowsize=1 2,arrowlength=1,arrowinset=0.25,tbarsize=0.7 5,bracketlength=0.15,rbracketlength=0.15}
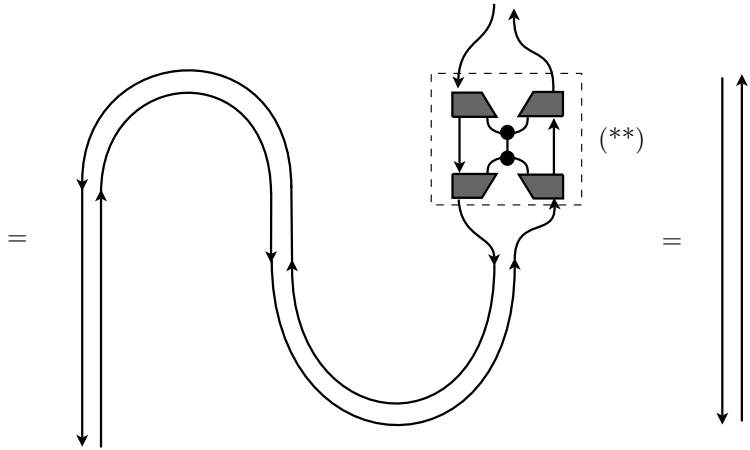
\begin{pspicture}(0,0)(98.1,62.5)
\psbezier{<-}(10.4,37.6)(10.4,57.85)(38.2,59.6)(38.2,38)
\psline{<-}(35.5,28)(35.53,37.38)
\psline(38.3,28.2)(38.2,38.4)
\psbezier(12.9,37.27)(12.9,54.54)(35.5,55.7)(35.5,37.27)
\psbezier{->}(35.6,28.4)(35.6,0.1)(67.9,-1.5)(67.9,28.7)
\psbezier{<-}(38.3,28.5)(38.3,4.4)(65.2,2.1)(65.2,27.8)
\psbezier(67.42,45.21)(69.62,45.66)(69.62,46.53)(69.62,47.41)
\psbezier(66.37,45.21)(64.27,45.66)(64.27,46.53)(64.27,47.41)
\psline(66.93,42.81)(66.93,45.05)
\newrgbcolor{userFillColour}{0.4 0.4 0.4}
\psline[fillcolor=userFillColour,fillstyle=solid](74.31,47.55)
(74.31,50.77)
(70.38,50.77)
(68.43,47.55)(74.31,47.55)
\newrgbcolor{userFillColour}{0.4 0.4 0.4}
\psline[fillcolor=userFillColour,fillstyle=solid](59.61,47.46)
(59.61,50.68)
(63.52,50.68)
(65.47,47.46)(59.61,47.46)
\psline{<-}(10.4,3.57)(10.4,37.9)
\psline{->}(12.9,3.47)(12.9,37.8)
\psbezier(67.44,42.07)(69.64,41.63)(69.64,40.77)(69.64,39.91)
\psbezier(66.39,42.07)(64.29,41.63)(64.29,40.77)(64.29,39.91)
\newrgbcolor{userFillColour}{0.4 0.4 0.4}
\psline[fillcolor=userFillColour,fillstyle=solid](74.33,39.77)
(74.33,36.6)
(70.4,36.6)
(68.45,39.77)(74.33,39.77)
\newrgbcolor{userFillColour}{0.4 0.4 0.4}
\psline[fillcolor=userFillColour,fillstyle=solid](59.63,39.86)
(59.63,36.69)
(63.54,36.69)
(65.49,39.86)(59.63,39.86)
\psline{->}(60.56,47.51)(60.56,39.99)
\psline{<-}(73.11,47.3)(73.11,39.77)
\rput(1.8,31){=}
\rput(88.8,30.7){=}
\psline{<-}(95.5,6.5)(95.5,52.7)
\psline{->}(98.1,6.94)(98.1,53.14)
\rput{90}(66.93,41.98){\psellipse[linestyle=none,fillstyle=solid](0,0)(1.02,0.98)}
\rput{90}(66.93,45.38){\psellipse[linestyle=none,fillstyle=solid](0,0)(1.02,0.98)}
\pspolygon[linewidth=0.15,linestyle=dashed,dash=1 1](56.8,53.24)(76.8,53.24)(76.8,35.74)(56.8,35.74)
\rput(82.1,44.34){(**)}
\psbezier{<-}(65.2,27.6)(65.1,31.7)(60.75,31.2)(60.45,36.45)
\psbezier{->}(67.9,28.46)(67.9,34.35)(73.4,31.34)(73.13,36.71)
\psbezier{->}(65.17,62.5)(65.07,57.03)(60.65,58.23)(60.35,51.22)
\psbezier{<-}(67.8,61.89)(67.8,54.02)(73.3,58.05)(73.03,50.87)
\end{pspicture}

\end{center}
where the annotated elements are:
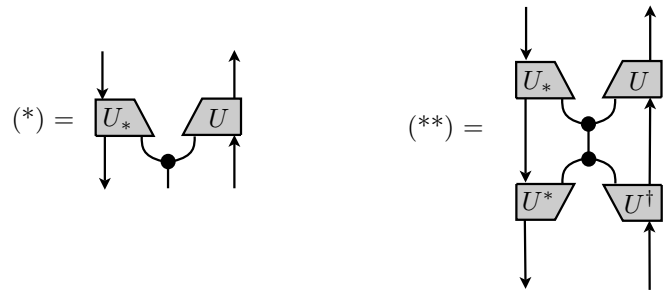
\begin{center}
\ifx\JPicScale\undefined\def\JPicScale{1}\fi
\psset{unit=\JPicScale mm}
\psset{linewidth=0.3,dotsep=1,hatchwidth=0.3,hatchsep=1.5,shadowsize=1,dimen=middle}
\psset{dotsize=0.7 2.5,dotscale=1 1,fillcolor=black}
\psset{arrowsize=1 2,arrowlength=1,arrowinset=0.25,tbarsize=0.7 5,bracketlength=0.15,rbracketlength=0.15}
\begin{pspicture}(0,0)(94.83,39.23)
\rput{90}(85.1,23.48){\psellipse[linestyle=none,fillstyle=solid](0,0)(1.03,1)}
\psbezier(85.74,23.5)(88.64,24.17)(88.64,25.48)(88.64,26.8)
\psbezier(84.36,23.5)(81.6,24.17)(81.6,25.48)(81.6,26.8)
\psline(85.1,19.1)(85.1,23.25)
\newrgbcolor{userFillColour}{0.8 0.8 0.8}
\psline[fillcolor=userFillColour,fillstyle=solid](94.8,27)
(94.8,31.83)
(89.64,31.83)
(87.07,27)(94.8,27)
\rput(91.87,29.23){$U$}
\newrgbcolor{userFillColour}{0.8 0.8 0.8}
\psline[fillcolor=userFillColour,fillstyle=solid](75.48,26.87)
(75.48,31.7)
(80.62,31.7)
(83.18,26.87)(75.48,26.87)
\rput(78.4,29.1){$U_*$}
\psline{<-}(76.73,1.51)(76.63,11.07)
\psline{->}(93.23,1.38)(93.2,10.6)
\psline{<-}(76.8,31.93)(76.8,39.03)
\psline{->}(93.3,31.9)(93.3,39.23)
\rput{90}(85.13,18.81){\psellipse[linestyle=none,fillstyle=solid](0,0)(1.02,-1)}
\psbezier(85.77,18.79)(88.67,18.13)(88.67,16.84)(88.67,15.54)
\psbezier(84.39,18.79)(81.63,18.13)(81.63,16.84)(81.63,15.54)
\newrgbcolor{userFillColour}{0.8 0.8 0.8}
\psline[fillcolor=userFillColour,fillstyle=solid](94.83,15.34)
(94.83,10.58)
(89.67,10.58)
(87.1,15.34)(94.83,15.34)
\rput(91.9,13.14){$U^\dagger$}
\newrgbcolor{userFillColour}{0.8 0.8 0.8}
\psline[fillcolor=userFillColour,fillstyle=solid](75.51,15.47)
(75.51,10.71)
(80.65,10.71)
(83.21,15.47)(75.51,15.47)
\rput(78.43,13.27){$U^*$}
\psline{->}(76.73,26.94)(76.73,15.66)
\psline{<-}(93.3,27)(93.23,15.34)
\rput(66.1,22.84){(**) =}
\rput{90}(29.2,18.48){\psellipse[linestyle=none,fillstyle=solid](0,0)(1.03,1)}
\psbezier(29.84,18.5)(32.74,19.17)(32.74,20.48)(32.74,21.8)
\psbezier(28.46,18.5)(25.7,19.17)(25.7,20.48)(25.7,21.8)
\psline(29.2,14.9)(29.2,18.25)
\newrgbcolor{userFillColour}{0.8 0.8 0.8}
\psline[fillcolor=userFillColour,fillstyle=solid](38.9,22)
(38.9,26.83)
(33.74,26.83)
(31.17,22)(38.9,22)
\rput(35.97,24.23){$U$}
\newrgbcolor{userFillColour}{0.8 0.8 0.8}
\psline[fillcolor=userFillColour,fillstyle=solid](19.58,21.87)
(19.58,26.7)
(24.72,26.7)
(27.28,21.87)(19.58,21.87)
\rput(22.5,24.1){$U_*$}
\psline{<-}(20.8,14.7)(20.8,21.8)
\psline{->}(38.02,14.9)(38.02,22)
\psline{<-}(20.48,26.65)(20.48,33.12)
\psline{->}(38.02,26.84)(38.02,33.3)
\rput(12.57,24.03){(*) =}
\end{pspicture}

\end{center}
The distinction of `quantum vs.~classical' is thus displayed as the distinction of `two wires vs.~one wire'.  Compared with the concise graphical representation of the same process in $\Wee \CCc^X$, this unfolded picture displays the details of the interaction between the classical and quantum data flow.  In particular, the equality 
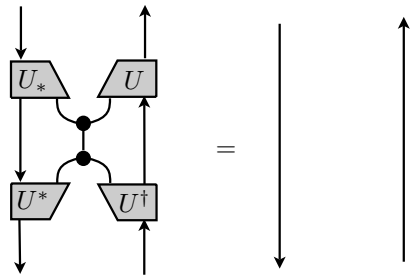
\begin{center}
\ifx\JPicScale\undefined\def\JPicScale{1}\fi
\psset{unit=\JPicScale mm}
\psset{linewidth=0.3,dotsep=1,hatchwidth=0.3,hatchsep=1.5,shadowsize=1,dimen=middle}
\psset{dotsize=0.7 2.5,dotscale=1 1,fillcolor=black}
\psset{arrowsize=1 2,arrowlength=1,arrowinset=0.25,tbarsize=0.7 5,bracketlength=0.15,rbracketlength=0.15}
\begin{pspicture}(0,0)(54.7,36.59)
\rput{90}(12.1,20.84){\psellipse[linestyle=none,fillstyle=solid](0,0)(1.03,1)}
\psbezier(12.74,20.86)(15.64,21.53)(15.64,22.84)(15.64,24.16)
\psbezier(11.36,20.86)(8.6,21.53)(8.6,22.84)(8.6,24.16)
\psline(12.1,17.26)(12.1,20.61)
\newrgbcolor{userFillColour}{0.8 0.8 0.8}
\psline[fillcolor=userFillColour,fillstyle=solid](21.8,24.36)
(21.8,29.19)
(16.64,29.19)
(14.07,24.36)(21.8,24.36)
\rput(18.87,26.59){$U$}
\newrgbcolor{userFillColour}{0.8 0.8 0.8}
\psline[fillcolor=userFillColour,fillstyle=solid](2.48,24.23)
(2.48,29.06)
(7.62,29.06)
(10.18,24.23)(2.48,24.23)
\rput(5.4,26.46){$U_*$}
\psline{<-}(3.7,0.79)(3.6,8.08)
\psline{->}(20.2,0.69)(20.2,8.1)
\psline{<-}(3.8,29.29)(3.8,36.39)
\psline{->}(20.3,29.49)(20.3,36.59)
\rput{90}(12.13,16.16){\psellipse[linestyle=none,fillstyle=solid](0,0)(1.02,-1)}
\psbezier(12.77,16.15)(15.67,15.49)(15.67,14.2)(15.67,12.9)
\psbezier(11.39,16.15)(8.63,15.49)(8.63,14.2)(8.63,12.9)
\newrgbcolor{userFillColour}{0.8 0.8 0.8}
\psline[fillcolor=userFillColour,fillstyle=solid](21.83,12.7)
(21.83,7.94)
(16.67,7.94)
(14.1,12.7)(21.83,12.7)
\rput(18.9,10.5){$U^\dagger$}
\newrgbcolor{userFillColour}{0.8 0.8 0.8}
\psline[fillcolor=userFillColour,fillstyle=solid](2.51,12.83)
(2.51,8.07)
(7.65,8.07)
(10.21,12.83)(2.51,12.83)
\rput(5.43,10.63){$U^*$}
\psline{->}(3.73,24.3)(3.73,13.02)
\psline{<-}(20.2,24.5)(20.23,12.7)
\rput(31.1,17.1){=}
\psline{<-}(38.2,1.5)(38.2,34.08)
\psline{->}(54.7,2.42)(54.7,35)
\end{pspicture}

\end{center}
is a consequence of unitarity of $U$ in $\Wee \CCc^X$, because the unitarity equation $U\circ_X U^{\adj_X}=\id$, unfolded in the graphical language of $\CCc$, is just
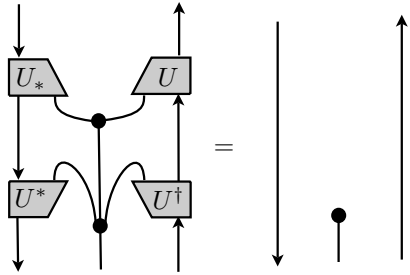
\begin{center}
\ifx\JPicScale\undefined\def\JPicScale{1}\fi
\psset{unit=\JPicScale mm}
\psset{linewidth=0.3,dotsep=1,hatchwidth=0.3,hatchsep=1.5,shadowsize=1,dimen=middle}
\psset{dotsize=0.7 2.5,dotscale=1 1,fillcolor=black}
\psset{arrowsize=1 2,arrowlength=1,arrowinset=0.25,tbarsize=0.7 5,bracketlength=0.15,rbracketlength=0.15}
\begin{pspicture}(0,0)(54.7,36.69)
\rput{90}(14.4,20.84){\psellipse[linestyle=none,fillstyle=solid](0,0)(1.03,1)}
\psbezier(15.1,20.96)(20.44,21.63)(20.44,22.94)(20.44,24.26)
\psbezier(13.5,20.86)(8.6,21.53)(8.6,22.84)(8.6,24.16)
\psline(14.7,1.1)(14.4,20.6)
\newrgbcolor{userFillColour}{0.8 0.8 0.8}
\psline[fillcolor=userFillColour,fillstyle=solid](26.6,24.46)
(26.6,29.29)
(21.44,29.29)
(18.87,24.46)(26.6,24.46)
\rput(23.67,26.69){$U$}
\newrgbcolor{userFillColour}{0.8 0.8 0.8}
\psline[fillcolor=userFillColour,fillstyle=solid](2.48,24.23)
(2.48,29.06)
(7.62,29.06)
(10.18,24.23)(2.48,24.23)
\rput(5.4,26.46){$U_*$}
\psline{<-}(3.7,0.79)(3.6,8.08)
\psline{->}(25,0.79)(25,8.2)
\psline{<-}(3.8,29.29)(3.8,36.39)
\psline{->}(25.1,29.59)(25.1,36.69)
\rput{90}(14.6,6.88){\psellipse[linestyle=none,fillstyle=solid](0,0)(1.02,-1)}
\psbezier(15.3,7.3)(15.7,13.6)(20.2,18.3)(20.47,13)
\psbezier(14,7.4)(13.5,14.4)(8.5,18)(8.5,13)
\newrgbcolor{userFillColour}{0.8 0.8 0.8}
\psline[fillcolor=userFillColour,fillstyle=solid](26.63,12.8)
(26.63,8.04)
(21.47,8.04)
(18.9,12.8)(26.63,12.8)
\rput(23.7,10.6){$U^\dagger$}
\newrgbcolor{userFillColour}{0.8 0.8 0.8}
\psline[fillcolor=userFillColour,fillstyle=solid](2.51,12.83)
(2.51,8.07)
(7.65,8.07)
(10.21,12.83)(2.51,12.83)
\rput(5.43,10.63){$U^*$}
\psline{->}(3.73,24.3)(3.73,13.02)
\psline{<-}(25,24.5)(25.03,12.8)
\rput(31.1,17.1){=}
\psline{<-}(38.2,1.5)(38.2,34.08)
\psline{->}(54.7,2.42)(54.7,35)
\rput{90}(46.3,8.3){\psellipse[linestyle=none,fillstyle=solid](0,0)(1.02,-1)}
\psline(46.3,8.2)(46.3,2.1)
\end{pspicture}

\end{center}

\subsection{Indexed mixed quantum states and operations}

The $\Cee$-construction of definition \ref{cpos}  can be viewed as the hull of the $\Wee$-construction in the sense that the canonical functor $\CCc\to \Cee\CCc$ factors
\[
\CCc \epi \Wee \CCc\inclusion\Cee\CCc\,. 
\]

\paragraph{Remark.}  Mixed states arise from pure states by mixing.  If $f\in \Wee \CCc^X$ then one can  show that $\omega_p(f)\in \Cee \CCc$ by relying on prop.~\ref{posIfCpos}. Conversely, one can also show that all mixed states arise in this manner. Hence, $\Cee \CCc(A)$ is the `abstract' convex closure of $\Wee \CCc(A)$. 

\medskip
Repeating the above but now for the $\Cee$-construction will allow us to extend our dictionary of quantum operations from pure to mixed ones. 

The category $\Cee \XC$ consists of
\begin{itemize}
\item the same objects as $\CCc_{\qo}$\,;
\item morphism $f\in \Cee\XC(A,B)$ are of the form 
\bear
f &= & (B^* \otimes_X(\varepsilon_{V^*})_X \otimes_X B)\circ_X(\varphi_{\ast_X}\otimes_X \varphi)\\
& = & A^*XA\tto{A\Delta_X A^*} A^*XXA \tto{\varphi_*\otimes \varphi}B^*V^*VB \tto{B^*\varepsilon_{V^*} B} B^*B
\eear
for some $\varphi \in \left(XA, VB\right)$, that is graphically:
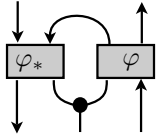
\begin{center}
\ifx\JPicScale\undefined\def\JPicScale{1}\fi
\psset{unit=\JPicScale mm}
\psset{linewidth=0.3,dotsep=1,hatchwidth=0.3,hatchsep=1.5,shadowsize=1,dimen=middle}
\psset{dotsize=0.7 2.5,dotscale=1 1,fillcolor=black}
\psset{arrowsize=1 2,arrowlength=1,arrowinset=0.25,tbarsize=0.7 5,bracketlength=0.15,rbracketlength=0.15}
\begin{pspicture}(0,0)(19.42,18.5)
\newrgbcolor{userFillColour}{0.8 0.8 0.8}
\pspolygon[fillcolor=userFillColour,fillstyle=solid](11.9,12.68)(19.42,12.68)(19.42,8.3)(11.9,8.3)
\newrgbcolor{userFillColour}{0.8 0.8 0.8}
\pspolygon[fillcolor=userFillColour,fillstyle=solid](-0.12,12.69)(7.4,12.69)(7.4,8.31)(-0.12,8.31)
\rput{90}(9.6,4.68){\psellipse[linestyle=none,fillstyle=solid](0,0)(1.04,1)}
\psbezier(10.24,4.7)(13.14,5.37)(13.14,6.68)(13.14,8)
\psbezier(8.86,4.7)(6.1,5.37)(6.1,6.68)(6.1,8)
\psline(9.6,1.1)(9.6,4.45)
\rput(16.37,10.43){$\varphi$}
\rput(2.9,10.3){$\varphi_*$}
\psline{<-}(1.3,13)(1.3,18.5)
\psline{<-}(1.2,0.9)(1.2,8)
\psline{->}(17.7,1.1)(17.7,8.2)
\psline{->}(17.8,12.9)(17.8,18.4)
\psbezier{->}(13.5,12.8)(13.5,17.6)(5.6,17.6)(5.6,12.8)
\end{pspicture}

\end{center}
\end{itemize}
The isomorphism $\XC\cong \CX$ induces the isomorphism $\Cee\XC\cong \Cee\CX$,  and the morphism $f^X \in \Cee\CX(A,B)$ depict graphically as:
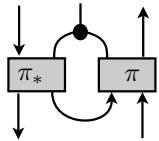
\begin{center}
\ifx\JPicScale\undefined\def\JPicScale{1}\fi
\psset{unit=\JPicScale mm}
\psset{linewidth=0.3,dotsep=1,hatchwidth=0.3,hatchsep=1.5,shadowsize=1,dimen=middle}
\psset{dotsize=0.7 2.5,dotscale=1 1,fillcolor=black}
\psset{arrowsize=1 2,arrowlength=1,arrowinset=0.25,tbarsize=0.7 5,bracketlength=0.15,rbracketlength=0.15}
\begin{pspicture}(0,0)(19.75,18.67)
\newrgbcolor{userFillColour}{0.8 0.8 0.8}
\pspolygon[fillcolor=userFillColour,fillstyle=solid](12.23,11.35)(19.75,11.35)(19.75,6.97)(12.23,6.97)
\newrgbcolor{userFillColour}{0.8 0.8 0.8}
\pspolygon[fillcolor=userFillColour,fillstyle=solid](0.21,11.36)(7.73,11.36)(7.73,6.98)(0.21,6.98)
\rput{90}(9.79,14.88){\psellipse[linestyle=none,fillstyle=solid](0,0)(1.1,-1)}
\psbezier(10.43,14.86)(13.33,14.15)(13.33,12.77)(13.33,11.37)
\psbezier(9.05,14.86)(6.29,14.15)(6.29,12.77)(6.29,11.37)
\psline(9.79,18.67)(9.79,15.13)
\rput(16.7,9.1){$\pi$}
\rput(3.23,8.97){$\pi_*$}
\psline{<-}(1.63,11.69)(1.63,18.37)
\psline{<-}(1.53,0.57)(1.53,6.7)
\psline{->}(18.03,0.74)(18.03,6.87)
\psline{->}(18.13,11.57)(18.13,18.25)
\psbezier{<-}(13.93,6.87)(13.93,2.04)(6.03,2.04)(6.03,6.87)
\end{pspicture}

\end{center}
for some  $\pi\in  \left(VA, XB\right)$.  

\begin{defn}\em
A \em mixed quantum state controlled by quantity $X$ \em is an element of $\Cee\CX(A)$ and a \em mixed quantum evolution controlled by quantity $X$ \em is a morphism in $\Cee\CX(A, B)$.
\end{defn}

For \em generalized measurements \em or \em PMVMs \em we refer the reader to \cite{Coecke-Paquette}.  The reader can easily come up with many variations on the same theme.  

Similarly as in the case of pure operations, protocols involving mixed operations can now be represented in $\Cee\CX$.

\subsection{Conjoining classical interfaces} \label{Conjoining}

In section \ref{Positivity} we introduced classical operations and in section \ref{Relativizing} we adjoined a classical interface to a quantum universe.  Each classical interface lived in a different category.  The task in this section is to conjoin all these classical interfaces of the quantum universe within one category, and to apply the  classical operations between the classical interfaces.  We will then be able to feed processed data obtained at one interface (= observable) as control data into another interface. 

A \em $\CCc$-indexed category \em is a contravariant functor from $\CCc$ into ${\sf CAT}$ \cite{Johnstone}. The collection of classical interfaces together with the functions between them can be packaged as a $\CclasG$-indexed category as follows:
\[
\WP: \CclasG^{op}  \to  {\sf CAT}
\]
with
\beqa
X & \stackrel{\WP}{\mapsto} & \Cee \XC\\
(X\tto{\varphi} Y)& \mapsto & \Cee \CCc_Y \tto{\fsharp} \Cee \XC
\eeqa
where each $\fsharp$ is an 
identity-on-objects-functor with the arrow part
\beqa
\Bigl(A^*YA\tto{g} B^*B\Bigr) &  \stackrel{\fsharp}{\mapsto} & \Bigl(A^*XA\tto{A^*\varphi A} A^*YA \tto{g} B^*B\Bigr)
\eeqa

In the graphical language of $\CCc$ we have:
\medskip
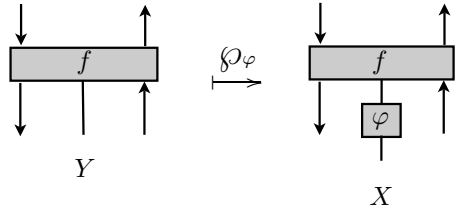
\begin{center}
\ifx\JPicScale\undefined\def\JPicScale{1}\fi
\psset{unit=\JPicScale mm}
\psset{linewidth=0.3,dotsep=1,hatchwidth=0.3,hatchsep=1.5,shadowsize=1,dimen=middle}
\psset{dotsize=0.7 2.5,dotscale=1 1,fillcolor=black}
\psset{arrowsize=1 2,arrowlength=1,arrowinset=0.25,tbarsize=0.7 5,bracketlength=0.15,rbracketlength=0.15}
\begin{pspicture}(0,0)(60,30.26)
\newrgbcolor{userFillColour}{0.8 0.8 0.8}
\pspolygon[fillcolor=userFillColour,fillstyle=solid](1.1,24.28)(20.62,24.28)(20.62,19.9)(1.1,19.9)
\psline(10.8,12.7)(10.7,20)
\psline{<-}(2.5,24.6)(2.5,30.1)
\psline{<-}(2.4,12.5)(2.4,19.6)
\psline{->}(18.9,12.7)(18.9,19.8)
\psline{->}(19,24.5)(19,30)
\rput(31.2,21.6){$\stackrel{\WP \varphi}{\mapsto}$}
\rput(11,8.3){$Y$}
\newrgbcolor{userFillColour}{0.8 0.8 0.8}
\pspolygon[fillcolor=userFillColour,fillstyle=solid](40.87,24.45)(60,24.45)(60,20.07)(40.87,20.07)
\psline(50.43,16.77)(50.43,20.12)
\psline{<-}(42.29,24.76)(42.29,30.26)
\psline{<-}(42.19,12.66)(42.19,19.76)
\psline{->}(58.69,12.86)(58.69,19.96)
\psline{->}(58.79,24.66)(58.79,30.16)
\newrgbcolor{userFillColour}{0.8 0.8 0.8}
\pspolygon[fillcolor=userFillColour,fillstyle=solid](47.84,16.8)(52.96,16.8)(52.96,12.42)(47.84,12.42)
\psline(50.4,9.25)(50.4,12.6)
\rput(50.3,14.5){$\varphi$}
\rput(50.4,4.5){$X$}
\rput(50.2,22.3){$f$}
\rput(10.9,22.3){$f$}
\end{pspicture}

\end{center}

The well-definedness of $\WP$ and the reason for restricting to functions is explained in the following proposition.

\be{prop}\label{prop:Groth}
Let $\varphi:X\to Y$ be a morphism between classical structures. With the above notations, 
 $\fsharp (g):A^*XA\to B^*B$ is in $\Cee\XC$ for every $g\in \Cee\XC$ if and only if $\varphi$ is a classical map, and  $\WP\varphi: \Cee \CCc_Y \to\Cee \XC$ is a functor if and only if $\varphi$ is a function.  
\ee{prop}

The first part of this proposition follows by proposition \ref{posIfCpos}.    For $\WP\varphi$ to preserve composition $\varphi$ needs to preserve $\Delta$, while for $\WP\varphi$ to preserve identities $\varphi$ needs to preserve $\top$, and the converse also holds. We invite the reader to draw the appropriate pictures to verify this.


Given any indexed category ${\cal X}$, its \em total category\em, which is obtained by applying the \em Grothendieck construction\em, pairs morphisms in the indexing category with morphisms 
in the indexed categories. In the  case of $\WP$, it pairs functions on classical data with controlled  completely positive maps on quantum data.  Explicitly, we obtain a category 
\[
\See\CCc=\int\WP
\]
that consists of
\bit 
\item pairs $\langle X, A\rangle$ with $X\in \CclasG$ and $A\in \WP X$ as objects, that is,
\[
|\See \CCc| = |\CCc_{f}|\times |\CCc_{\qo}|\,;
\]
\item pairs 
\[
\Bigl\langle \varphi:X\to Y\ ,\ g:A\to \fsharp(B)\Bigr\rangle
\]
as morphisms of type $\langle X, A\rangle\to \langle Y, B\rangle$,  with $\varphi\in\CclasG(X,Y)$ and $g\in \WP X(A,B)$ --- since $\WP$ is 
an identity-on-objects-functor and hence we have $\fsharp(B)=B$ --- i.e. 
\[
\See\CCc\big(\langle X,A\rangle,\langle Y,B\rangle\big) = \CclasG(X,Y)\times \Cee\XC(A,B)\,;
\]
\item 
$\id_{\langle X, A\rangle}=\bigl\langle X\tto{X}X\,,\, A^*XA\tto{A^*\top_X A}A^*A\bigr\rangle$\,;
\item 
composition
\[
\langle \nu, g\rangle\circ_{\See\CCc}\langle \varphi, f\rangle=\bigl\langle \nu\circ \varphi\,,\, \fsharp(g)\circ_X f\bigr\rangle
\]
for $\langle \varphi, f\rangle:\langle X, A\rangle\to\langle Y, B\rangle$ and 
$\langle \nu, g\rangle:\langle Y, B\rangle\to\langle Z, C\rangle$.
\eit

In the graphical calculus of $\CCc$ the morphism $\fsharp(g)\circ_X f$ is:
\begin{center}
\ifx\JPicScale\undefined\def\JPicScale{1}\fi
\psset{unit=\JPicScale mm}
\psset{linewidth=0.3,dotsep=1,hatchwidth=0.3,hatchsep=1.5,shadowsize=1,dimen=middle}
\psset{dotsize=0.7 2.5,dotscale=1 1,fillcolor=black}
\psset{arrowsize=1 2,arrowlength=1,arrowinset=0.25,tbarsize=0.7 5,bracketlength=0.15,rbracketlength=0.15}
\begin{pspicture}(0,0)(24.82,37.59)
\newrgbcolor{userFillColour}{0.8 0.8 0.8}
\pspolygon[fillcolor=userFillColour,fillstyle=solid](8.64,31.78)(24.56,31.78)(24.56,27.4)(8.64,27.4)
\rput{90}(16.78,7.78){\psellipse[linestyle=none,fillstyle=solid](0,0)(1.03,1)}
\psbezier(16.04,7.8)(18.26,8.47)(18.26,9.78)(18.26,11.1)
\psline(16.78,4.2)(16.78,7.55)
\rput(16.56,29.39){$g$}
\psline{<-}(10.06,32.09)(10.06,37.59)
\psline{<-}(10.12,15.28)(10.12,27.06)
\psline{->}(23.32,15.61)(23.32,27.39)
\psline{->}(23.26,31.99)(23.26,37.49)
\newrgbcolor{userFillColour}{0.8 0.8 0.8}
\pspolygon[fillcolor=userFillColour,fillstyle=solid](8.9,15.38)(24.82,15.38)(24.82,11)(8.9,11)
\rput(16.82,12.99){$f$}
\newrgbcolor{userFillColour}{0.8 0.8 0.8}
\pspolygon[fillcolor=userFillColour,fillstyle=solid](1.7,15.38)(6.82,15.38)(6.82,11)(1.7,11)
\rput(4.16,13.08){$\varphi$}
\psbezier(4.76,15.48)(4.86,19.88)(16.56,18.48)(16.76,27.38)
\psline{<-}(10.16,4.51)(10.16,11.09)
\psline{->}(23.36,4.39)(23.36,10.97)
\rput(17.1,20.4){$Y$}
\rput(16.96,0.68){$X$}
\psbezier(16.14,7.7)(12.66,7.68)(4.66,8.28)(4.66,10.98)
\end{pspicture}

\end{center}

\be{prop}
$\See\CCc$ is a symmetric monoidal category. 
\ee{prop}

\paragraph{Remark.}
Since $\CCc_f$ is not a dagger category neither is $\See\CCc$.

\bigskip
From the composition law in $\See\CCc$ it easily follows that we need to interpret a pair of morphisms $\langle \varphi:X\to Y, g \rangle$ as follows:
\bit
\item The components are a classical operation $\varphi$ and a $X$-controlled quantum operation $g$,  applied in parallel, hence each consuming  one copy of the initially available classical data of type $X$.  After this joint process the available classical data is now of type $Y$.
\eit

\paragraph{Example.}
The ability to vary classical data types in particular enables \em erasure \em of classical data.  For example, erasure of all classical data is just the cartesian map to $I$ as the final object of $\CCc_f$
\[
\langle \top_X, \id_A\rangle:\langle X,A\rangle\to \langle I,A\rangle
\]
while the erasure of part of it is just the cartesian projection
\[
\langle \id_X\times\top_Y, \id_A\rangle:\langle X\times Y,A\rangle\to \langle X,A\rangle\,.
\]

\paragraph{Example.}
We can write down pairs consisting of measurement data obtained in a measurement and the corresponding $\langle bra|$ e.g. 
\[
\bigl\langle +_X, \langle+_X| \bigr\rangle:\langle I,Q\rangle\to \langle X,I\rangle
\]
for a $+$-outcome in a destructive measurement of a qubit along the $X$-axis. Somewhat more involved is the following example taken from one-way quantum computing \cite{RBB}.  We can represent consecutive measurements on the first three qubits of a four qubit cluster state followed by an operation $Q^*XYZQ\tto{f}Q^*Q$ which performs an operation
on the last qubit which depends on all the measured data:
 \[
\xymatrix@=0.44in{ 
\langle I,Q\rangle\\
\langle XYZ,Q\rangle\ar[u]^{\bigl\langle \top_{XYZ}, f\bigr\rangle}\\
\langle XY,QQ\rangle\ar[u]^{\bigl\langle XY-_Z, \langle-_Z| Q \bigr\rangle}\\
\langle X,QQQ\rangle\ar[u]^{\bigl\langle X+_Y, \langle+_Y| QQ \bigr\rangle}\\
\langle I,QQQQ\rangle\ar[u]^{\bigl\langle +_X, \langle+_X| QQQ \bigr\rangle} \\
\langle I,I\rangle\ar[u]^{\langle I,\Psi_{cluster}\rangle} 
}
\] 
where we can rely on the categorical axiomatisation of unbiased observables in \cite{Coecke-Duncan,Coecke-Paquette-Perdrix} to provide explicit categorical semantics for the $X$-, $Y$- and $Z$-measurements. In a picture we have:
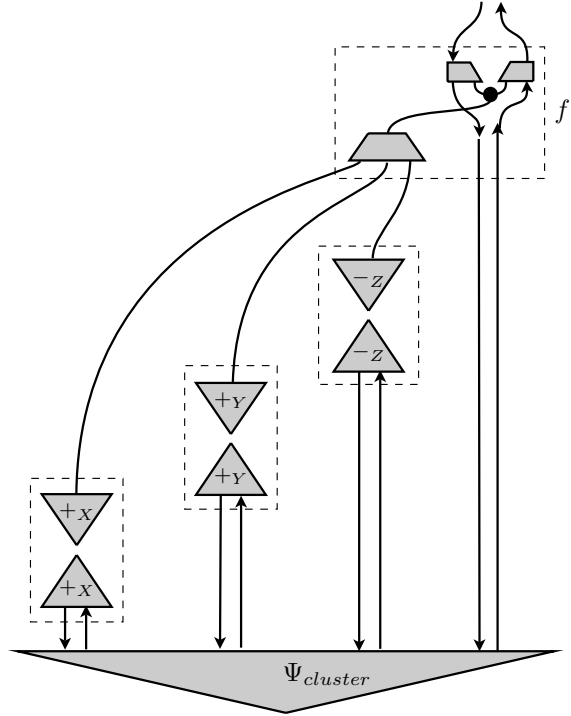
\begin{center}
\ifx\JPicScale\undefined\def\JPicScale{1}\fi
\psset{unit=\JPicScale mm}
\psset{linewidth=0.3,dotsep=1,hatchwidth=0.3,hatchsep=1.5,shadowsize=1,dimen=middle}
\psset{dotsize=0.7 2.5,dotscale=1 1,fillcolor=black}
\psset{arrowsize=1 2,arrowlength=1,arrowinset=0.25,tbarsize=0.7 5,bracketlength=0.15,rbracketlength=0.15}
\begin{pspicture}(0,0)(75.75,95.31)
\rput{90}(63.19,82.99){\psellipse[linestyle=none,fillstyle=solid](0,0)(1.03,0.99)}
\psbezier(63.55,82.91)(65.27,83.26)(65.27,83.95)(65.27,84.65)
\psbezier(62.74,82.91)(61.11,83.26)(61.11,83.95)(61.11,84.65)
\newrgbcolor{userFillColour}{0.8 0.8 0.8}
\psline[fillcolor=userFillColour,fillstyle=solid](68.9,84.76)
(68.9,87.3)
(65.86,87.3)
(64.34,84.76)(68.9,84.76)
\newrgbcolor{userFillColour}{0.8 0.8 0.8}
\psline[fillcolor=userFillColour,fillstyle=solid](57.5,84.69)
(57.5,87.23)
(60.53,87.23)
(62.04,84.69)(57.5,84.69)
\psline{<-}(61.75,8.9)(61.65,77.05)
\psline{->}(64.15,8.9)(64.15,79.2)
\psbezier(47.55,61.2)(47.62,64.94)(52.48,67.3)(52.55,74.5)
\psline{<-}(45.75,8.9)(45.75,46.1)
\psline{->}(48.55,9.22)(48.55,46.3)
\pspolygon[linewidth=0.15,linestyle=dashed,dash=1 1](42.55,89.31)(70.4,89.31)(70.4,71.8)(42.55,71.8)
\psbezier{<-}(61.75,77.3)(61.68,80.85)(58.42,80.07)(58.2,84.61)
\psbezier{->}(64.45,78.46)(64.45,83)(68.2,80.68)(68.01,84.81)
\psbezier{->}(62.01,95.31)(61.94,91.6)(58.43,92.08)(58.2,87.31)
\psbezier{<-}(64.6,95.31)(64.6,89.97)(68.29,92.29)(68.1,87.41)
\newrgbcolor{userFillColour}{0.8 0.8 0.8}
\psline[fillcolor=userFillColour,fillstyle=solid](36,0.7)
(71.65,8.9)
(0.35,8.9)(36,0.7)
\rput(41.5,5.9){$\Psi_{cluster}$}
\newrgbcolor{userFillColour}{0.8 0.8 0.8}
\psline[fillcolor=userFillColour,fillstyle=solid](28.7,37.8)
(33.35,44.6)
(24.05,44.6)(28.7,37.8)
\newrgbcolor{userFillColour}{0.8 0.8 0.8}
\psline[fillcolor=userFillColour,fillstyle=solid](28.7,36.6)
(33.35,29.7)
(24.05,29.7)(28.7,36.6)
{\footnotesize\rput(28.75,42.3){$+_Y$}}
{\footnotesize\rput(28.75,32.2){$+_Y$}}
\pspolygon[linewidth=0.15,linestyle=dashed,dash=1 1](22.55,46.9)(34.85,46.9)(34.85,27.8)(22.55,27.8)
\newrgbcolor{userFillColour}{0.8 0.8 0.8}
\psline[fillcolor=userFillColour,fillstyle=solid](8.2,23)
(12.85,29.8)
(3.55,29.8)(8.2,23)
\newrgbcolor{userFillColour}{0.8 0.8 0.8}
\psline[fillcolor=userFillColour,fillstyle=solid](8.2,21.7)
(12.85,14.8)
(3.55,14.8)(8.2,21.7)
{\footnotesize\rput(8.25,27.5){$+_X$}}
{\footnotesize\rput(8.25,17.3){$+_X$}}
\pspolygon[linewidth=0.15,linestyle=dashed,dash=1 1](2.05,31.9)(14.35,31.9)(14.35,12.8)(2.05,12.8)
\newrgbcolor{userFillColour}{0.8 0.8 0.8}
\psline[fillcolor=userFillColour,fillstyle=solid](47,54.2)
(51.65,61)
(42.35,61)(47,54.2)
\newrgbcolor{userFillColour}{0.8 0.8 0.8}
\psline[fillcolor=userFillColour,fillstyle=solid](47,53)
(51.65,46.1)
(42.35,46.1)(47,53)
{\footnotesize\rput(47.05,58.5){$-_Z$}}
{\footnotesize\rput(47.05,48.6){$-_Z$}}
\pspolygon[linewidth=0.15,linestyle=dashed,dash=1 1](40.35,44.4)(53.65,44.4)(53.65,62.8)(40.35,62.8)
\newrgbcolor{userFillColour}{0.8 0.8 0.8}
\psline[fillcolor=userFillColour,fillstyle=solid](46.95,77.8)
(51.95,77.8)
(54.45,74.2)
(44.45,74.2)(46.95,77.8)
\psbezier(28.95,44.6)(29.65,68.16)(49.29,68.88)(49.45,73.9)
\psbezier(8.15,29.8)(8.55,67.3)(45.85,72.3)(45.85,74.1)
\psbezier(49.55,77.7)(49.85,81.8)(63.15,79.9)(63.35,82.5)
\psline{<-}(27.25,9.2)(27.35,29.6)
\psline{->}(30.05,9.38)(30.05,29.6)
\psline{<-}(6.65,9.1)(6.65,14.82)
\psline{->}(9.45,9.15)(9.45,15)
\rput(72.75,80.7){$f$}
\end{pspicture}

\end{center}
where the bra's and $\Psi_{cluster}$ each in fact correspond to two triangles:
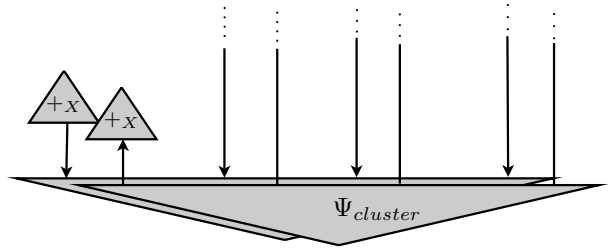
\begin{center}
\ifx\JPicScale\undefined\def\JPicScale{1}\fi
\psset{unit=\JPicScale mm}
\psset{linewidth=0.3,dotsep=1,hatchwidth=0.3,hatchsep=1.5,shadowsize=1,dimen=middle}
\psset{dotsize=0.7 2.5,dotscale=1 1,fillcolor=black}
\psset{arrowsize=1 2,arrowlength=1,arrowinset=0.25,tbarsize=0.7 5,bracketlength=0.15,rbracketlength=0.15}
\begin{pspicture}(0,0)(78.07,32.8)
\newrgbcolor{userFillColour}{0.8 0.8 0.8}
\psline[fillcolor=userFillColour,fillstyle=solid](36.32,1.4)
(71.97,9.6)
(0.67,9.6)(36.32,1.4)
\psline(72.12,8.82)(72.12,27.6)
\psline(51.62,8.9)(51.62,27.45)
\psline(35.32,8.7)(35.32,26.84)
\psline{<-}(7.27,9.6)(7.42,16.9)
\newrgbcolor{userFillColour}{0.8 0.8 0.8}
\psline[fillcolor=userFillColour,fillstyle=solid](6.97,23.9)
(11.62,17)
(2.32,17)(6.97,23.9)
{\footnotesize\rput(7.02,19.5){$+_X$}}
\psline{<-}(66.02,9.79)(65.92,28.51)
\psline{<-}(45.87,9.75)(45.87,28.3)
\newrgbcolor{userFillColour}{0.8 0.8 0.8}
\psline[fillcolor=userFillColour,fillstyle=solid](43.5,0.7)
(78.07,8.7)
(8.92,8.7)(43.5,0.7)
\rput(49.2,5.5){$\Psi_{cluster}$ }
\psline{<-}(28.3,9.69)(28.3,26.9)
\psline{->}(14.82,8.7)(14.82,14.82)
\newrgbcolor{userFillColour}{0.8 0.8 0.8}
\psline[fillcolor=userFillColour,fillstyle=solid](14.62,21.7)
(19.27,14.8)
(9.97,14.8)(14.62,21.7)
{\footnotesize\rput(14.67,17.3){$+_X$}}
\psline[linestyle=dotted](72.12,28.09)(72.12,31.6)
\psline[linestyle=dotted](65.97,29.29)(65.97,32.8)
\psline[linestyle=dotted](51.62,28.41)(51.62,31.55)
\psline[linestyle=dotted](45.92,28.97)(45.92,32.12)
\psline[linestyle=dotted](28.32,28.38)(28.32,32.4)
\psline[linestyle=dotted](35.32,27.76)(35.32,31.78)
\end{pspicture}
\qquad\qquad\qquad
\end{center}

\section{Embedding measurements and controls,\\ limitations and final remarks}\label{Challenge}

Unfortunately, due to Proposition \ref{prop:Groth} which prohibits probabilistic classical data operations in ${\See\CCc}$, we cannot represent stochastic variables within $\See\CCc$.  For this reason,  we also cannot represent the quantum measurements except for in post-selected fashion as we did above.  More concretely, measurement of classical data type $Y$ `would be' a composite
\[
\xymatrix@=1.58in{
A^*X A\ar[r]^{\langle \id_{Y\otimes X},m\rangle\circ_{\See\CCc} \langle \bot_Y\otimes \id_X, \id_A\rangle\ \ }& A^*X Y A
}
\]
where $m\in\Cee\CCc_Y(A,A)$ corresponds via  
\[
\Wee\CCc^Y\simeq\ \Wee\CCc_Y\inclusion \Cee\CCc_Y
\]
to a pure quantum measurement in the sense of Definition \ref{def:PureMeasurement}, if it wasn't for the fact that \em $\bot_Y:I\to Y$ is not a function\em. Indeed:
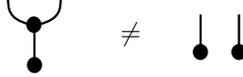
\begin{center}
\ifx\JPicScale\undefined\def\JPicScale{1}\fi
\psset{unit=\JPicScale mm}
\psset{linewidth=0.3,dotsep=1,hatchwidth=0.3,hatchsep=1.5,shadowsize=1,dimen=middle}
\psset{dotsize=0.7 2.5,dotscale=1 1,fillcolor=black}
\psset{arrowsize=1 2,arrowlength=1,arrowinset=0.25,tbarsize=0.7 5,bracketlength=0.15,rbracketlength=0.15}
\begin{pspicture}(0,0)(34.71,11.41)
\rput{90}(6.44,7.87){\psellipse[linestyle=none,fillstyle=solid](0,0)(1.1,-1.01)}
\psbezier(5.8,7.89)(3,8.59)(3,10)(3,11.41)
\psbezier(7.2,7.89)(10,8.59)(10,10)(10,11.41)
\rput{90}(6.51,2.48){\psellipse[linestyle=none,fillstyle=solid](0,0)(1.07,1.01)}
\psline(6.51,7.47)(6.51,2.58)
\rput(19.3,6.6){$\not=$}
\rput{90}(28.6,4.2){\psellipse[linestyle=none,fillstyle=solid](0,0)(1.07,1.01)}
\psline(28.6,9.18)(28.6,4.3)
\rput{90}(33.7,4.22){\psellipse[linestyle=none,fillstyle=solid](0,0)(1.07,1.01)}
\psline(33.7,9.2)(33.7,4.32)
\end{pspicture}

\end{center}

\subsection{Classicality = decoherence}

The characterisation in Proposition \ref{posIfCpos} of classical information flows in terms of decoherence indicates a type-wise distinction between classical and quantum data flows by requiring the former to be decoherent.  In other words: we will characterise classical data by the fact that it is invariant under the application of corresponding decoherences. 

\be{prop}\label{posIfCposBis}
For a morphism $f:A^*XA\to B^*YB$ where $X$ and $Y$ are classical structures the following are equivalent\,{\rm:}
\bit
\item $f_\Xi^{X,Y}=A^*XXA\tto{A^*\nabla A}A^*XA\tto{f}B^*YB\tto{B^*\Delta B}B^*YYB$
is a completely positive map\,{\rm;}
\item  
$f=g_\classsub^{X,Y}\!\!=A^*XA\tto{A^*\Delta A}A^*XXA\tto{g}B^*YYB\tto{B^*\nabla B}B^*YB$ where $g:A^*XXA\to B^*YYB$ is a  completely positive map which is `decoherent in the classical data types', that is,
\[
(B^*\,\Xi_Y B)\circ g=g\circ (A^*\,\Xi_X A)= g\,.
\]
\eit
Moreover, there is an identity-on-objects isomorphism of categories{\rm:}  
\[
\xymatrix@=0.64in{
\CCc_{\classsub q} \ar@/^0.8em/[r]^{(-)_\Xi} & \CCc_{\Xi q} \ar@/^0.8em/[l]^{(-)_\classsub}
}
\]
where the categories $\CCc_{\classsub q}$ and $\CCc_{\Xi q}$ both have the same objects as $\See\CCc$
and where the morphisms of type $\langle X, A\rangle\to\langle Y, B\rangle$ are respectively those of form   $g_\classsub^{X,Y}$ and $f_\Xi^{X,Y}$ subject to the conditions stipulated above.
\ee{prop}

These morphisms are graphically represented by:
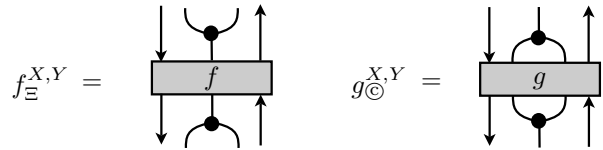
\begin{center}
\ifx\JPicScale\undefined\def\JPicScale{1}\fi
\psset{unit=\JPicScale mm}
\psset{linewidth=0.3,dotsep=1,hatchwidth=0.3,hatchsep=1.5,shadowsize=1,dimen=middle}
\psset{dotsize=0.7 2.5,dotscale=1 1,fillcolor=black}
\psset{arrowsize=1 2,arrowlength=1,arrowinset=0.25,tbarsize=0.7 5,bracketlength=0.15,rbracketlength=0.15}
\begin{pspicture}(0,0)(77.06,19.5)
\psline(25.53,3.98)(25.52,7.7)
\psline{<-}(62.36,11.6)(62.36,19.11)
\psline{->}(75.56,11.81)(75.56,19.32)
\newrgbcolor{userFillColour}{0.8 0.8 0.8}
\pspolygon[fillcolor=userFillColour,fillstyle=solid](61.14,11.7)(77.06,11.7)(77.06,7.32)(61.14,7.32)
\rput(69.06,9.31){$g$}
\psline{<-}(62.4,0.64)(62.4,7.41)
\psline{->}(75.6,0.52)(75.6,7.29)
\rput{90}(68.94,4.01){\psellipse[linestyle=none,fillstyle=solid](0,0)(1.04,-1.01)}
\psbezier(68.3,4.03)(65.5,4.69)(65.5,6.03)(65.5,7.36)
\psbezier(69.7,4.03)(72.5,4.69)(72.5,6.03)(72.5,7.36)
\psline(69.01,3.64)(69.04,0.42)
\rput{-0}(68.89,15.04){\psellipse[linestyle=none,fillstyle=solid](0,0)(1.01,1.01)}
\psbezier(68.24,15.02)(65.44,14.38)(65.44,13.08)(65.44,11.79)
\psbezier(69.64,15.02)(72.44,14.38)(72.44,13.08)(72.44,11.79)
\psline(68.95,15.4)(68.94,19.12)
\rput(2.66,8.88){$f_{\Xi}^{X,Y}$}
\rput(9.56,8.88){=}
\psline{<-}(18.72,11.78)(18.72,19.29)
\psline{->}(31.92,11.99)(31.92,19.5)
\newrgbcolor{userFillColour}{0.8 0.8 0.8}
\pspolygon[fillcolor=userFillColour,fillstyle=solid](17.5,11.88)(33.42,11.88)(33.42,7.5)(17.5,7.5)
\rput(25.46,9.68){$f$}
\psline{<-}(18.76,0.82)(18.76,7.59)
\psline{->}(31.96,0.7)(31.96,7.47)
\rput{90}(25.37,15.62){\psellipse[linestyle=none,fillstyle=solid](0,0)(1.04,-1.01)}
\psbezier(24.72,15.64)(21.92,16.3)(21.92,17.63)(21.92,18.97)
\psbezier(26.12,15.64)(28.92,16.3)(28.92,17.63)(28.92,18.97)
\psline(25.43,15.24)(25.47,12.03)
\rput{-0}(25.47,3.62){\psellipse[linestyle=none,fillstyle=solid](0,0)(1.01,1.01)}
\psbezier(24.82,3.6)(22.02,2.96)(22.02,1.66)(22.02,0.37)
\psbezier(26.22,3.6)(29.02,2.96)(29.02,1.66)(29.02,0.37)
\rput(47.76,8.97){$g_\classsub^{X,Y}$}
\rput(54.66,8.97){=}
\end{pspicture}

\end{center}

\be{prop}
The category $\CCc_{\classsub q}$ is dagger compact.
\ee{prop}

The category $\See\CCc$ canonically embeds in $\CCc_{\classsub q}$ via the 
identity-on-objects-functor
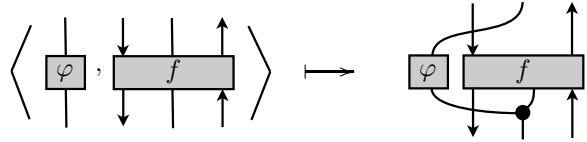
\begin{center}
\ifx\JPicScale\undefined\def\JPicScale{1}\fi
\psset{unit=\JPicScale mm}
\psset{linewidth=0.3,dotsep=1,hatchwidth=0.3,hatchsep=1.5,shadowsize=1,dimen=middle}
\psset{dotsize=0.7 2.5,dotscale=1 1,fillcolor=black}
\psset{arrowsize=1 2,arrowlength=1,arrowinset=0.25,tbarsize=0.7 5,bracketlength=0.15,rbracketlength=0.15}
\begin{pspicture}(0,0)(78.02,19.3)
\rput{90}(69.98,4.59){\psellipse[linestyle=none,fillstyle=solid](0,0)(1.03,1)}
\psbezier(69.24,4.6)(71.46,5.27)(71.46,6.58)(71.46,7.9)
\psline(69.98,1)(69.98,4.35)
\psline{<-}(63.32,12.08)(63.32,19.1)
\psline{->}(76.52,12.28)(76.52,19.3)
\newrgbcolor{userFillColour}{0.8 0.8 0.8}
\pspolygon[fillcolor=userFillColour,fillstyle=solid](62.1,12.18)(78.02,12.18)(78.02,7.8)(62.1,7.8)
\rput(70,10.1){$f$}
\newrgbcolor{userFillColour}{0.8 0.8 0.8}
\pspolygon[fillcolor=userFillColour,fillstyle=solid](54.9,12.18)(60.02,12.18)(60.02,7.8)(54.9,7.8)
\rput(57.36,9.88){$\varphi$}
\psbezier(57.96,12.2)(58.2,16.8)(69.76,13.99)(69.96,19.29)
\psline{<-}(63.36,1.31)(63.36,7.89)
\psline{->}(76.56,1.19)(76.56,7.77)
\psbezier(69.34,4.5)(65.86,4.48)(57.86,5.08)(57.86,7.78)
\psline(23.48,2.5)(23.4,7.72)
\psline{<-}(16.82,11.98)(16.82,17.16)
\psline{->}(30.02,12.12)(30.02,17.3)
\newrgbcolor{userFillColour}{0.8 0.8 0.8}
\pspolygon[fillcolor=userFillColour,fillstyle=solid](15.6,12.08)(31.52,12.08)(31.52,7.7)(15.6,7.7)
\rput(23.5,10){$f$}
\newrgbcolor{userFillColour}{0.8 0.8 0.8}
\pspolygon[fillcolor=userFillColour,fillstyle=solid](6.64,12.1)(11.76,12.1)(11.76,7.72)(6.64,7.72)
\rput(9.1,9.8){$\varphi$}
\psline{<-}(16.86,2.74)(16.86,7.79)
\psline{->}(30.06,2.65)(30.06,7.7)
\psline(23.4,12.14)(23.32,17.15)
\psline(9.3,2.58)(9.22,7.8)
\psline(9.22,12.22)(9.14,17.23)
\rput(13.6,10){,}
\psline(33.9,3.6)(36.4,10)
\psline(36.4,10)(33.5,16.09)
\psline(4.41,3.7)(2,10.1)
\psline(2,10.11)(4.6,16.2)
\rput(44.3,10){$\mapsto$}
\end{pspicture}

\end{center}
Probabilistic operations $\Cclas$ also live in $\CCc_{\classsub q}$ via the functor
\[
\Cclas\to \CCc_{\classsub q}::
\left\{\begin{array}{l}
X\mapsto \langle X, I\rangle\\
\varphi\mapsto \id_I\otimes\varphi\otimes \id_I
\end{array}\right.
\]
and one easily verifies that all measurements and any other operations defined in this paper all live within $\CCc_{\classsub q}$.  

\paragraph{Example.} 
In the previous example we can now replace the `conditional' measurements by genuine measurements.  We can also represent  controlled measurements which depend on outcomes of other measurements:
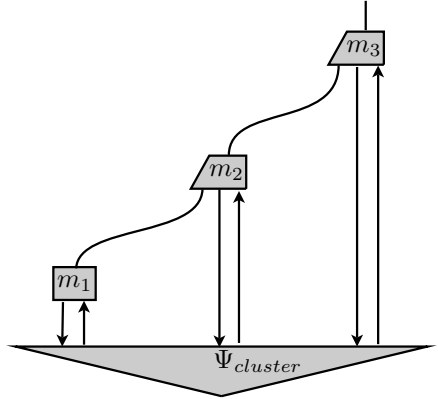
\begin{center}
\ifx\JPicScale\undefined\def\JPicScale{1}\fi
\psset{unit=\JPicScale mm}
\psset{linewidth=0.3,dotsep=1,hatchwidth=0.3,hatchsep=1.5,shadowsize=1,dimen=middle}
\psset{dotsize=0.7 2.5,dotscale=1 1,fillcolor=black}
\psset{arrowsize=1 2,arrowlength=1,arrowinset=0.25,tbarsize=0.7 5,bracketlength=0.15,rbracketlength=0.15}
\begin{pspicture}(0,0)(55.95,53.4)
\newrgbcolor{userFillColour}{0.8 0.8 0.8}
\psline[fillcolor=userFillColour,fillstyle=solid](31.86,28.4)
(31.86,32.8)
(26.95,32.8)
(24.5,28.4)(31.86,28.4)
\psline{<-}(46.6,7.4)(46.6,44.6)
\psline{->}(49.4,7.72)(49.4,44.8)
\newrgbcolor{userFillColour}{0.8 0.8 0.8}
\psline[fillcolor=userFillColour,fillstyle=solid](28.57,0.8)
(55.95,7.4)
(1.2,7.4)(28.57,0.8)
\rput(34,5.5){$\Psi_{cluster}$ }
\psline{<-}(28.3,7.3)(28.3,28.3)
\psline{->}(30.9,7.88)(30.9,28.1)
\psline{<-}(7.4,7.4)(7.5,13.32)
\psline{->}(10.3,7.65)(10.3,13.5)
\newrgbcolor{userFillColour}{0.8 0.8 0.8}
\pspolygon[fillcolor=userFillColour,fillstyle=solid](6.22,17.97)(11.85,17.97)(11.85,13.6)(6.22,13.6)
\rput(9.05,15.8){$m_1$}
\rput(29.25,30.6){$m_2$}
\newrgbcolor{userFillColour}{0.8 0.8 0.8}
\psline[fillcolor=userFillColour,fillstyle=solid](50.16,44.9)
(50.16,49.3)
(45.25,49.3)
(42.8,44.9)(50.16,44.9)
\rput(47.55,47.1){$m_3$}
\psbezier(29.55,32.9)(29.65,39.8)(44.05,36.3)(44.15,44.8)
\psbezier(9.25,17.8)(9.55,23.7)(25.75,21.8)(26.05,28.3)
\psline(47.75,49.5)(47.75,53.4)
\end{pspicture}

\end{center}
as well as the measurements which depend on a coin-toss: 
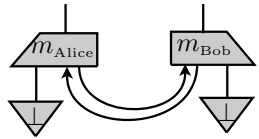
\begin{center}
\ifx\JPicScale\undefined\def\JPicScale{1}\fi
\psset{unit=\JPicScale mm}
\psset{linewidth=0.3,dotsep=1,hatchwidth=0.3,hatchsep=1.5,shadowsize=1,dimen=middle}
\psset{dotsize=0.7 2.5,dotscale=1 1,fillcolor=black}
\psset{arrowsize=1 2,arrowlength=1,arrowinset=0.25,tbarsize=0.7 5,bracketlength=0.15,rbracketlength=0.15}
\begin{pspicture}(0,0)(34.91,18.8)
\psline(31.3,5.2)(31.3,10.92)
\newrgbcolor{userFillColour}{0.8 0.8 0.8}
\psline[fillcolor=userFillColour,fillstyle=solid](31.22,1.7)
(34.7,6.3)
(27.75,6.3)(31.22,1.7)
\rput(31.2,4.4){$\bot$}
\psline(5.9,4.7)(5.9,10.42)
\newrgbcolor{userFillColour}{0.8 0.8 0.8}
\psline[fillcolor=userFillColour,fillstyle=solid](14.01,10.5)
(14.01,14.9)
(6.4,14.9)
(2.6,10.5)(14.01,10.5)
\newrgbcolor{userFillColour}{0.8 0.8 0.8}
\psline[fillcolor=userFillColour,fillstyle=solid](5.82,1.2)
(9.3,5.8)
(2.35,5.8)(5.82,1.2)
\rput(9.4,12.8){$m_{\mbox{\tiny Alice}}$}
\psbezier{->}(11.6,10.5)(11.6,3)(25.7,3.1)(25.7,10.8)
\rput(5.7,3.9){$\bot$}
\newrgbcolor{userFillColour}{0.8 0.8 0.8}
\psline[fillcolor=userFillColour,fillstyle=solid](23.82,10.7)
(23.82,15.1)
(31.22,15.1)
(34.91,10.7)(23.82,10.7)
\rput(28.3,13){$m_{\mbox{\tiny Bob}}$}
\psline(28.1,14.9)(28.1,18.8)
\psline(9.8,14.9)(9.8,18.8)
\psbezier{<-}(10,10.43)(10,1.1)(27.3,1.22)(27.3,10.8)
\end{pspicture}

\end{center}
that are used in quantum key distribution \cite{Ekert}.

\subsection{${\mbox{\large ?} \over \mbox{${\See\CCc}$}}={\mbox{broadcasting} \over \mbox{cloning}}$}

If the ultimate goal of categorical quantum mechanics is a full categorical description of the interaction between classical and quantum information flows, then the category $\See\CCc$ built in this paper is not big enough. Indeed, it does not capture probabilistic operations on classical data as well as on quantum measurements.  On the other hand,  $\CCc_{\classsub q}$ is `not satisfactory' for a number of reasons, including:
\bit
\item $\CCc_{\classsub q}$ does not encode the properties of classical data flow in a structural manner in a similar manner  in which co-Kleilsli  composition in $\See\CCc$ embodies clone-ability and delete-ability of classical data.
\item $\CCc_{\classsub q}$ lacks the clear separation between classical operations and classically controlled quantum operations of $\See\CCc$. 
\eit

So why do the constructions in Section \ref{Relativizing} prohibit probabilistic classical data?  The answer is simply that while deterministic classical data can be cloned, probabilistic classical data can't. Moreover, the composition mechanism underlying Kleisli indexing --- and consequently the interaction between classical and quantum data in $\See\CCc$ --- allows us to distinguish classical data from quantum data because only the former can be cloned. 

If we want to separate probabilistic classical data from quantum data we should not appeal to their distinct behaviour under cloning but to their distinct behaviour under \em broadcasting \em \cite{Broadcast}.  Solving the equation in the title of this section remains an open challenge. 



\end{document}